\documentclass[structuredabstract]{aa}
\usepackage{graphicx}
\usepackage{psfrag}
\usepackage{natbib}
\usepackage{txfonts}
\usepackage{longtable}

\begin{document}
\def\lsun{L_{\sun}}
\def\msun{M_{\sun}}
\def\scana{{\tt Scanamorphos} }
\title{The Earliest Phases of Star Formation (EPoS): \\
A Herschel\thanks{\emph{Herschel} is 
an ESA space observatory with science instruments provided by European-led
 Principal Investigator consortia and with important participation from NASA.} key program}
\subtitle{The precursors to high mass stars and clusters}

\author{
S. Ragan\inst{\ref{mpia}}, 
Th. Henning\inst{\ref{mpia}}, 
O. Krause\inst{\ref{mpia}}, 
J. Pitann\inst{\ref{mpia}}, 
H. Beuther\inst{\ref{mpia}}, 
H. Linz\inst{\ref{mpia}}, 
J. Tackenberg\inst{\ref{mpia}}, 
Z. Balog\inst{\ref{mpia}}, 
M. Hennemann\inst{\ref{saclay}}, 
R. Launhardt\inst{\ref{mpia}}, 
N. Lippok\inst{\ref{mpia}}, 
M. Nielbock\inst{\ref{mpia}}, 
A. Schmiedeke\inst{\ref{mpia},\ref{unikoln}}, 
F. Schuller\inst{\ref{mpifr}}, 
J. Steinacker\inst{\ref{ipag},\ref{mpia}}, 
A. Stutz\inst{\ref{mpia}}, 
T. Vasyunina\inst{\ref{uva}}
}

\authorrunning{S. Ragan et al.} 
\titlerunning{Herschel EPoS: high-mass overview}
 
\institute{
Max-Planck-Institute for Astronomy, K\"onigstuhl 17, D-69117 Heidelberg, Germany, 
\email{ragan@mpia.de} \label{mpia}
\and
AIM Paris-Saclay, CEA/DSM/IRFU -- CNRS/INSU -- Universit\'e Paris Diderot, CEA Saclay, 91191 Gif-sur-Yvette cedex, France \label{saclay}
\and
Universit\"at zu K\"oln, Z\"ulpicher Strasse 77, 50937 K\"oln, Germany \label{unikoln}
\and
Max-Planck-Institut f\"ur Radioastronomie, Auf dem H\"ugel 69, D-53121 Bonn, Germany \label{mpifr}
\and
Institut de Plan{\'e}tologie et d$'$Astrophysique de Grenoble, 414 Rue de la Piscine, 38400 St-Martin d$'$H{\`e}res, France \label{ipag}
\and
Department of Chemistry, Astronomy and Physics, University of Virginia, Charlottesville, VA 22904, USA \label{uva}
}
 
\date{Received 16 March 2012 / Accepted 27 July 2012}

\abstract
{Stars are born deeply embedded in molecular clouds. In the earliest embedded phases, protostars emit the bulk of their radiation in the far-infrared wavelength range, where Herschel is perfectly suited to probe at high angular resolution and dynamic range. In the high-mass regime, the birthplaces of protostars are thought to be in the high-density structures known as infrared-dark clouds (IRDCs). While massive IRDCs are believed to have the right conditions to give rise to massive stars and clusters, the evolutionary sequence of this process is not well-characterized.}
{As part of the Earliest Phases of Star formation (EPoS) Herschel Guaranteed Time Key Program, we isolate the embedded structures within IRDCs and other cold, massive molecular clouds. We present the full sample of 45 high-mass regions which were mapped at PACS 70, 100, and 160\,$\mu$m and SPIRE 250, 350, and 500\,$\mu$m. In the present paper, we characterize a population of cores which appear in the PACS bands and place them into context with their host molecular cloud and investigate their evolutionary stage.}
{We construct spectral energy distributions (SEDs) of 496 cores which appear in all PACS bands, 34\% of which lack counterparts at 24\,$\mu$m. From single-temperature modified blackbody fits of the SEDs, we derive the temperature, luminosity, and mass of each core. These properties predominantly reflect the conditions in the cold, outer regions. Taking into account optical depth effects and performing simple radiative transfer models, we explore the origin of emission at PACS wavelengths.}
{The core population has a median temperature of 20\,K and has masses and luminosities that span four to five orders of magnitude. Cores with a counterpart at 24\,$\mu$m are warmer and bluer on average than cores without a 24\,$\mu$m counterpart. We conclude that cores bright at 24\,$\mu$m are on average more advanced in their evolution, where a central protostar(s) have heated the outer bulk of the core, than 24\,$\mu$m-dark cores.
The 24\,$\mu$m emission itself can arise in instances where our line of sight aligns with an exposed part of the warm inner core.  About 10\% of the total cloud mass is found in a given cloud's core population. We uncover over 300 further candidate cores which are dark until 100\,$\mu$m. These are possibly starless objects, and further observations will help us determine the nature of these very cold cores.}
{}


\maketitle

\section{Introduction}
\label{s:intro}

\begin{figure*}
\includegraphics[width=\textwidth]{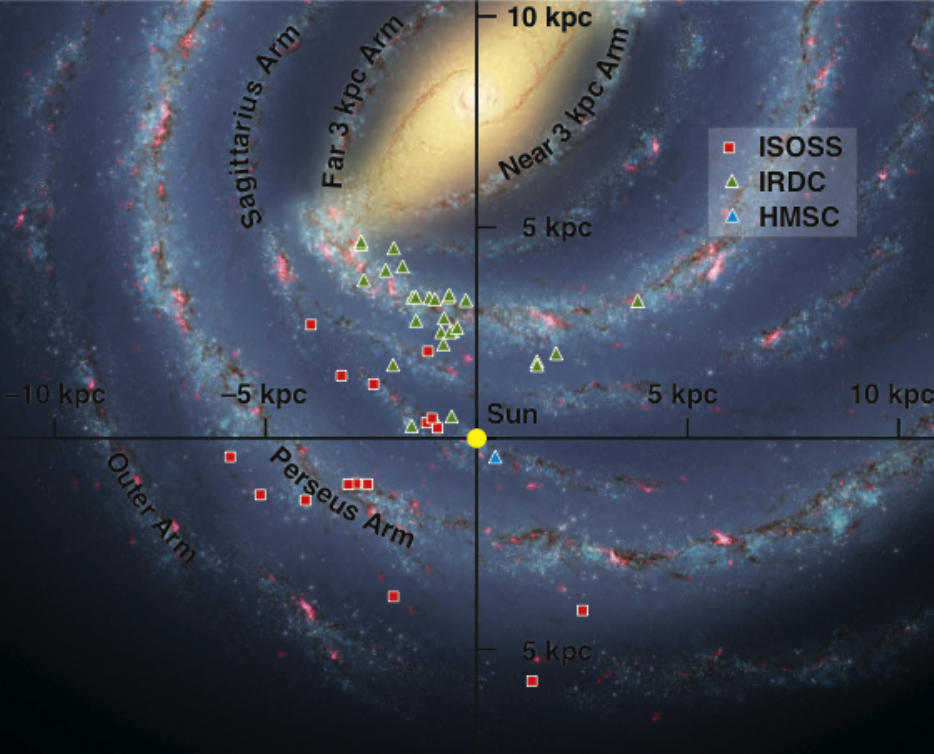}
\caption{Face-on schematic view of the distribution of IRDCs (green triangles) and ISOSS sources (red squares) and the HMSC 07029 (blue triangle) in the Milky Way. The background image is an artist's impression of the Milky Way based on the GLIMPSE survey, credit R. Hurt [SSC-Caltech], adapted by MPIA graphics department. The kinematic distances to each object are derived using the \citet{Reid2009} model.  \label{fig:gal_distrib2}}
\end{figure*}

Star formation is a critical ingredient in a broad range of astrophysical phenomena, yet there are fundamental components of the process -- particularly in the early stages -- that remain poorly understood. Over the past decades, a basic framework for the formation of low-mass stars has developed beginning with gravitationally bound pre-stellar cores \citep{WardThompson2002}, evolving into Class 0 and Class I protostars then Class II and Class III pre-main sequence stars \citep{ShuAdamsLizano1987,Andre1993}. Such a sequence for the formation of high-mass stars has not yet been established. Several good candidates for massive young cores have been identified using the 170\,$\mu$m ISOPHOT Serendipity Survey (ISOSS) \citep{Lemke1996,Krause2003,Krause2004,Birkmann2006} or sensitive millimeter surveys \citep[e.g.][]{Klein2005,Sridharan:2005}. However, upon further investigation, most have been found to already host (deeply embedded) low- to intermediate-mass protostars \citep[e.g.][]{Motte_cygX,Hennemann2008,BeutherHenning2009}. It is the stage previous to the onset of massive protostar formation that continues to elude observers.

Many gaps remain in our understanding of how massive stars and clusters form, beginning with the elusive initial conditions. Do massive stars result from the gravitational collapse of cold, very massive cores \citep{Evans2002,Beuther_ppv} like their low-mass siblings? What role does the environment play in determining the ultimate fate of such cores?  As a (massive) protostar evolves, how drastically do its properties change, and what impact does it have on its surroundings? In order to study these very early, embedded phases of (massive) star formation, access to the far-infrared wavelength regime, where the peak of the cold dust radiation ($T_{dust} \sim$ 10-20\,K) is located, is critical. In addition, high angular resolution is needed to study massive star-forming regions which usually reside several kiloparsecs from the Sun.  The {\em Herschel} far-infrared satellite \citep{A&ASpecialIssue-Herschel} drastically improves our ability to peer deep into the dense regions where such young cores are embedded.

The Earliest Phases of Star formation (EPoS) Guaranteed Time Key Program \citep[P.I. O. Krause; ][]{A&ASpecialIssue-Henning,A&ASpecialIssue-Linz,A&ASpecialIssue-Beuther,A&ASpecialIssue-Stutz} is a PACS and SPIRE photometric mapping survey which targets objects known to be in the cold early phases of star formation. There are two main components to the EPoS sample: 15 isolated, low-mass globules at various evolutionary stages, from starless to Class I, and 45 high-mass regions, which are mostly larger, high density molecular cloud complexes containing a range of objects within their boundaries. In \citet{A&ASpecialIssue-Stutz}, Nielbock et al. (2012, submitted), and a forthcoming comprehensive study by Launhardt et al. (in prep.), the low-mass part of the EPoS sample is investigated. In this overview, we focus on the high-mass star-forming regions.

Our sample is comprised of objects known as infrared-dark clouds (IRDCs), which were first discovered in silhouette against the bright Galactic background in the mid-infrared with the ISOCAM instrument \citep{Perault1996} and the MSX satellite \citep{egan_msx} at 15 and 8\,$\mu$m, respectively. Several surveys in millimeter and sub-millimeter continuum and spectral lines have followed \citep[e.g.][]{carey_msx, carey_submmIRDC, Teyssier2002, Pillai_ammonia, rathborne2006, ragan_msxsurv, Vasyunina2009} and have established that massive IRDCs harbor the precursors and early phases of massive star and cluster formation. We select 29 IRDCs, most of which are in the inner quadrants of the Galaxy (see Figure~\ref{fig:gal_distrib2}). Assuming the IRDCs lie on the near side of the Milky Way (see Section~\ref{ss:distance}), they coincide with the Scutum-Centraurus spiral arm, which (in the first quadrant) overlaps with the Molecular Ring \citep{Jackson_galdistr_IRDCs}. Also part of our sample are sources discovered in the ISO Serendipity Survey (ISOSS) at 170\,$\mu$m, which are seen to harbor cold, massive clumps at large Galactocentric distances. 

We present the first comprehensive results of the {\em Herschel} PACS and SPIRE imaging survey of 45 massive targets as part of the EPoS survey.  The goals of this study are as follows: (1) give a general characterization of the sample based on {\em Herschel} data in concert with existing complementary datasets; (2) characterize point sources embedded within the targeted clouds; and (3) connect the point source properties to the overall cloud structure and environment.  

\section{Sample description}
\label{s:sample}

\subsection{Target selection}
\label{ss:targets}

The high-mass portion of EPoS GT Key Program sample was compiled from several different surveys with varying approaches, but they are unified in that insofar as the existing data show, they are all candidates to be in the earliest stages of high-mass star formation.  Table \ref{tab:sample} summarizes our targets, with their positions, distances, and mass estimates from various Galactic plane survey data which we will detail below.  Appendix A describes the selection strategy and each of the individual sources in greater detail.   

\begin{figure}
\begin{center}
\includegraphics[scale=0.4,angle=90]{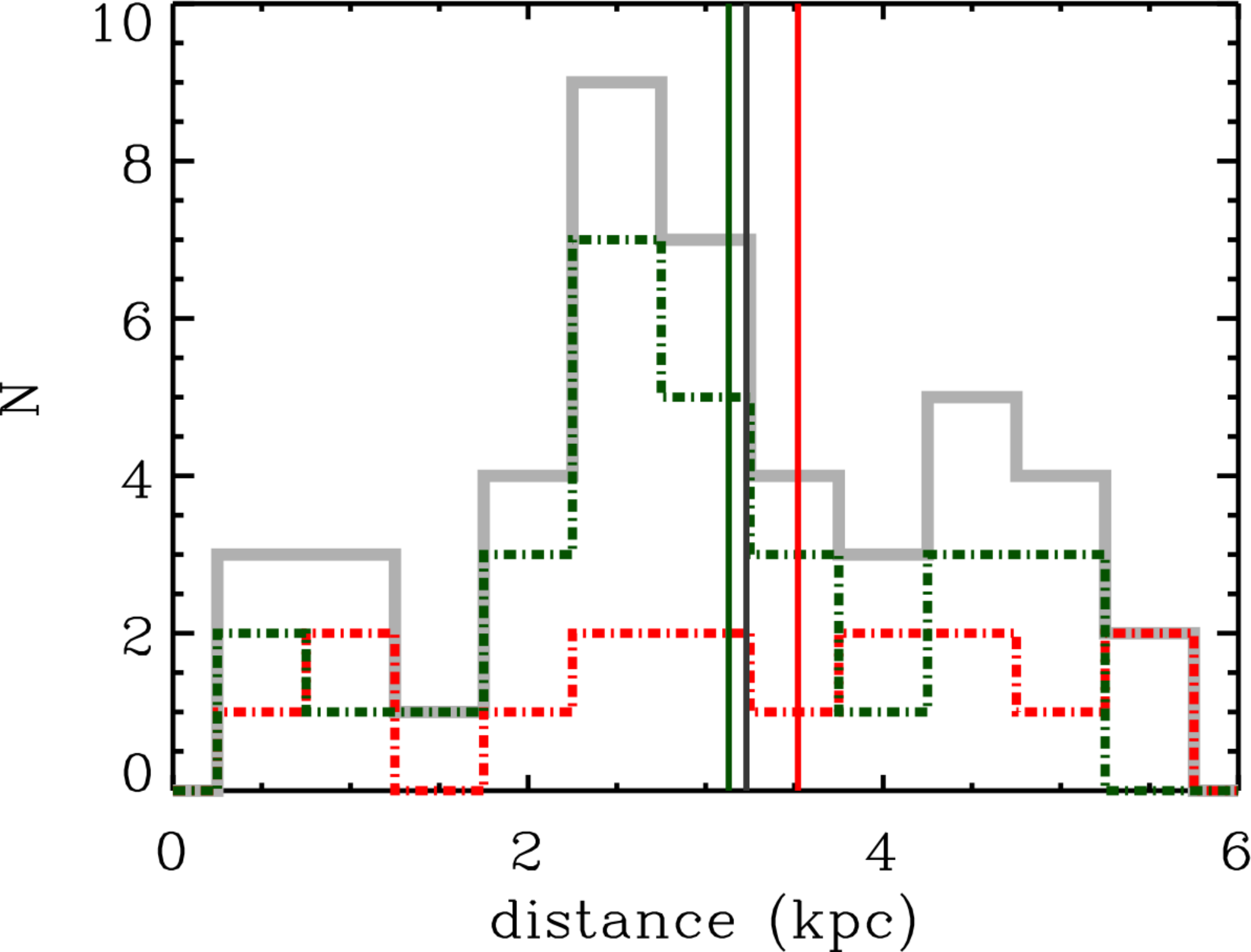}
\caption{Histogram of distances of the 45 clouds in the sample in 0.5\,kpc bins. The full sample is plotted in the solid gray histogram, the histogram for just the IRDCs is plotted in the green dot-dashed line, and the ISOSS targets are plotted in the red histogram. The median distance of the full sample, 3.2\,kpc, is plotted in the solid gray line, and the median for the IRDC-only (3.1\,kpc) and ISOSS-only (3.5\,kpc) are shown with solid green and red vertical lines. \label{fig:distances}}
\end{center}
\end{figure}

\begin{figure*}
\begin{center}
\includegraphics[scale=0.7,angle=90]{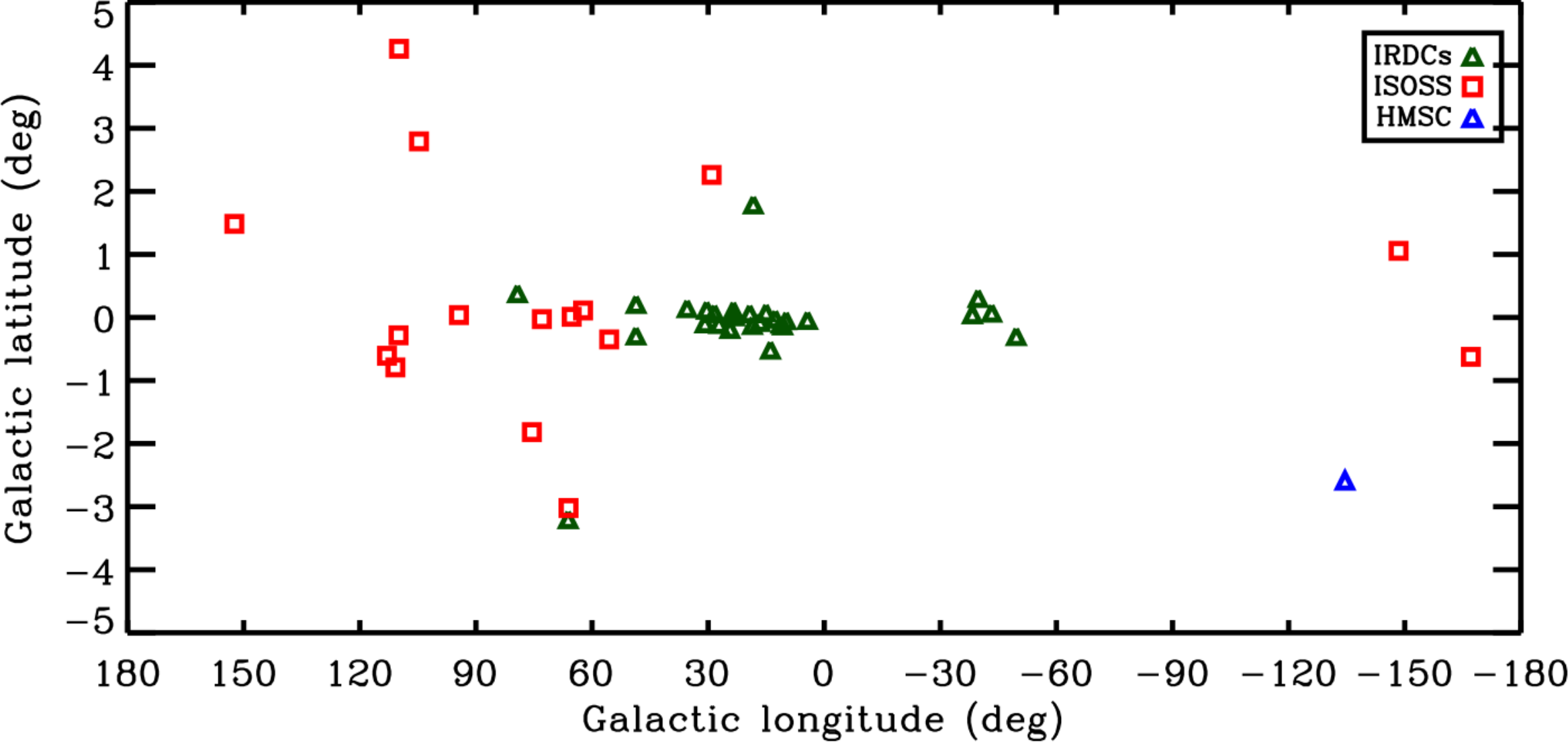}
\caption{Galactic distribution of EPoS high-mass sample in Galactic latitude and longitude.  Green triangles represent IRDCs and red squares represent ISOSS sources. The blue triangle is HMSC 07029. \label{fig:gal_distrib}}
\end{center}
\end{figure*}

\subsection{Distances and Galactic distribution}
\label{ss:distance}

We derive kinematic distances for each of the IRDCs following the method put forward in \citet{Reid2009}, adopting the updated distance to the Galactic Center of $R_0 = 8.4$\,kpc and a circular rotation speed of $\Theta_0 = 254$\,km s$^{-1}$. Table~\ref{tab:sample} lists the $v_{\rm lsr}$ that was used, the reference from which this velocity was obtained, and the resulting distances and uncertainties, assuming the recommended 7 km s$^{-1}$ uncertainty in $v_{\rm lsr}$, the source of which is listed in the References column of Table~\ref{tab:sample}.  We show the distribution of distances in Figure~\ref{fig:distances}. The overall median distance is 3.2\,kpc. We also show that the median distance to ``classical'' IRDCs (3.1\,kpc) is only slightly smaller than that of the ISOSS portion of the sample, 3.5\,kpc. Considering the typical distance uncertainties, these subsamples occupy equivalent distance ranges.

We show the spatial distribution of the targets throughout the Galaxy in Figures~\ref{fig:gal_distrib} and \ref{fig:gal_distrib2}.  Only a small fraction of the sources lie outside $\pm1^{\circ}$ in Galactic latitude, and we see in Figure~\ref{fig:gal_distrib} that the farthest outliers tend to be the ISOSS sources. As shown in Figure~\ref{fig:gal_distrib2}, the majority of the well-studied IRDCs lie in the first quadrant, coincident with the Scutum-Centaurus spiral arm. On the other hand, the ISOSS sources appear to coincide with the Sagittarius arm (in the first quadrant) or the Perseus arm (in the outer Galaxy). 

\subsection{Ancillary data}
\subsubsection{Mid-infrared archival data}

We make use of the {\em Spitzer}/MIPSGAL \citep{MIPSGAL} survey data at 24\,$\mu$m which covers 27 objects in our sample.  MIPS 24\,$\mu$m data for 11 other objects are available from other projects (see Table~\ref{tab:obs}). In general, MIPS 24\,$\mu$m data provide point source sensitivity down to $\sim$2 mJy.  Above 2-3 Jy, the MIPS detector saturates. Where available, we use the IRAS and MSX point source catalogs to estimate fluxes for bright mid-infrared sources.

\subsubsection{Sub-millimeter survey data and boundary definition}
\label{ss:atlasgal}

Based on previous observations of the targets in our sample, the approximate dimension of each cloud is qualitatively known and was the basis of the selection of the mapped area indicated in Table~\ref{tab:obs}.  In order to proceed analyzing each cloud in detail, however, we must first consistently define their boundaries. For this task, we choose to use sub-millimeter maps, originating from the optically-thin emission from cold dust, to draw the boundary of the cloud and determine the total cloud masses.

The ATLASGAL survey \citep{ATLASGAL} covers 24 of the IRDCs in the EPoS sample at 870\,$\mu$m and are listed in Table~\ref{tab:sample}.  For 17 other sources (15 ISOSS sources plus HMSC07029-1215 and IRDC073.31+0.36) limited area maps at 850\,$\mu$m \citep{SCUBA_legacy, carey_submmIRDC, Krause2003, Birkmann2006, Birkmann2007, Hennemann2008} are in the SCUBA archive, and for three other sources (IRDC\, 321.73+0.05, IRDC\, 18151, and IRDC\, 20081) we use published MAMBO 1.2\,mm maps from \citet{Beuther2002} and \citet{Vasyunina2009}. In all, for all but one source, ISOSSJ06114+1726, we have continuum maps with which we define the boundary of a cloud and estimate the total cloud mass as described below.

For the portion of the sample covered by ATLASGAL, we determine the cloud boundary from regions of sub-millimeter emission based on smoothed ATLASGAL 870\,$\mu$m maps created in the following manner:

\noindent (1) As our {\em Herschel} maps do not always include the full extent of a given cloud complex, e.g. focusing instead on the infrared-dark portion, there are cases in which sub-millimeter emission extends beyond the region of uniform coverage in our {\em Herschel} maps. We therefore crop the sub-millimeter map to match the area of uniform coverage based on a smoothed (to $\sim$1$'$) SPIRE 500\,$\mu$m coverage map. \\
(2) The cropped sub-millimeter map is then convolved by a $\sim$1$'$ gaussian.\\
(3) We compute column density maps using the standard formulation \citep{Hildebrand1983}:

\begin{equation}
N_{H2} = \frac{R_{gd} S_{\nu}}{B_{\nu}(T_{dust})~\kappa_{\nu} m_{H2} \Omega}
\end{equation}

\noindent where $R_{gd}$ is the gas-to-dust ratio, assumed to be 100, $S_{\nu}$ is the specific flux at 850\,$\mu$m, 870\,$\mu$m, or 1.2\,mm enclosed within the IRDC boundary, $B_{\nu}(T_{dust})$ is the Planck function at dust temperature, $T_{dust}$, assumed to be 20\,K\footnote{We note that recent observations \citep{Peretto2010,Beuther_18454} have shown that IRDCs may exhibit temperature gradients, especially when situated near a star forming region. In future work, we shall investigate the degree of this effect throughout the sample making use of more detailed dust temperature maps.}. We use the dust model for intermediate volume densities and thin ice mantles by \citet[][Table 1, column 5]{ossenkopf_henning}, and we adopt a value for the dust mass absorption coefficient $\kappa_{850{\mu}m}$ and $\kappa_{875{\mu}m}$ of 1.85 cm$^2$ g$^{-1}$ and $\kappa_{1.2mm}$ of 1.0cm$^2$ g$^{-1}$.  We use the automated {\tt clumpfind} algorithm \citep{williams_clumpfind} to identify pixels in the smoothed sub-millimeter map which lie above a 8 $\times$ 10$^{20}$ cm$^{-2}$, or $A_V \sim 0.8$ mag, threshold. The resulting boundary is plotted in the figures in Appendix B. 

The smoothing performed in steps (1) and (2) is necessary to avoid ``over-clipping,'' i.e. exclusion of clearly associated regions, in cases where the ATLASGAL data are noisy. Due to the nature of single-dish emission maps, often the region surrounding a bright source shows artifacts (e.g. ``bowls'' of reduced or negative pixel values). By smoothing, we avoid erroneously excluding these neighboring regions of interest.  

For objects not covered by ATLASGAL (see Table~\ref{tab:sample}), we rely on SCUBA archival data at 850\,$\mu$m and MAMBO 1.2\,mm maps to determine boundaries and masses. Unlike the ATLASGAL wide-area maps, the SCUBA and MAMBO data were obtained in targeted fashion, such area of the sub-millimeter map is smaller than the {\em Herschel} map size, rendering step 1 of the above boundary definition method unnecessary. We therefore draw a circular boundary to include the significant sub-millimeter emission.

\begin{figure}
\begin{center}
\includegraphics[scale=0.4,angle=90]{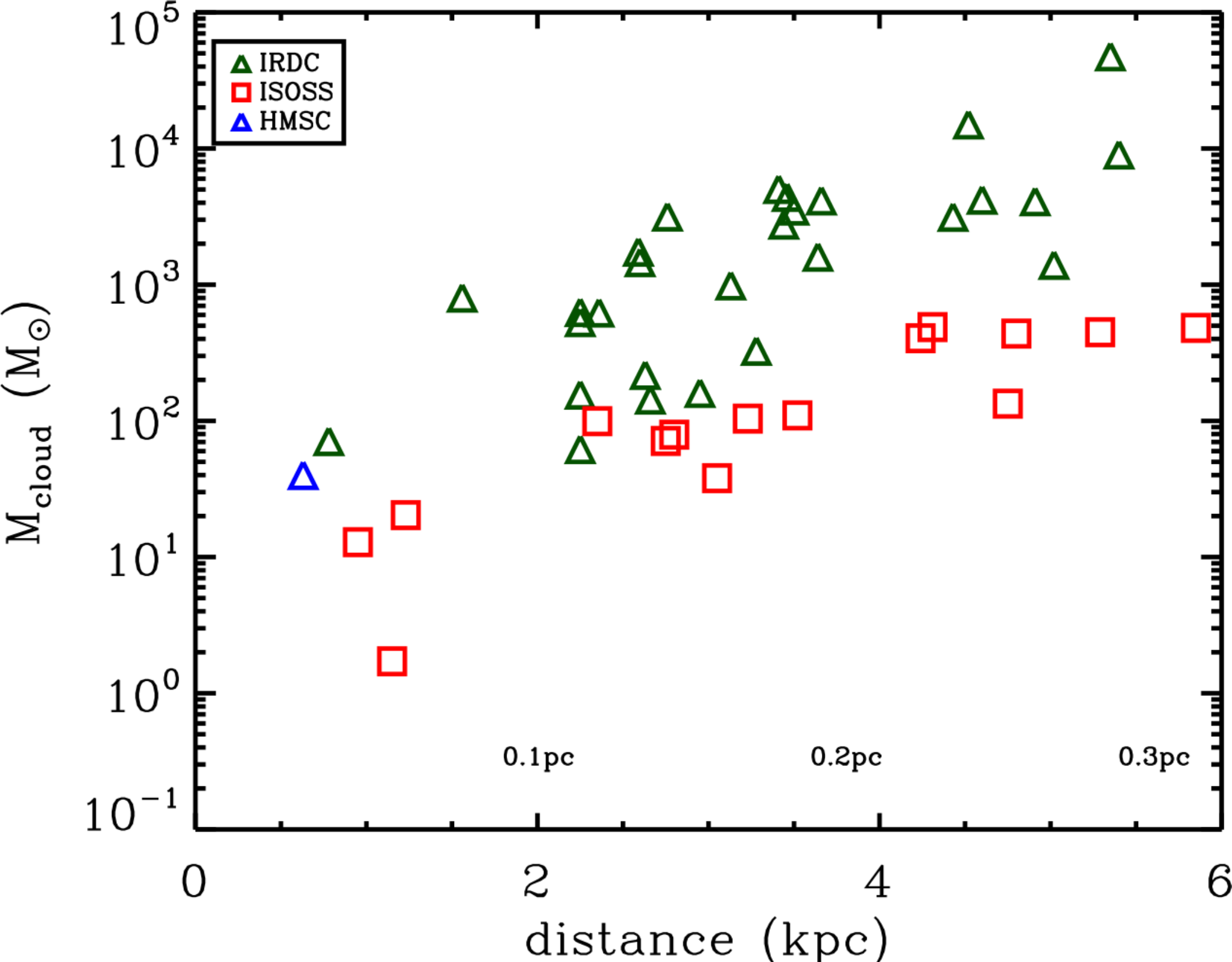}
\end{center}
\caption{\label{fig:totmass_vs_dist} Cloud mass derived from ATLASGAL, SCUBA, or MAMBO continuum maps plotted as a function of distance. Green triangles are IRDCs, red squares are ISOSS sources, and the blue triangle is HMSC\,07029. Toward the bottom, the core size we can resolve with our PACS maps is shown as a function of distance. See Section~\ref{ss:atlasgal} for the relevant assumptions.}
\end{figure}

\begin{figure}
\begin{center}
\includegraphics[scale=0.4,angle=90]{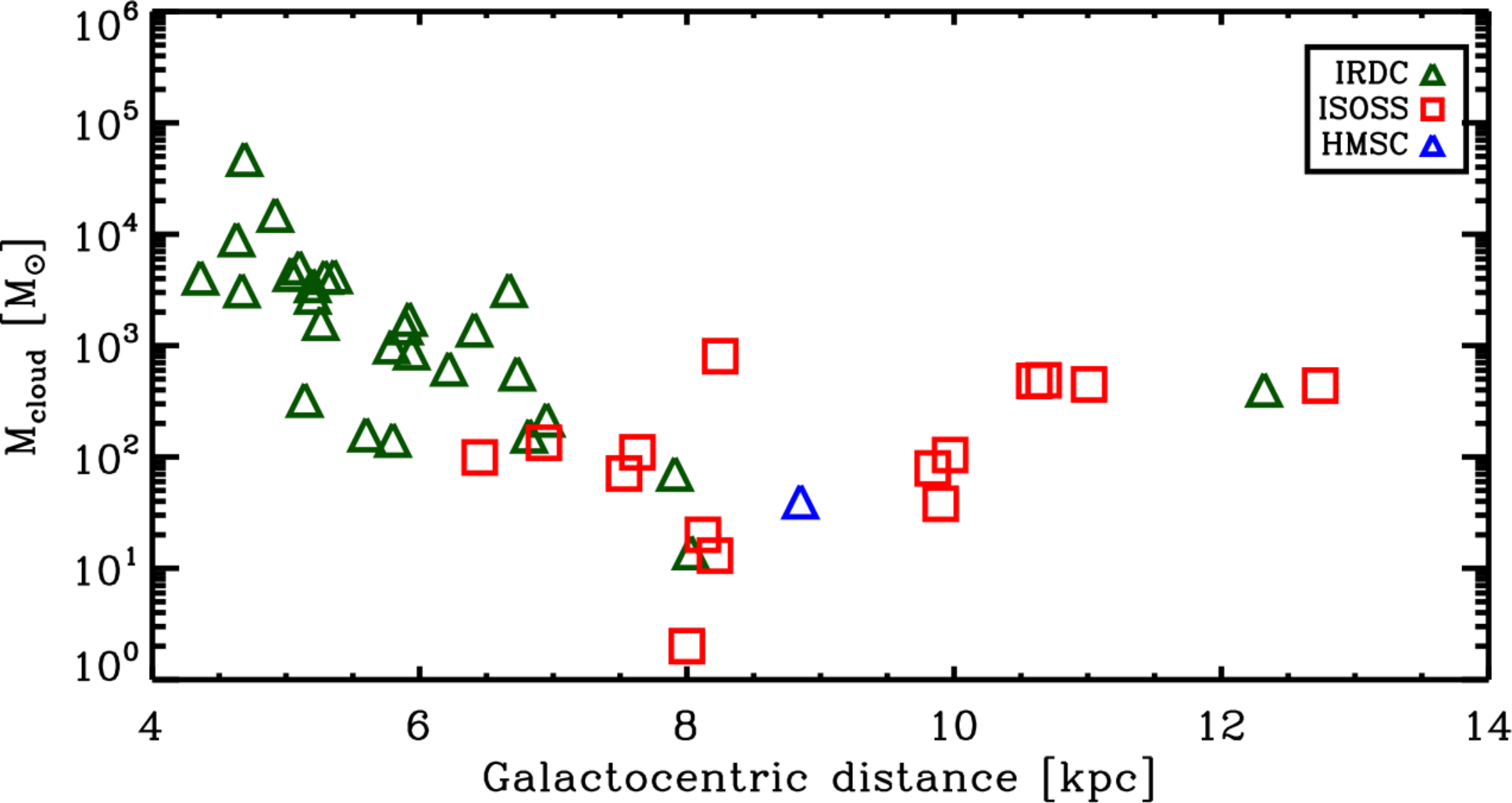}
\caption{Cloud mass as a function of Galactocentric distance.  Green triangles are IRDCs, red squares are ISOSS sources, and the blue triangle is HMSC\,07029. \label{fig:agmass_galcentricdist}}
\end{center}
\end{figure}

We also estimate the total mass of each cloud assuming optically thin emission and following the method in \citet{Hildebrand1983} 

\begin{equation}
M_{cloud} = \frac{R_{gd}~d^2~S_{\nu}}{B_{\nu}(T_{dust})~\kappa_{\nu}}
\end{equation}

\noindent with distance, $d$, considering the flux density, $S_{\nu}$ which is 3-$\sigma$ above the average rms in the region.  We list the masses in Table~\ref{tab:sample}.  We plot the cloud mass as a function of distance in Figure~\ref{fig:totmass_vs_dist} and as a function of Galactocentric distance in Figure~\ref{fig:agmass_galcentricdist}.  Given the uncertainties in dust opacity, distance, dust temperature, and flux measurements, these mass estimates are reliable to within a factor of two.

\section{Herschel observations and data reduction}
\label{s:obs}
\subsection{PACS}

We observed 45 sources at 70\,$\mu$m,
100\,$\mu$m, and 160 $\mu$m with the Photodetector Array Camera and
Spectrometer \citep[PACS; ][]{A&ASpecialIssue-PACS}. 
We obtained two perpendicular scan directions to minimize striping in the
final combined maps.  The scan speed for all observations was
20$\arcsec$~sec$^{-1}$, and a total of 6 repetitions were obtained for
each scan direction for the 70\,$\mu$m, and 160 $\mu$m simultaneous
observations, while 3 repetitions were obtained for each scan direction
for the 100\,$\mu$m, and 160 $\mu$m simultaneous observations.  Map
sizes vary according to the target extent and are listed in Table~\ref{tab:obs}.  
The PACS data for our sources were processed to
level 1 using HIPE \citep{HIPE}.  In this step we apply ``second level
de-glitching'' to remove bad data (``glitches'') for a given position in the map.
 This is done by applying a clipping algorithm (based on the median absolute deviation)
to all flux measurements in the data stream that will ultimately contribute
to the respective map pixel. For our dataset, we use the {\it 'timeordered'} option
with an {\it 'nsigma'} value of 25, which removes obvious glitches
satisfactorily while keeping the core of stronger point sources intact.

After producing level 1 data, we have generated final level 2 maps
using \scana \citep{Roussel2012}.  Because our sources generally
fill the scan maps with variable and bright emission, the highpass
median--window subtraction method implemented in HIPE for the Level 2 processing
produces images which can suffer
from variable missing flux levels and stripping at increased levels
relative to the \scana maps.  For this reason we choose
\scana as our benchmark level 2 data product.  The final level
2 maps were processed using various \scana v.9 options; we find
the ``galactic'' option to be best suited for our data--set and
scientific goals.  Additionally, these data were processed including
the non--zero--acceleration telescope turn--around data. The 
resultant maps are presented in the image gallery in Appendix B.

\subsection{SPIRE}

We obtained maps at 250, 350, and 500\,$\mu$m using the Spectral and Photometric Imaging Receiver \citep[SPIRE; ][]{A&ASpecialIssue-SPIRE}. The map dimensions are listed in Table~\ref{tab:obs}. The data were processed until level-1 with HIPE (developer build 5.0, branch 1892, calibration tree 5.1) using the standard photometer script ({\tt POF5\_pipeline.py}, 02.03.2010) provided by the SPIRE ICC team. The level-1 maps were further processed using \scana \citep{Roussel2012}, version 9 (patched, dated 08.03.2011), which included a de-striping algorithm implemented for maps with fewer than 3 scan legs per scan. We used the ``galactic'' option and included the non-zero-acceleration telescope turn-around data. The resultant maps are presented in the image gallery in Appendix B.

\begin{figure}
\includegraphics[angle=90,scale=0.4]{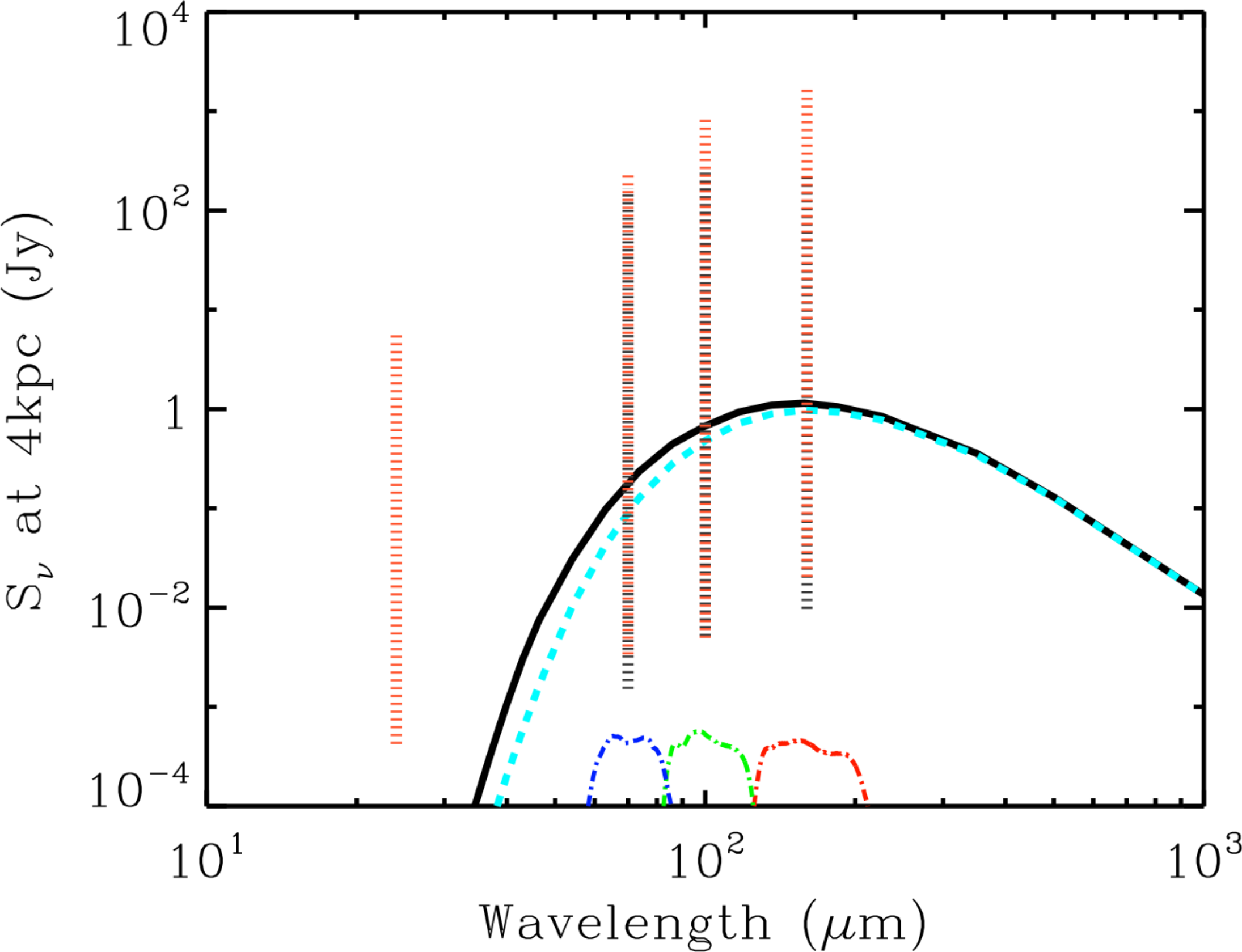}
\caption{\label{fig:SEDaverage} Example SED for a core with the median properties of the ``near'' sample: $M = 2\,\msun$ and $T = 20\,K$. The solid black curve depicts the SED assuming optically thin emission over the full spectrum.  The cyan dashed curve shows an SED with the same parameters, but without an assumption of optically thin emission.  The blue, green, and red dash-dotted curves show the filter response function ($\times 10^{-3}$) for the 70\,$\mu$m, 100\,$\mu$m, and 160\,$\mu$m bands, respectively. The full range of fluxes at each wavelength for 24\,$\mu$m-bright (red dotted line) and 24\,$\mu$m-dark (gray dotted line) are shown. All flux densities have been normalized to the fiducial 4\,kpc distance.}
\end{figure}
 
\begin{figure*}
\begin{center}
\includegraphics[scale=0.8,angle=90]{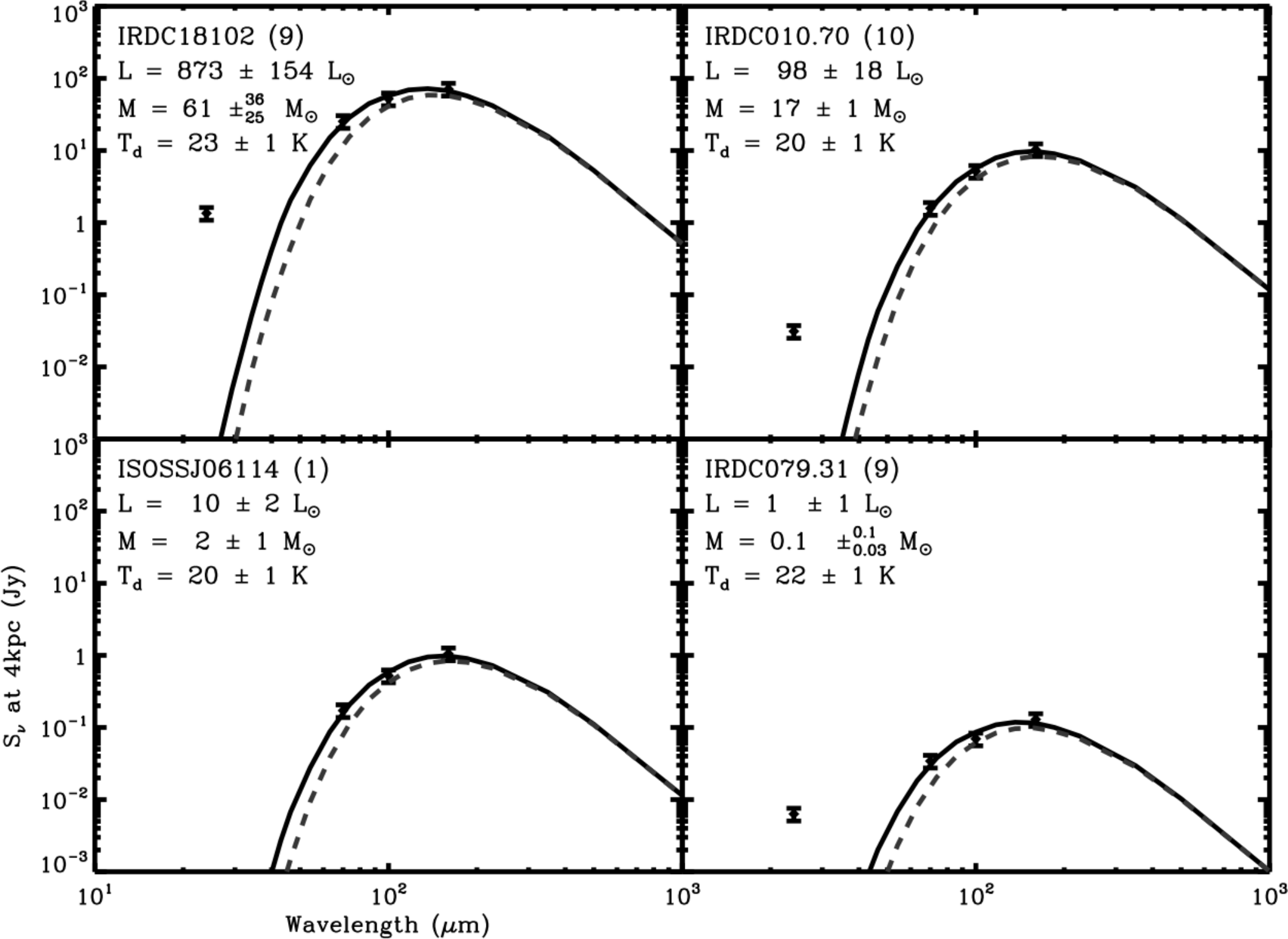}
\end{center}
\caption{\label{fig:SEDrange} Example SEDs of objects at a range of luminosities: 873$\lsun$ (top left), 98$\lsun$ (top right), 10$\lsun$ (bottom left), and 1$\lsun$ (bottom right) cores, all normalized to 4\,kpc distance. The cloud name and core ID number of the object are given in each  panel, along with the mass and temperature, and all associated uncertainties. The dashed line in each panel represents the SED without the optically thin assumption.}
\end{figure*}

\section{Point source extraction and analysis}
\label{s:ptsrc}
\subsection{Source extraction and photometry}
\label{ss:extract}

We first gridded each PACS image to a uniform 1 arcsecond per pixel scale. Because of the diffuse, variable background emission in the PACS wavelength regime, the extraction of point sources, in particular for sources deeply embedded in regions of high extinction or with compact background emission components, is not straightforward with a standard algorithm designed for low-background source extraction \citep[e.g. {\tt starfinder}][]{starfinder}. To remedy this, we apply an unsharp-mask filter to each PACS image in order to remove the large-scale emission, i.e. we smooth each image and subtract the smoothed image from the original science frame. We use empirically-determined smoothing filter sizes of 9, 9, and 20 pixels for the 70, 100, and 160\,$\mu$m bands, respectively. The larger 160\,$\mu$m filter size is needed because the diffuse emission from the cold material tends to vary on larger scales and this wavelength.

We perform the source extraction with {\tt starfinder} on the difference image in order to acquire the {\em positions} of point sources. These positions are given as a prior for a second iteration of {\tt starfinder} on the original science frame, on which we perform PSF photometry and extract fluxes on objects which are 5-$\sigma$ above the median background of the image.  We used an empirical PSF of Vesta (Observation Day 160) and rotate to account for the scan direction of a given observation. 

Compact source candidates are evaluated internally by {\tt starfinder} on to which degree they agree with a given instrument PSF. To do this, {\tt starfinder} computes a correlation coefficient, which is a normalized sum of products of pixel values in the science image and the PSF reference image, taken as a function of spatial offsets between image and PSF.  The {\tt starfinder} algorithm then tries to find the maximum of this coefficient by slightly varying the mentioned spatial offsets. In order to achieve higher accuracy for this estimate, a cubic convolution interpolation method \citep{ParkSchowengerdt1983} is implemented to compute the correlation coefficient on a sub-pixel level. The explicit form of the correlation coefficient is given in section 3.4 of \citet{Diolaiti2000}, and was itself taken from the book of \citet{Gonzalez1992}. Since the coefficient is a normalized quantity, a perfect agreement between a source in the science frame and the PSF would result in a correlation coefficient of 1.0. As mentioned in \citet{Diolaiti2000}, a correlation coefficient between 0.7 and 0.8 is still acceptable in case of noisier data to consider a found source as an unresolved source resembling the PSF. We tested a range of required correlation coefficient values and determined by-eye that a value of 0.79 was most suitable.

Where available, we use 24\,$\mu$m images from the {\em Spitzer}/MIPS archive, mainly from the MIPSGAL survey \citep{MIPSGAL}. We indicate which of our targets were covered by MIPSGAL or other MIPS 24\,$\mu$m observations in Table~\ref{tab:obs}. We extract the point source flux from the MIPS 24\,$\mu$m image using the {\tt tinytim} PSF. We then match all catalogs using a world coordinate system-based algorithm \citep[see][]{Gutermuth_ngc1333} with a position-matching tolerance of 11$''$ (the FWHM of the PACS 160\,$\mu$m PSF), and we report the position (from the 100\,$\mu$m source position) and flux densities of all sources detected in at least the three PACS bands in Table~\ref{pointsrc}. The positions of the point sources are marked on each panel in the image gallery figures (Appendix B).  The uncertainties in extracted flux densities are of order 15, 20, and 20\% for PACS 70, 100, and 160\,$\mu$m, respectively, and 15\% for the MIPS 24\,$\mu$m flux densities. 

We cross correlate our PACS point source catalog with the IRAS and MSX point source catalogs and note matches in Table~\ref{pointsrc}. In cases where the MIPS 24\,$\mu$m image is saturated, we list the IRAS 25\,$\mu$m or MSX 21\,$\mu$m flux density. We note that the IRAS and MSX fluxes are integrated over much larger apertures than MIPS and PACS, so in some cases there are multiple source resolved with {\em Herschel} or {\em Spitzer} which appeared as a single source in IRAS or MSX. Some sources in Table~\ref{pointsrc} have the same associated IRAS or MSX source.

In all, we extract 496 point sources which have counterparts at all PACS wavelengths. Upon by-eye inspection, we identified 51 point sources which appear point-like in the PACS bands but did not meet our criteria for significance and/or PSF-correlation coefficient. On the other hand, 152 of the 496 sources were deemed ``unlikely'' to be identified by the human eye. In summary, 70\% of the reported detections are reliable when checked by eye, and an additional 10\% could be missing from the sample. These figures do not include the differing effects of distance on our ability to resolve substructure within the psf; such considerations will be addressed in forthcoming work on individual objects which can be supplemented when high-resolution data become available \citep[e.g.][]{BeutherHenning2009}.

\subsection{Spectral Energy Distribution Fitting}
\label{ss:sedfits}

We fit a single temperature modified Planck function to the spectral energy distribution (SED) comprised of the three PACS flux densities ($S_{\nu}$) at 70, 100, 160\,$\mu$m at the position of each extracted point source. At these wavelengths, we assume the dust is optically thin. We exclude the SPIRE data in order to preserve the high angular resolution scales probed with just the PACS data.  Furthermore, for the vast majority of point sources, no clear counterpart is possible to extract from SPIRE maps due to increasing beam dilution.  We also exclude the 24\,$\mu$m data point in our fitting. \citet{A&ASpecialIssue-Beuther} have shown that a second blackbody contribution, with a higher temperature and lower mass, is typically needed to account for the 24\,$\mu$m emission. This second blackbody represents the relatively small fraction of warm dust which lies near the central protostar, presumably because this material has been heated by the protostar itself. We revisit this interpreation in Section~\ref{s:sed}.

The automated fitting is performed with simulated annealing \citep{Kirkpatrick1983}. The model takes into account the frequency-dependent, optically thin dust opacity, $\kappa_{\nu}$, using the model computed by \citet{ossenkopf_henning}, assuming a density of 10$^6$ cm$^{-3}$ and thin ice mantles on the grains. The  SED comprised of the extracted flux densities are fit iteratively with the following function:

\begin{equation}
\label{eq:bbfit}
S_{\nu} = \frac{ B_{\nu}(T_{dust})~\kappa_{\nu} } {R_{gd}~d^2}~M
\end{equation}

\noindent where $B_{\nu}$ is the Planck function at a temperature, $T_{dust}$, $R_{gd}$ is the gas-to-dust ratio (assumed to be 100), $d$ is the distance to the source, and $M$ is the gas mass fit for a given object. We stress that $T_{dust}$ represents a single, average temperature of the unresolved point source, and here we do not attempt to model the temperature gradient on smaller scales.

\subsection{Sensitvity considerations}
\label{ss:sensitivity}

The PACS photometer has in general shown a good performance regarding point
source detection sensitivities. In the scan map release
note\footnote{http://herschel.esac.esa.int/twiki/pub/Public/PacsCalibrationWeb/
PhotMiniScan\_ReleaseNote\_20101112.pdf}
5-$\sigma$ detection limits of around 5, 5, and 11 mJy are mentioned for the
70, 100, and 160 $\mu$m filter, respectively. This performance has been
achieved on mini-maps (having a similar number of scan repetitions as our
science maps, but better coverage in the central part of such a map). However,
we cannot expect identical detection thresholds for our science maps. First,
the benchmark results have been obtained in observations of field stars {\em
without} high extended background emission. Furthermore, these mini-maps have
been processed by utilizing the standard approach for drift and 1/$f$ noise
removal: an aggressive high--pass filtering of the data time lines with small
median windows, which robustly pushes the noise in the map to very low levels.
Such an approach is not possible for our science maps since it would severely
corrupt the extended emission present in our maps, and would at the same time
affect the point source fluxes. Instead of high--pass filtering we use the
{\sc Scanamorphos} program (see Sec. 3.1) as a well--balanced compromise for
retaining both the correct point source fluxes and the extended emission, at
the cost of slightly higher noise levels.
Because the objects in our sample exist in a wide variety of environments,
and the foreground and background emission varies not only from region to
region but also within a given map, the overall point source sensitivity is
difficult to quantify.  In general, the minimum fluxes we detect are 0.03 Jy at 70 $\mu$m, 0.03-0.1 Jy at 100 $\mu$m, and 0.1-0.3 Jy at 160 $\mu$m.  Naturally, the sensitivity worsens in regions of bright, diffuse emission (e.g. IRDC\,18102 and 18454).

\begin{figure*}
\includegraphics[scale=0.8,angle=90]{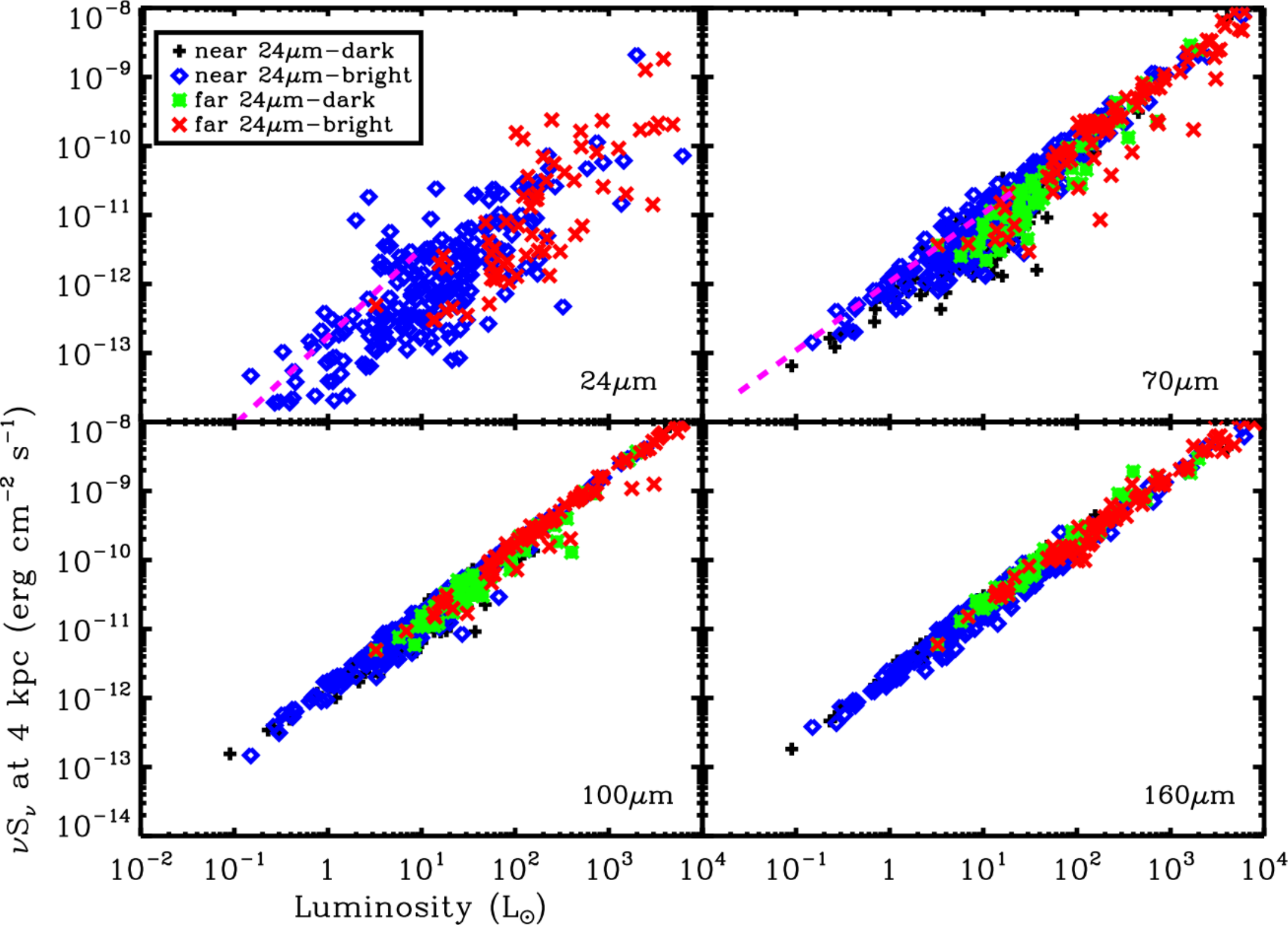}
\caption{\label{lumfall} Flux at 24\,$\mu$m (upper-left), 70\,$\mu$m (upper-right), 100\,$\mu$m (lower-left), and 160\,$\mu$m (lower-right), normalized to 4\,kpc distance, as a function of luminosity from SED fits. The black crosses and blue diamonds represent the 24\,$\mu$m-dark and -bright (respectively) cores nearer than 4\,kpc. The green squares and red triangles represent the 24\,$\mu$m-dark and -bright (respectively) cores further than 4\,kpc. The magenta dashed line in the two upper panels represents the relationship shown in \citet{Dunham2008} for embedded protostars from the {\em Spitzer} c2d survey.}
\end{figure*}

\begin{figure}
\includegraphics[scale=0.45,angle=90]{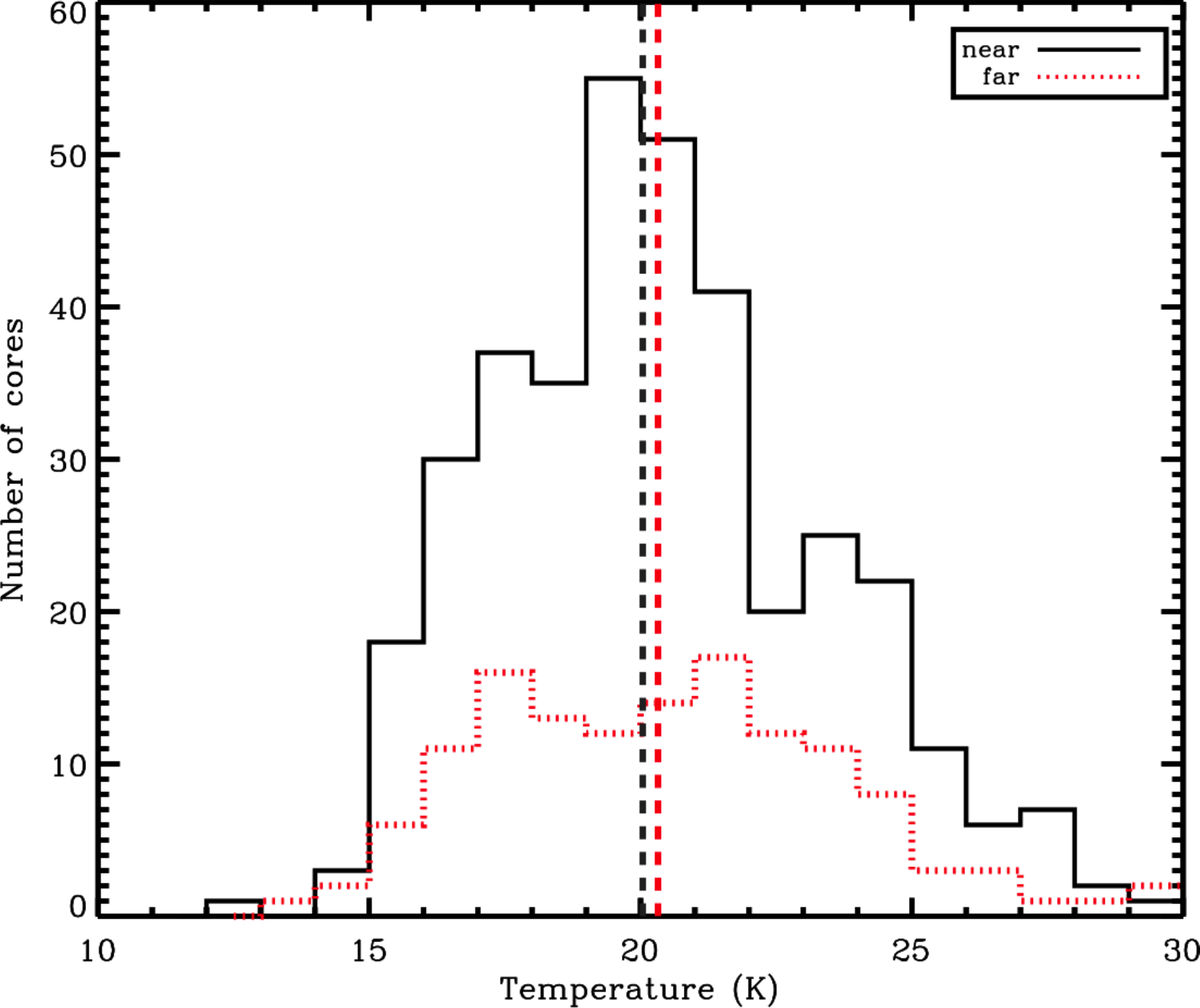}
\includegraphics[scale=0.45,angle=90]{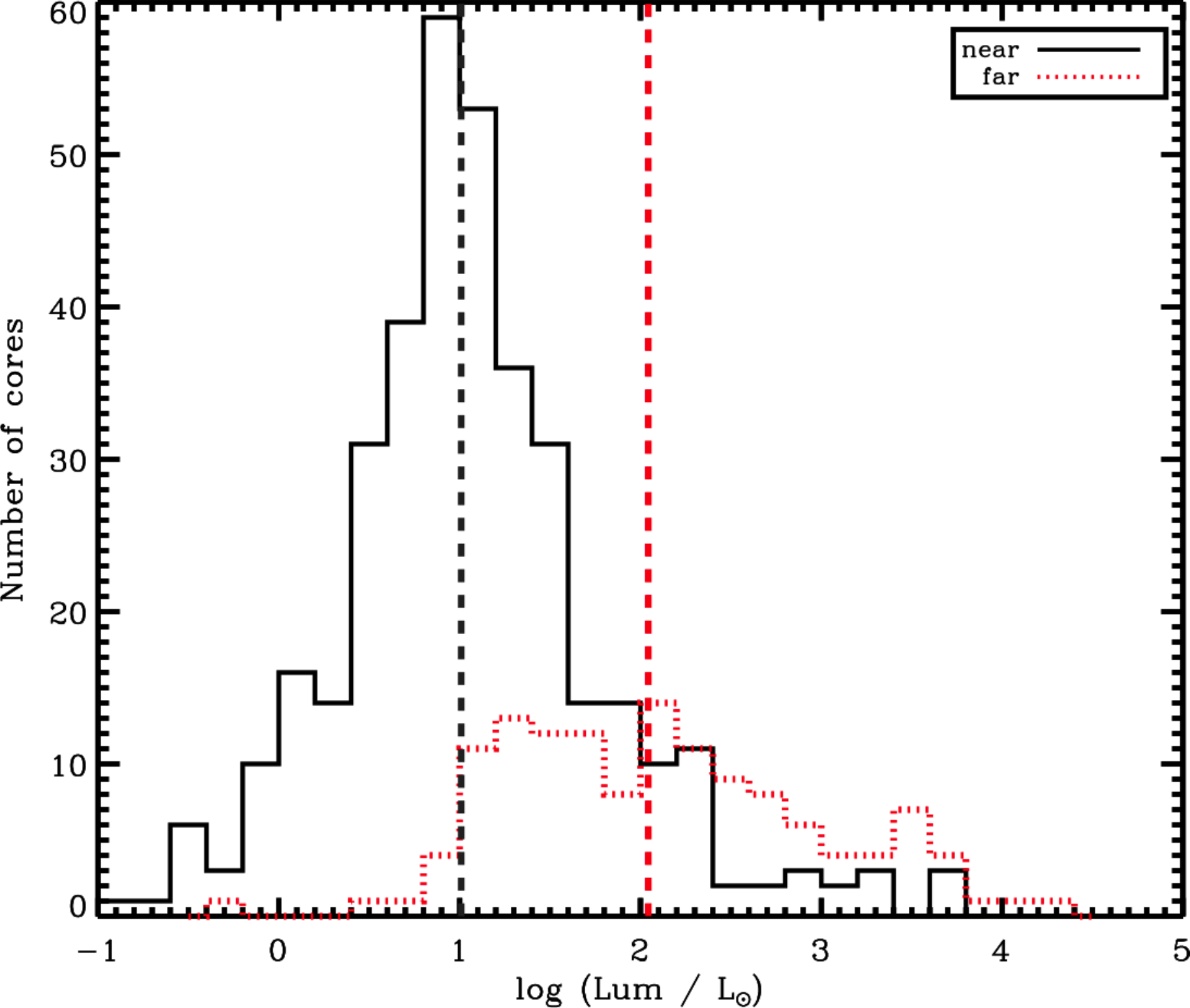}
\includegraphics[scale=0.45,angle=90]{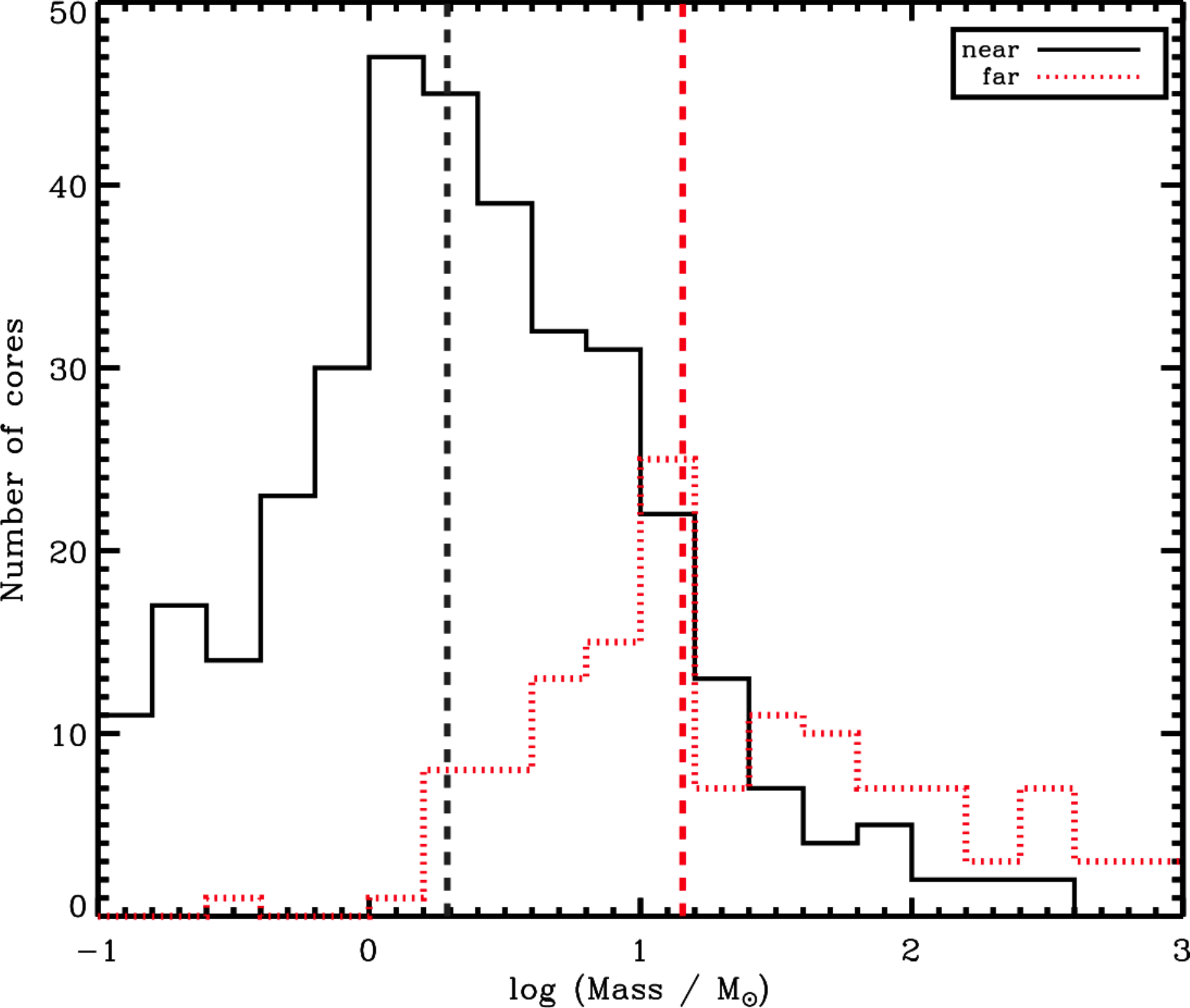}
\caption{Distribution of point source properties within ``near'' (black histogram) and ``far'' (red dotted histogram) IRDCs.  The top panel shows the core temperature distribution; the center panel shows the core luminosity distribution, and the bottom panel shows the core mass distribution. The median values for the near (black) and far (red) distributions are plotted with the vertical dashed lines. \label{propdistrib2}}
\end{figure}

\begin{figure}
\includegraphics[scale=0.4,angle=90]{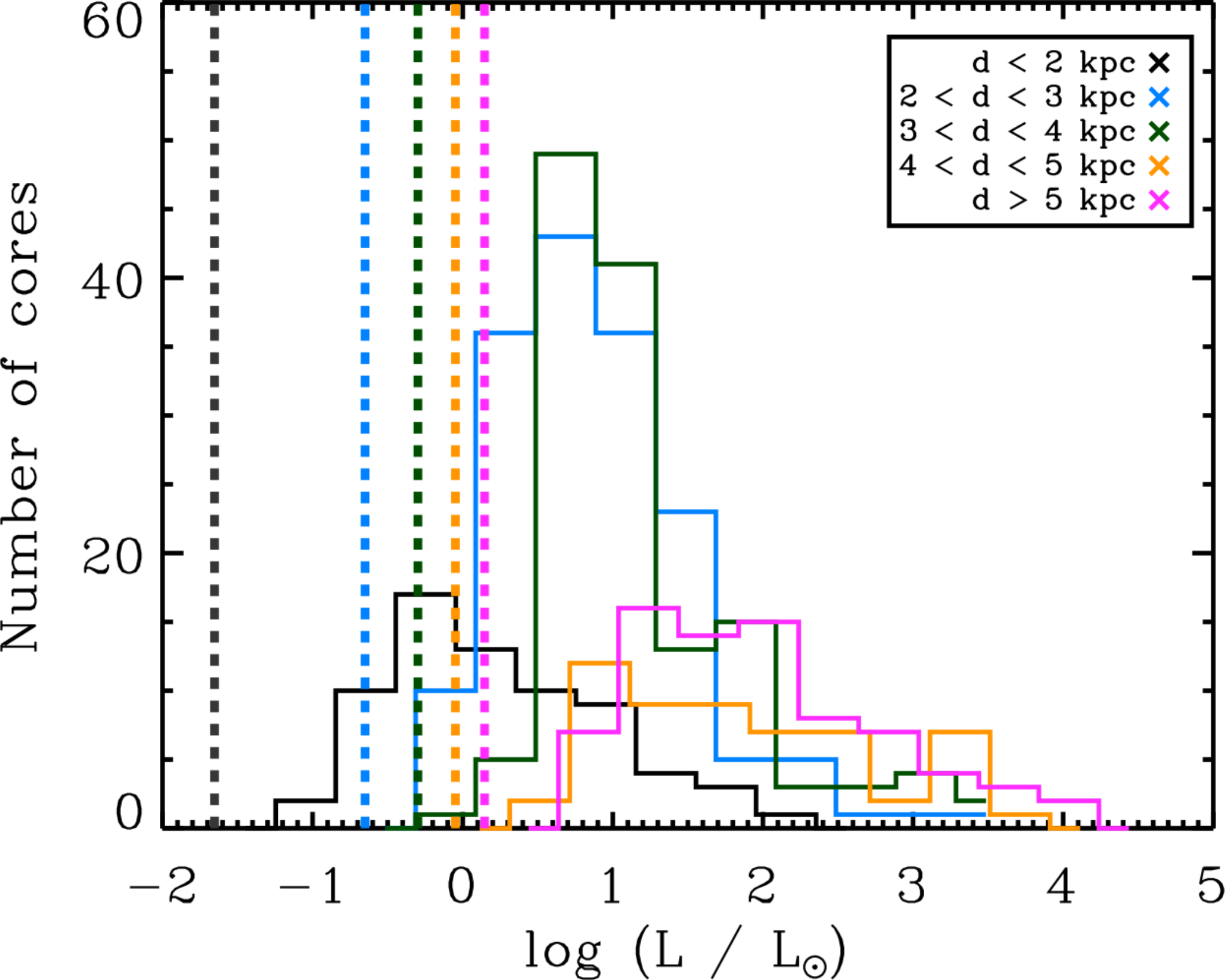}
\caption{\label{fig:senshistogram} Histograms of core luminosities in distance bins $d < 2\,kpc$ (black), $2 < d < 3\,kpc$ (blue), $3 < d < 4\,kpc$ (green), $4 < d < 5\,kpc$ (orange), and $d > 5\,kpc$ (magenta). The vertical dashed lines represent the luminosity sensitivity for each corresponding color distance bin. }
\end{figure}
 
\section{Spectral Energy Distributions}
\label{s:sed}

In Table~\ref{pointsrc}, we present a catalog of all point sources found in the 45 targets, requiring detections in all three PACS bands, and report their positions and flux densities. We enforce strict extraction and matching criteria (see Section~\ref{s:ptsrc}) in order to compile the most homogeneous, flux-limited catalog  possible with these data. There are a total of 496 cores which fit this description. Where available (see Table~\ref{tab:obs}), we include the MIPS 24\,$\mu$m flux density when a counterpart is present. For bright sources, the MIPS detector is often saturated, in which case we report an IRAS 25\,$\mu$m flux density or MSX 21\,$\mu$m flux density, when one was available. We note that in such cases, the IRAS and MSX fluxes are measured over much larger beam areas, thus the reported flux density will certainly be an overestimate of that due to the {\em Herschel} point source.

The clouds in our sample span a range of distances from 0.6 to 5.9\,kpc (see Figure~\ref{fig:distances}).  One consequence of the large range of distances is that the definition of a ``point source'' corresponds to different physical scales throughout the sample. For example, in the nearest IRDC at 0.63\,kpc, the 11.4$''$ angular resolution of the 160$~\mu$m band corresponds to 7200 AU, or 0.03 pc, which resolves what is considered a typical core scale \citep{BerginTafalla_ARAA2007}.  On the other hand, a point source in the most distant IRDC at 5.9\,kpc corresponds to a physical scale almost ten times larger, which is closer to the ``clump'' scale, an object which could be comprised of many ``cores.''  As a result, we are more sensitive to the low-mass, low-luminosity population of cores in nearer IRDCs than the more distant ones.  Therefore, throughout our analysis, we distinguish between point sources within ``near'' and ``far,'' using 4\,kpc as the defining distance.  A point source in the ``near'' category has a characteristic physical scale of 0.1 to 0.2~pc, and a ``far'' point source is between 0.2 and 0.3~pc. The range of physical sizes to which our point sources correspond is illustrated in Figure~\ref{fig:totmass_vs_dist}.  Because most clouds reside in the near regime, we will, for simplicity throughout the remainder of the paper, refer to the point sources as ``cores,'' the physical interpretation of these objects.  When relevant for comparing sample properties, we normalize flux to a distance of 4\,kpc so the properties of the entire sample can be fairly compared. 

We construct SEDs for each source having detections in at least the three PACS bands. In Table~\ref{pointsrc}, we list the flux densities at 70, 100 and 160\,$\mu$m, as well as the 24\,$\mu$m flux density when available\footnote{Alternatively, the IRAS 25\,$\mu$m flux density or MSX 21\,$\mu$m flux density is given in Table~\ref{pointsrc} if MIPS 24\,$\mu$m is saturated at the position.}. Figure~\ref{fig:SEDaverage} shows the full range of flux densities, normalized to 4\,kpc distance as well as the fit (from Equation~\ref{eq:bbfit}) SED of a  core with the median properties of the sample.  The solid curve shows the fit assuming optically-thin emission, and the dashed curve shows a Planck function with the same physical properties but without the optically-thin assumption (but assuming $N_{H_2} \sim 10^{23} {\rm cm}^{-2}$). The total luminosity differs by only $\sim$10\%.

The SEDs peak in the range between 130 and 190\,$\mu$m, and the median peak flux density (of the 4\,kpc distance-normalized SED) is about 1~Jy at 160\,$\mu$m. Figure~\ref{fig:SEDrange} shows example SEDs for individual objects selected to show the full range in luminosities of the sample with the best fit modified blackbody with (solid line) and without (dashed line) the optically thin assumption. 

The 24\,$\mu$m flux density typically can not be accounted for in the single component, single temperature fit to the PACS flux densities, but instead requires a second ``warm'' component to be added. We opt to exclude the value of the 24\,$\mu$m data point in the analysis that follows for several reasons. First, shortward of 70\,$\mu$m, the optically thin assumption is less reliable, rendering the 24\,$\mu$m flux density a poor constraint of the warm component to extract meaningful physical information from a second component.  Secondly, as the point sources are unresolved, probing scales between 0.05 to 0.3\,pc, the enclosed volume is (by mass) dominated by the passively-heated outer core \citep{vandishoeck_wishoverview}.  A complete treatment of the core radiative transfer is beyond the scope of this paper. Recent work by \citet{Stamatellos2010} and \citet{Pavlyuchenkov2012} demonstrate the need for observations such as those presented here to constrain the flux on all scales in order to refine models.

Each SED fit using Equation~\ref{eq:bbfit} has a best-fit temperature and mass, and the integrated area under the function gives the total luminosity, all of which are reported in Table~\ref{pointsrc}. As we are probing on scales of 0.05 to 0.3\,pc (``cores''), these quantities represent the average values over the enclosed region.  While we do not include the value of the flux density at 24\,$\mu$m, we instead at times indicate whether a source is ``24\,$\mu$m-bright'' or ``24\,$\mu$m-dark,'' meaning that there is or is not (respectively) a counterpart at a given position in the 24\,$\mu$m image.  We return to the significance of the 24\,$\mu$m detections later in the Discussion.

\subsection{The source of emission at PACS wavelengths}
\label{ss:heating}

In Figure~\ref{lumfall} we plot the relation between the flux ($\nu S_{\nu}$) at each band and the corresponding core luminosity.  There is a direct correlation at each wavelength which tightens at longer wavelengths. \citet{Dunham2006,Dunham2008} find that for low-mass protostars, the 70\,$\mu$m flux (from {\em Spitzer} in their case) is key to directly probing the protostellar luminosity while not being heavily influenced by external heating or the disk geometry, which more strongly impacts the 24\,$\mu$m emission. The upper-right panel of Figure~\ref{lumfall} shows this correlation extends to the higher-luminosity cores we find in IRDCs. We also show in Figure~\ref{lumfall} that the 100\,$\mu$m, and moreso the 160\,$\mu$m, fluxes are excellent proxies for the core luminosities. But what is the main contributor to the 100 and 160\,$\mu$m flux in IRDC cores: the protostellar luminosity or the external heating? 

\citet{Dunham2006,Dunham2008} argue that, for low-mass protostars, at wavelengths of 100\,$\mu$m and longer, the external radiation field becomes important, however because cores in IRDCs are generally more massive and more luminous, and by deduction home to more massive and luminous protostars, the balance between internal heating from a protostar and external heating from the interstellar radiation field (ISRF) and/or local sources may be different.  Further complicating the picture is that the dense environments of IRDCs, the cores we detect are embedded in high column density material \citep[$A_V \sim 2-20$][]{Kainulainen2011}, which may serve shield the core material from significant heating from external sources. While \citet{Pavlyuchenkov2012} have shown that the stochastic heating caused by external UV field contributes insignificantly to the SED of IRDC cores longward of 100\,$\mu$m, further radiative transfer experiments are needed to quantify the impact of external radiation on the core heating and resultant SED of more massive cores at PACS wavelengths. 

As a simple test of this question, we perform one-dimensional radiative transfer calculations using the TRANSPHERE-1D\footnote{Available from www.ita.uni-heidelberg.de/$\sim$dullemond/software/transphere/} (vers. 3, C. Dullemond), a spherically-symmetric version of the code presented in \citet{Dullemond2002}, which is specially suited for high optical depth scenarios like our massive cores. We assume the core has a density ($\rho$) profile, $\rho(r) \propto r^{-2}$, where $r$ is the radius.  The average central density was 10$^{6-7}$\,cm$^{-3}$ over the integrated radius range between 200\,AU and 0.1\,pc, which are similar to the values used in recent radiative transfer calculations of IRDC cores \citep{Stamatellos2010,Wilcock2011}. At the assumed distance, 4\,kpc, the core would be 10$''$ in diameter, which would be unresolved in our PACS 160\,$\mu$m observations.

We assume a hybrid ISRF with the \citet{black94} model for $\lambda > 0.36$\,$\mu$m and \citet{draine78} for $\lambda < 0.36$\,$\mu$m, as was done in \citet{evans_td}, and we implement dust opacities from \citet{ossenkopf_henning} for $\lambda > 1$\,$\mu$m and \citet{mathis_isrf} for $\lambda < 1$\,$\mu$m. We test cases for a 10\,$\msun$ core, both with and without an internal protostar. For the protostellar case, we assumed a protostellar temperature of 8000\,K, mass of 1\,$\msun$, and radius of 5.3\,$R_{\odot}$, motivated by models of massive protostars by \citet{HosokawaOmukai2009} taking into account both the protostar and accretion for a $L_{tot} \sim 2\,\lsun$. Our simplified test assumes a single central source, but indeed most massive stars form in systems. Assuming that the sources are tightly clustered in a heavily embedded central region, the total, reprocessed emission coming out of the optically thick central core region will be most relevant for these calculations.

\begin{figure}
\includegraphics[angle=90,scale=0.45]{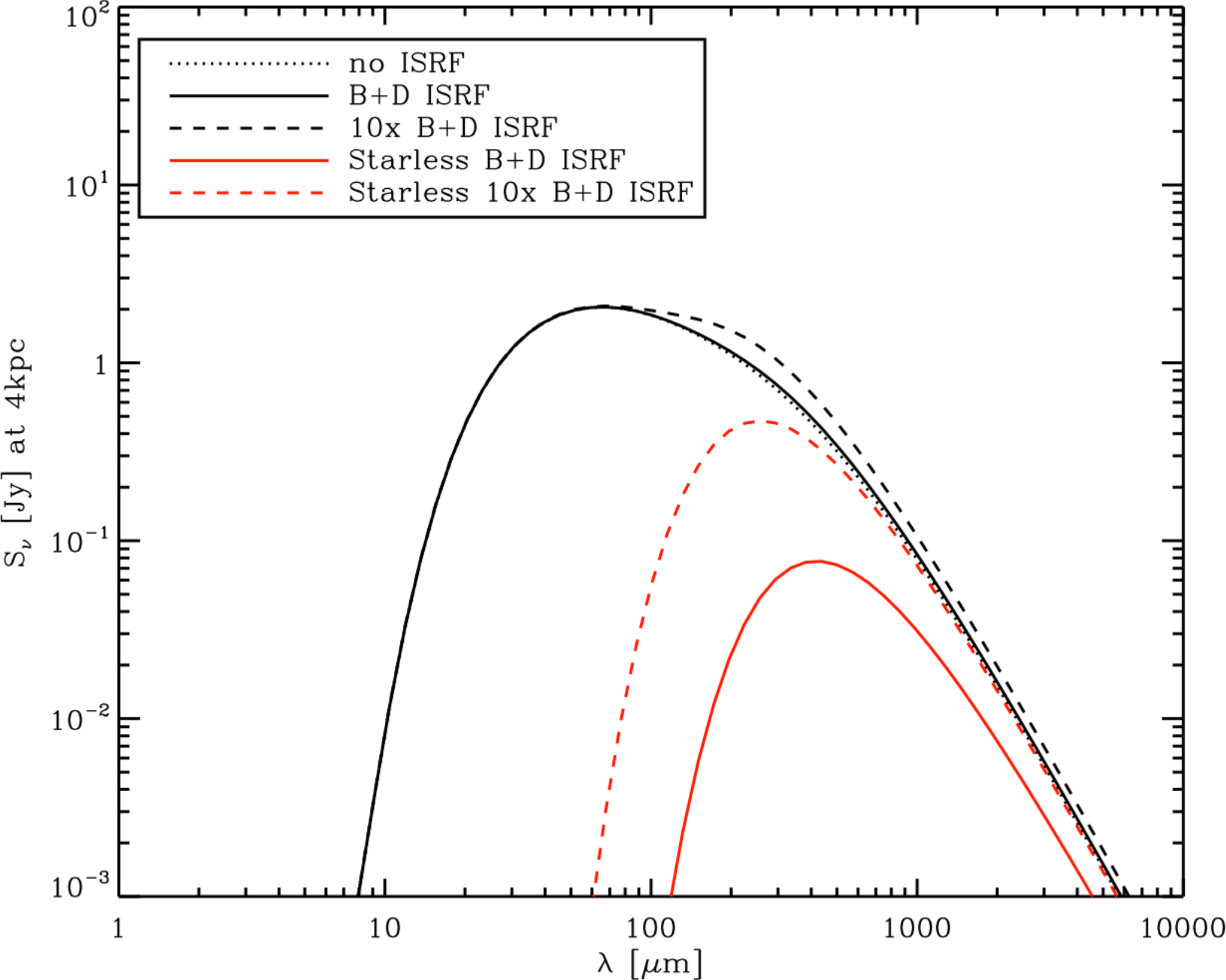}
\caption{\label{fig:radtrans} Results of TRANSPHERE-1D radiative transfer calculations for a 10\,$\msun$ core. The black lines represent the protostellar cases with varying levels of the B+D ISRF (see text). The dotted line has no ISRF, the solid line shows a ``standard'' B+D ISRF, and the dashed line shows the case for a ten times (uniformly) amplified B+D ISRF. The green lines represent the case for starless cores, and are for the same ISRF cases.}
\end{figure}

In Figure~\ref{fig:radtrans}, we show the resultant SED for a 10\,$\msun$ core with and without an internal protostar, and with varying levels of external radiation field, from no ISRF, a Black \& Draine (B+D) ISRF, and a ten times amplified B+D ISRF. The SED is normalized to a distance of 4\,kpc.  In the protostellar case, indeed we only begin to see a difference due to the ISRF at wavelengths longward of 100\,$\mu$m, and only marginally there. The bigger difference is seen when the protostellar SEDs are compared to the starless cases, where the only heating source is external. As the ISRF level is increased, the SED is significantly enhanced at wavelengths longer than 60\,$\mu$m. With enough enhancement of the ISRF, an externally headed starless core could mimic the FIR SED of a protostellar core. However, we note that even in the extreme case of a ten times amplified ISRF, no appreciable 24\,$\mu$m flux is produced in the starless case.  In future work, we shall explore the parameter space of core environments and the temperature structure. 

\begin{figure}
\includegraphics[angle=90,scale=0.45]{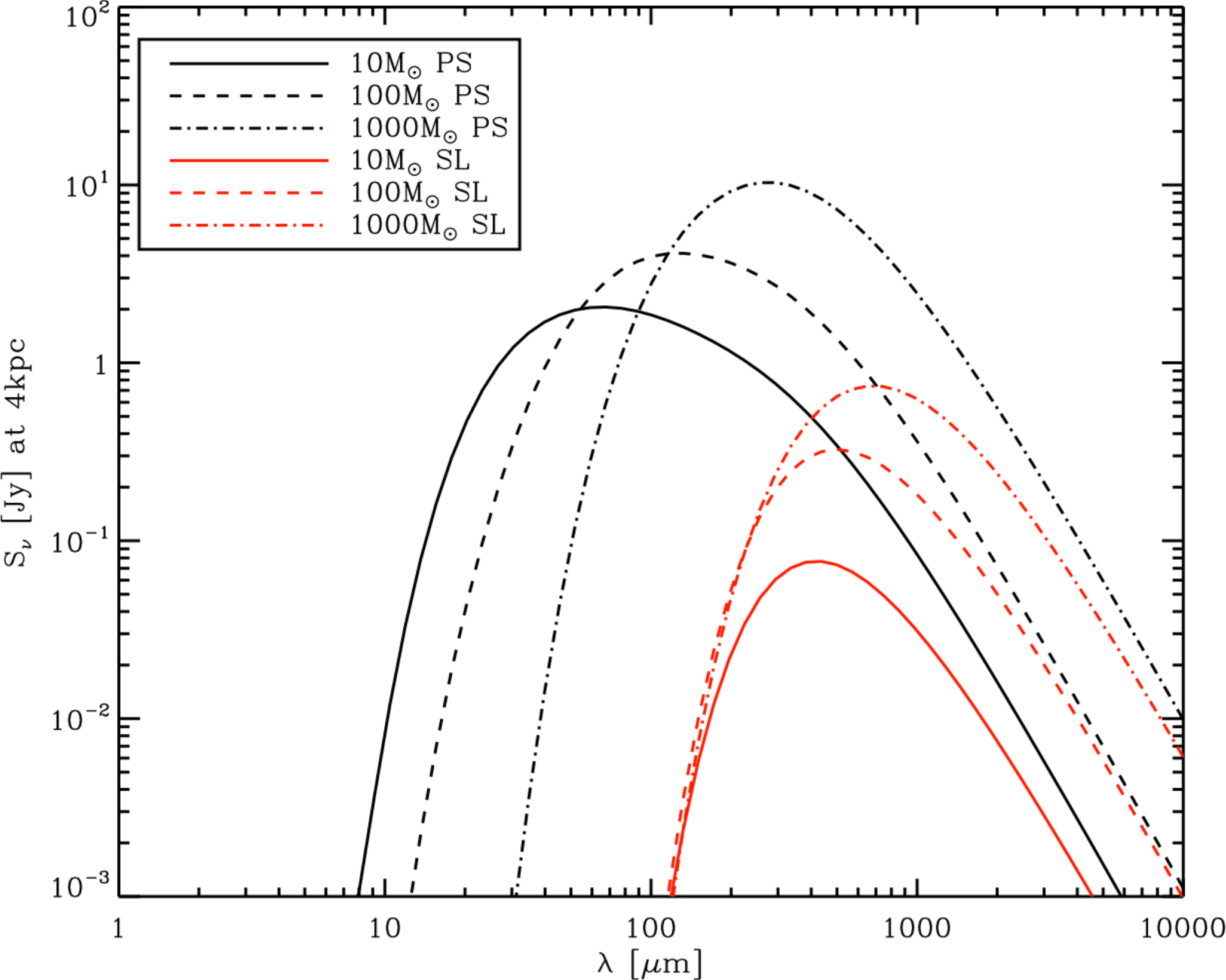}
\caption{\label{fig:radtrans2} Results of TRANSPHERE-1D radiative transfer calculations for a 10, 100, 1000\,$\msun$ protostellar (PS) and starless (SL) cores with the standard B+D hybrid model for the ISRF. The black lines represent the protostellar cases for 10\,$\msun$ (solid line), 100\,$\msun$ (dashed line), and 1000\,$\msun$ (dashed-dotted line). The green lines represent the same masses but without a central protostar.}
\end{figure}

\subsection{24\,$\mu$m emission}

Figure~\ref{fig:radtrans} shows that a 10\,$\msun$ core with a protostar would be detectable at 24\,$\mu$m, but as the core mass increases, the cores become more optically thick, thus the energy radiated at short wavelengths is reprocessed and re-emitted at the longer wavelengths. Here we emphasize that while such a trend is not surprising for the spherical symmetric cases. At these wavelengths the core geometry likely plays a strong role, so our simple 1D model breaks down. The central region is optically thick at 24\,$\mu$m, thus other effects are needed to account for the 24\,$\mu$m flux.

There are various ways of accounting for the emission seen at 24\,$\mu$m. We have discussed emission from externally, isotropically heated outer core regions and emission from inner core regions.  The emission from the inner core region generally would be reabsorbed by the outer core, but the inner core can be exposed by outflow activity when the jet/outflow alignment allows for it or the inhomogeneous distribution of clumpy material around the warm inner core \citep{Indebetouw2005}.  A core which is heated (anisotropically) by the radiation from a near neighbor can bring about excess emission as well, but we show in Figure~\ref{fig:radtrans2} that even in the case of the most enhanced ISRF, the amplification of flux does not appear at short wavelengths.  Further studies will be needed to examine for single cores if nearby stellar sources can contribute to the heating of the core and thus to the flux at 24\,$\mu$m. But given the numerous detection of 24 micron excess, it seems unlikely that all of them have stars close enough to impact the SED and which can illuminate the core unattenuated. 

Substantial 24\,$\mu$m flux would also be expected in the case when the entire core has been heated by the central source enough that the optical depth at 24\,$\mu$m becomes low enough that this emission can leave the core un-extincted. Because our observed SEDs can be fitted best with a single modified blackbody spectrum with low temperatures,  we conclude that the 24\,$\mu$m emission visible in some cases is not due to a global warming of the core but is rather emission from heated inner regions exposed by outflow activity. 

\section{Sample property distributions}
\label{ss:props}

We plot the distribution of fit parameters for the full set of 496 cores in our sample in Figure~\ref{propdistrib2}.  We distinguish between ``near'' and ``far'' sources using 4\,kpc as the dividing distance. At 4\,kpc and nearer, we probe scales of 0.2~pc and below. In total, there are 364 cores within near IRDCs and 132 cores within far IRDCs. We note that there are two factors which will reduce the number of cores detected in ``far'' IRDCs: (1) low-mass cores will fall below our detection limit (to quantify), and (2) due to coarser resolution of physical scales, cores which are separated by less than 0.2-0.3~pc will not be resolved.

The core temperatures have a 18\,K spread, and both the near and far cores have median values of 20\,K.  The median mass for near cores is 2\,$\msun$, and the luminosity distribution is strongly peaked around the median 10\,$\lsun$. For cores in far IRDCs, the median of the distribution are 14\,$\msun$ and 109\,$\lsun$ for the mass and luminosity, respectively, however our sensitivity to the low mass and low luminosity populations in far objects is limited. We show in Figure~\ref{fig:senshistogram} the luminosity distribution broken down into more refined distance bins.  From Figure~\ref{lumfall}, a core at our flux density limit at 160\,$\mu$m (0.1~Jy) corresponds to a 1\,$\lsun$ core at 4\,kpc. At nearer distances, we can detect sources down to 0.25\,$\lsun$ at 2\,kpc, and our sensitivity worsens to 2.3\,$\lsun$ at 6\,kpc. As was discussed in \citet{A&ASpecialIssue-Henning}, assuming flux uncertainties for the 70, 100, and 160\,$\mu$m bands of 15\%, 20\%, and 20\%, respectively, the uncertainties to the fits are 4\% in the temperature, 20\% in the luminosity, and 50\% in the mass. These uncertainties do not include the effects of optical depth, which, for example, can lead to underestimates of the mass.

\subsection{Core mass function}
\label{ss:massfun}

\begin{figure}
\includegraphics[scale=0.45,angle=90]{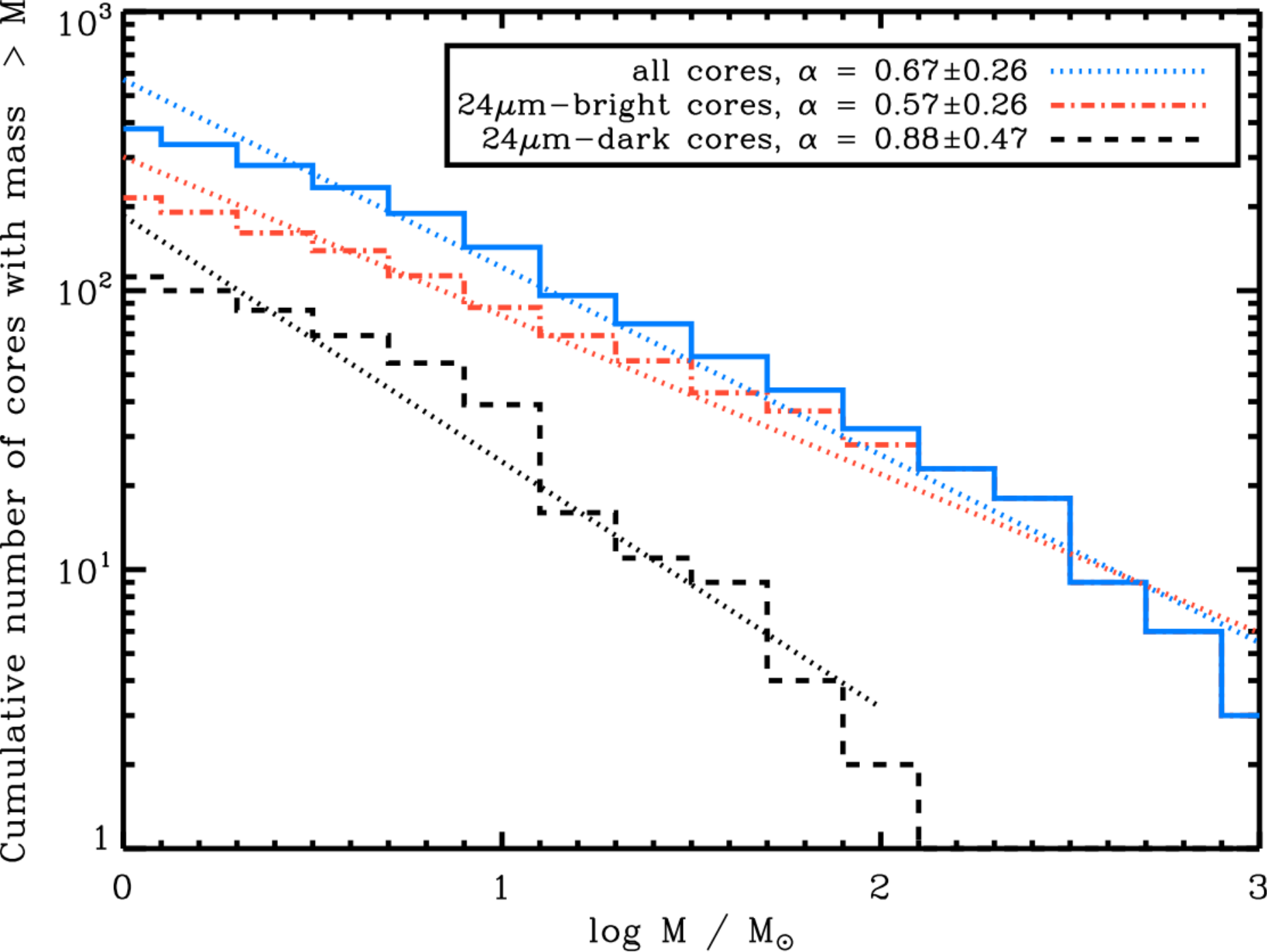}
\caption{\label{fig:massfun} The cumulative mass function of cores more massive than 1\,$\msun$. The total distribution of all cores is shown in blue solid line; the distribution of 24\,$\mu$m-bright cores is shown in the red dash-dotted line; the distribution of 24\,$\mu$m-dark cores is shown in black dashed line. A simple linear regression was fit to each distribution, taking the functional form $N (>M) \propto M^{\alpha}$. For reference, on this scale, a Salpeter slope is $\alpha = 1.35$. }
\end{figure}

In Figure~\ref{fig:massfun} we show the cumulative mass function of cores more massive than 1\,$\msun$, including the total distribution, then the distributions split in terms of the 24\,$\mu$m counterpart status of a given core. We fit the distributions with the function $N (>M) \propto M^{-\alpha}$. We find that the slope of the full distribution is $\alpha \sim 0.67 \pm 0.26$, while the 24\,$\mu$m-bright cores have a slightly shallower distribution ($\alpha \sim 0.57 \pm 0.26$) and the 24\,$\mu$m-dark cores have a steeper distribution ($\alpha \sim 0.88 \pm 0.47$).  

Figure~\ref{fig:massfun} immediately confirms what has been shown in the past in other high-mass regions \citep[e.g.][]{Motte_cygX} is also true in IRDCs:  very massive cores (with $M > 100\msun$) are rare, and those that we do find are quite likely to be mid-infrared bright. While the sampling is not uniform, thus the errors on the fit are quite large, we find general agreement of the clump mass functions massive regions in M17 from \citet[][$\alpha \sim 0.72,$]{StutzkiGusten1990} and IRDCs in \citet[][$\alpha \sim 0.76$]{Ragan_spitzer}, though it shallower than the log-normal function to the IRDC fragments in \citet{PerettoFuller2010} which follows a $\alpha \sim$ 2 slope in this mass regime.

\subsection{Mass - luminosity relationship}
\label{ss:masslum}

\begin{figure*}
\begin{center}
\includegraphics[scale=0.8]{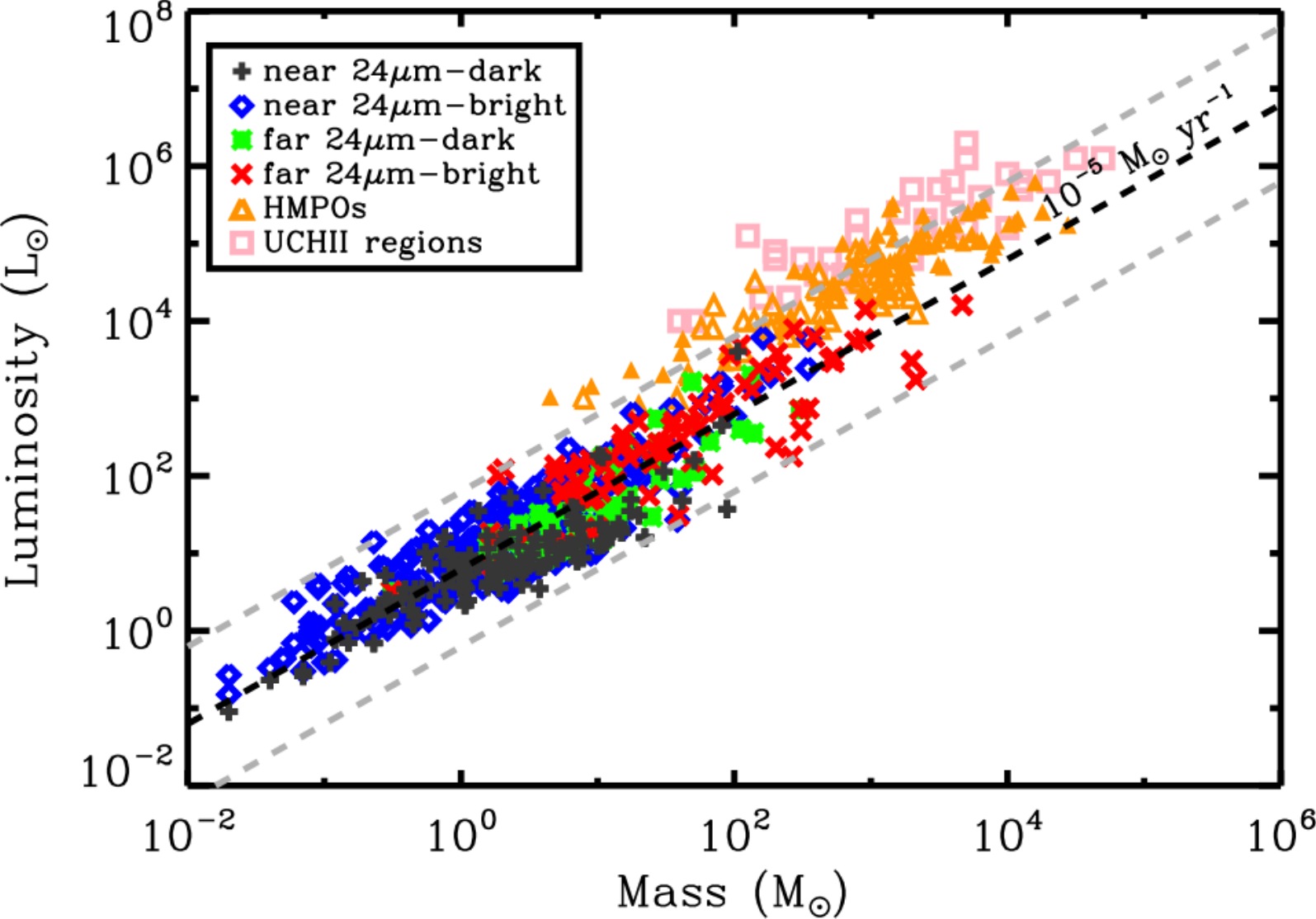}
\caption{\label{fig:masslum} 
Core luminosity versus core mass. We make four categories of sources: ``near'' sources, which are closer than 4\,kpc, with either a 24\,$\mu$m counterpart (``bright'') or no 24\,$\mu$m counterpart (``dark''), and ``far'' sources with the same bright/dark distinction. For comparison, we plot in orange triangles the \citet{Beuther2002,Beuther_erratum} sample of high-mass protostellar objects (HMPOs) and the \citet{Hunter2000} sample of ultra-compact HII regions in pink squares. With the black dashed line, we plot a line of constant accretion luminosity, $L_{acc} = G {\dot M} M_{star} / R_{star} $, of 10$^{-5} \msun$ yr$^{-1}$, assuming $M_{star} = 0.1 M_{core}$ and $R_{star} = 5\,R_{\odot}$, and the range indicated by the grey dashed lines on either side show $\pm$1 dex variation. See Section~\ref{ss:masslum} for details.}
\end{center}
\end{figure*}

 At the earliest embedded stages, accretion luminosity is believed to dominate the total luminosity of a core, and with time, the contribution of the protostellar luminosity grows \citep{HosokawaOmukai2009}.  In Figure~\ref{fig:masslum}, we show the relation between core mass and total luminosity of the population of cores presented in this work. For comparison, we also plot the positions more evolved objects: ultra-compact HII (UCHII) regions from the \citet{Hunter2000} sample and high-mass protostellar objects \citep[HMPOs][]{Beuther2002,Beuther_erratum}, believed to be the precursor to UCHII regions. HMPOs and UCHII regions occupy the highest mass and luminosity part of the diagram, but also overlap with the most luminous objects in our sample. We note that the masses and luminosities for the UCHII regions and HMPOs were measured over larger apertures than the PACS cores, thus their masses and luminosities are integrated over a larger region that includes both the protostar and its surroundings, potentially inflating the estimates of the mass and luminosity of the driving protostar.

Based on empirical accretion models by \citet{Saraceno1996} and \citet{Molinari2008}, objects on this diagram move from the lower right to the upper left of this plot as they evolve.  In this picture, our {\em Herschel} sources appear less evolved than HMPOs and UCHII regions, though the distinction is marginal. In Figure~\ref{fig:masslum}, we show a line of constant accretion rate, 10$^{-5} \msun$ yr$^{-1}$, and show the associated accretion luminosity, $L_{acc} = G {\dot M} M_{star} / R_{star} $, where we assume the radius of the accreting protostar, $R_{star}$, is 5\,$R_{\odot}$ and the accreting protostellar mass, $M_{star}$, is 0.1 times a given core mass (the quantity plotted on the x-axis). We show a spread of an order of magnitude in either direction, which can arise from higher accretion rates or variations in the assumed protostellar mass or radius. In addition, the luminosity of more evolved objects (such as HMPOs and UCHII regions) will include a larger contribution from the protostar than the youngest objects, which is not included in our simple calculation. Detailed models of non-spherical accretion are needed \citep[e.g.][]{Krumholz2009, Kuiper2010} to disentangle the growth of the most massive protostars.

\begin{figure}
\includegraphics[scale=0.44, angle=90]{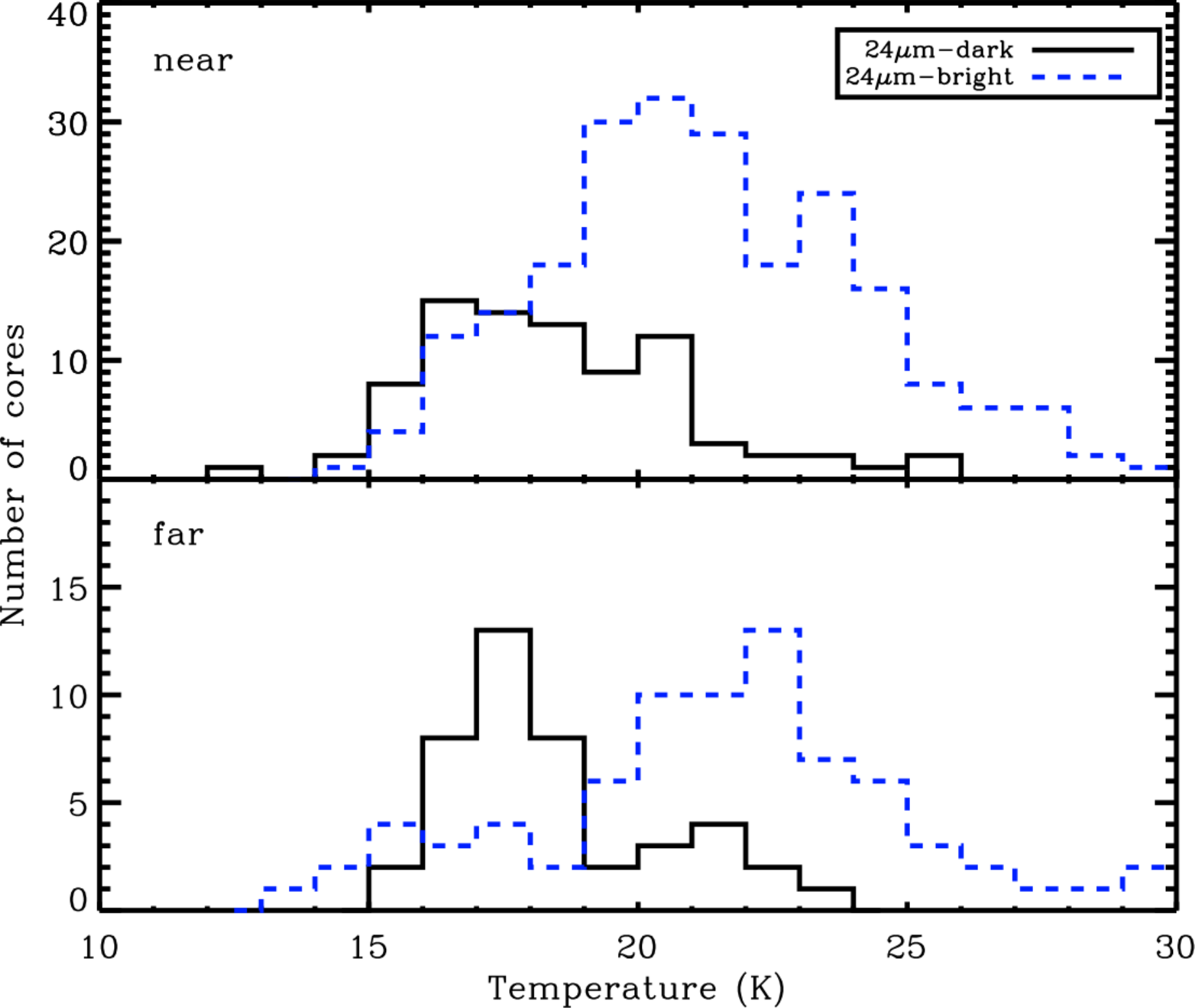}
\includegraphics[scale=0.44, angle=90]{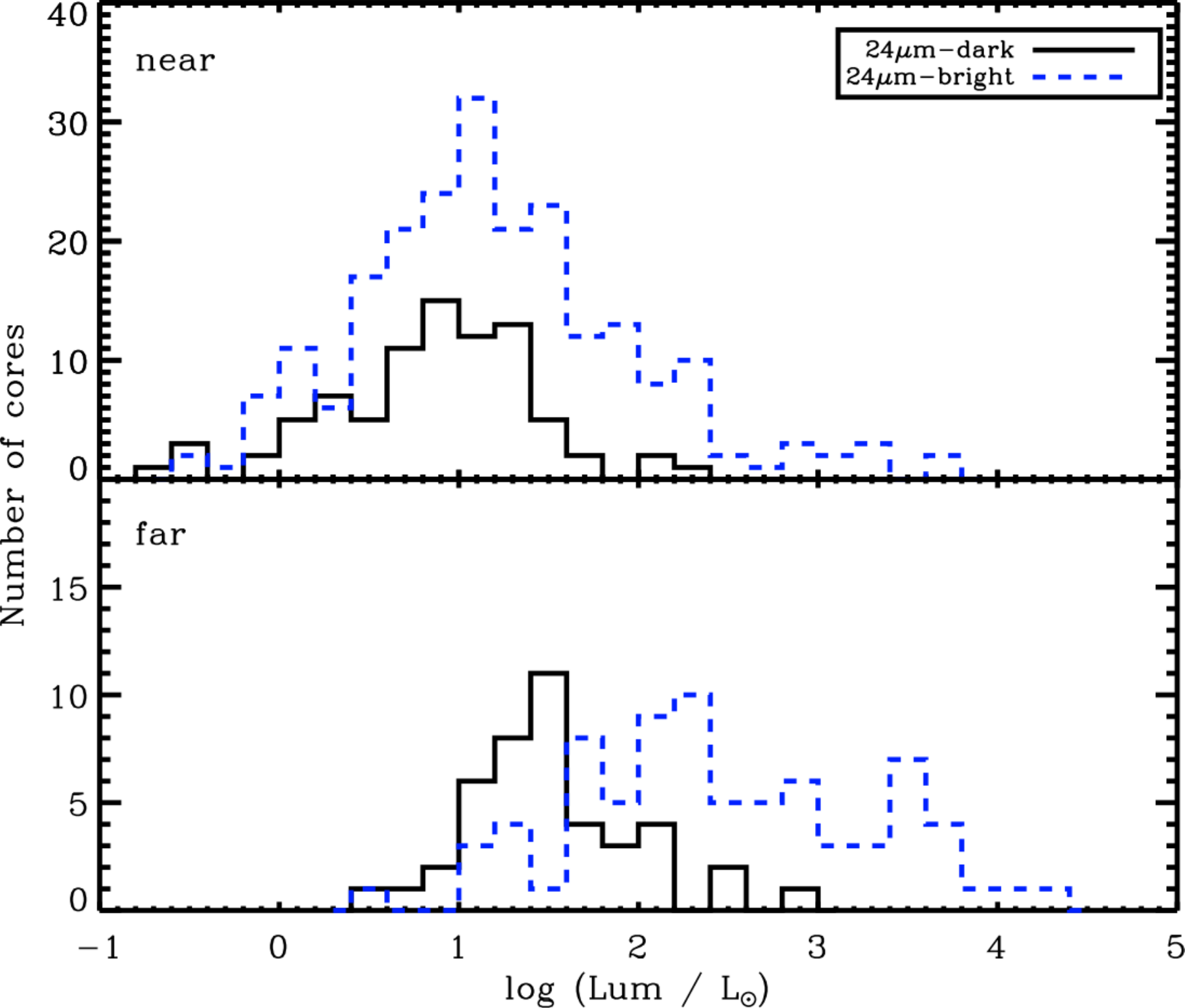}
\includegraphics[scale=0.44, angle=90]{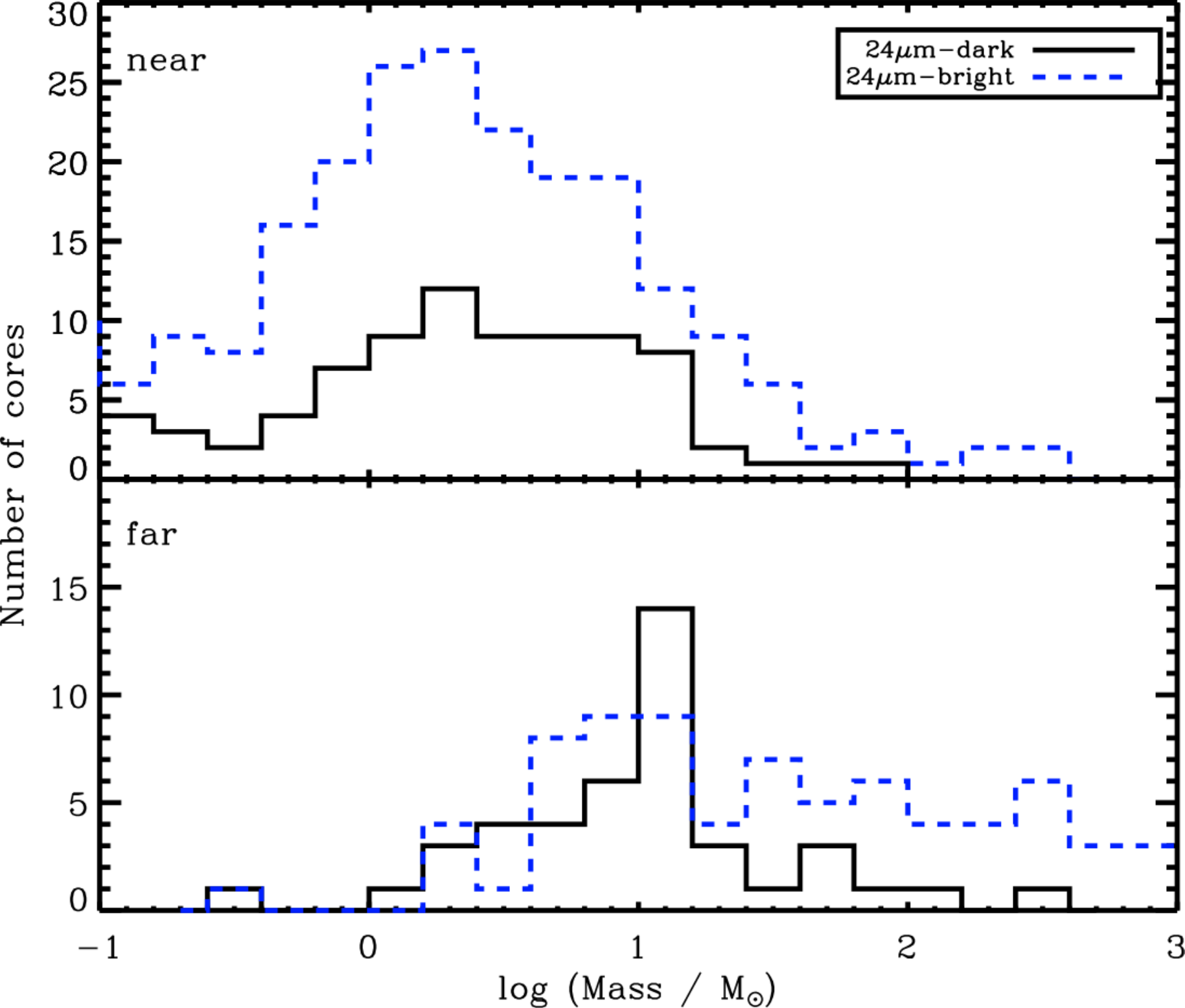}
\caption{Distribution of point source properties within ``near'' and ``far'' IRDCs (top and bottom of each panel, respectively).  The top panel shows the temperature distribution; the center panel shows the luminosity distribution, and the bottom panels show the mass distribution. The black histogram shows objects which have no 24\,$\mu$m counterpart, and the blue dashed histograms show the distribution of objects with a 24\,$\mu$m counterpart. \label{propdistrib}}
\end{figure}

\begin{figure*}
\begin{center}
\includegraphics[scale=0.8]{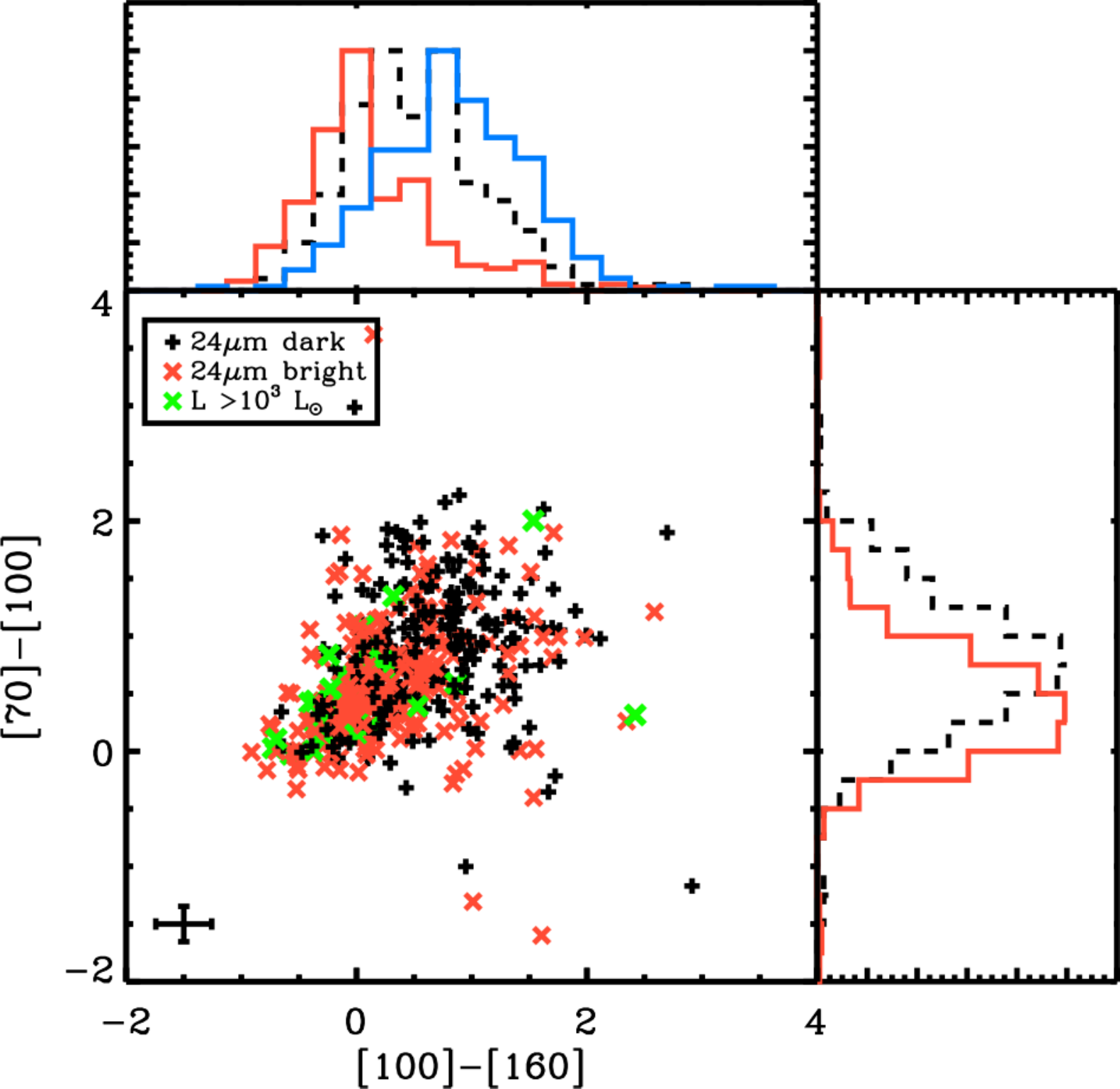}
\end{center}
\caption{PACS color-color diagram showing the 24\,$\mu$m-dark cores (red) and 24\,$\mu$m-bright cores (black). All sources with high luminosities ($L > 10^3\,\lsun$) are shown in green. The typical error is plotted in the lower-left corner.  The right panel shows the distribution of [70] - [100] colors, and the upper panel shows the distribution of [100] - [160] colors for the 24\,$\mu$m-bright cores (red), 24\,$\mu$m-dark cores (black), and (in blue) cores for which no counterpart at 70\,$\mu$m was detected.  \label{fircolor}}
\end{figure*}

\subsection{Evolutionary stage of cores}
\label{ss:sfsignposts}

\subsubsection{24\,$\mu$m counterparts}
\label{sss:compare24}

Out of the 496 cores in our sample, 422 have complementary 24\,$\mu$m data available. Of those 422 cores, 278 (66\%) have counterparts at 24\,$\mu$m. The appearance of a 24\,$\mu$m counterpart has, to-date, been interpreted as a signpost for a local heating source \citep{Beuther_protostars_IRDC,A&ASpecialIssue-Beuther}, but the mechanism by which 24\,$\mu$m escapes had not been determined.  As we discussed above, unless an outflow of some kind clears away some of the outer dense core material, the high optical depth at 24\,$\mu$m would prevent us from detecting a counterpart. In the following we explore the connection between the outer core properties, as probed by our PACS observations, and the presence of a 24\,$\mu$m counterpart. We note that the non-detection of a 24\,$\mu$m counterpart does not necessarily mean that there is no protostar present because the geometry may be such that high optical depth blocks the emission from the heated cavity. Therefore, some overlap between the ``dark'' and ``bright'' populations can be expected.

In Figure~\ref{propdistrib}, we examine the property distributions taking into account whether or not the core has a counterpart at MIPS 24\,$\mu$m.  The median core temperature for near (far) dark cores is 18.6\,K (19.5\,K) and 21.1\,K (21.9\,K) for bright cores. In addition, MIPS-bright cores tend to be more luminous, with the median value of 13\,$\lsun$ (173\,$\lsun$) for near (far) MIPS-bright cores compared to 9\,$\lsun$ (56\,$\lsun$) for near (far) MIPS-dark cores. The same trend holds for the masses, where near (far) MIPS-bright cores have a median mass of 3\,$\msun$ (16\,$\msun$) compared to the 2\,$\msun$ (13\,$\msun$) median for MIPS-dark cores. 

Figure~\ref{fircolor} shows a PACS color-color diagram for each core detected in the three PACS bands. The color based on wavelengths $A$ and $B$ is calculated as follows: 

\begin{equation}
[A] - [B] = -2.5~log \big( \frac{\nu_{A}~S_{A}}{\nu_{B}~S_{B}} \big)
\end{equation}

\noindent We differentiate between cores with and without MIPS 24\,$\mu$m counterparts in Figure~\ref{fircolor}. In both colors, [70]~-~[100] and [100]~-~[160], 24\,$\mu$m-dark cores are redder (larger numbers in color space) than 24\,$\mu$m-bright cores, though the two populations occupy the same ranges in color space independently of distance. In Figure~\ref{fircolor} we also plot the color distribution for an additional population of 312 candidate cores from our EPoS sample which appear only at 100 and 160\,$\mu$m. Since these cores only have two data points in their SEDs, we can not model their properties to the extent we have for cores detected in all three PACS bands. However, we have computed their [100]~-~[160] color, and we find that the colors are consistent with a yet colder (on average) population of cores than the 24\,$\mu$m and 70\,$\mu$m-bright cores.

The uniformity of PACS colors for all cores confirms the idea that in the far-infrared, the emission is dominated by the cold core ``envelope.'' Whether or not a core is detected at 24\,$\mu$m may be a reflection of the more advanced evolutionary stage of the 24\,$\mu$m-bright cores: one where the evolving protostar launches an outflow, exposing the warm inner regions. cores dark at 24\,$\mu$m can also be protostellar, but aligned such optically thick material in the core obscures the 24\,$\mu$m emission from our view, or they may be pre-stellar/starless with heating from an external sources (see Figure~\ref{fig:radtrans}), or they are more deeply embedded and therefore obscured from view. 

The presence of a 24\,$\mu$m counterpart coincides with the warmer, more luminous, and more massive cores of the population, though the effect appears small. We see in Figure~\ref{propdistrib2} that while the distance to a source affects our ability to detect low-mass and low-luminosity cores, we see no effective difference in the temperature range as a function of distance. If the existence of a 24\,$\mu$m counterpart is indeed the consequence of the later evolutionary state of the core, we expect that in more evolved cores, the outer core gas would have experienced the heating from the internal source, an effect which is also seen for the dense gas in IRDCs reported in \citet{Ragan2011a}. Since we do not resolve the cores, we derive only the line-of-sight average core temperature, but the 1.5 to 2\,K increase in temperature in 24\,$\mu$m-bright cores is similar to the gas measurements. To fully characterize the evolutionary differences in detail likely requires the inclusion of the near to mid-infrared data points with a full treatment of the optical depth effects. Such an undertaking is beyond the scope of this paper but will be addressed in forthcoming work.

\begin{figure}
\begin{center}
\includegraphics[scale=0.55]{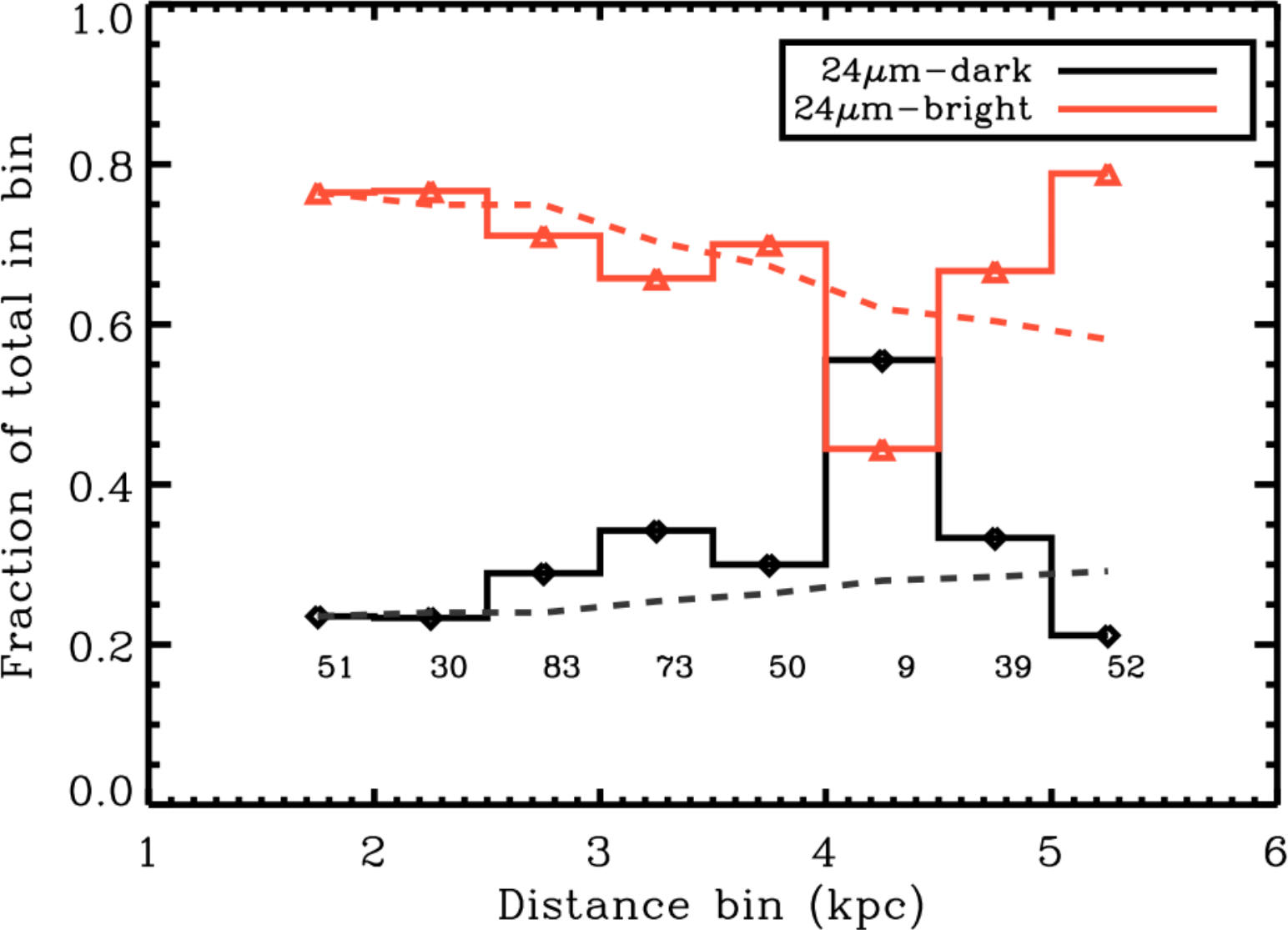}
\end{center}
\caption{\label{fig:frac24} In solid lines, we plot the fraction of cores which have a 24\,$\mu$m counterpart (red solid line, triangles) and those without 24\,$\mu$m counterparts (black solid line, diamonds). We plot this fraction only for targets which have available MIPS 24\,$\mu$m data, as indicated in Table~\ref{tab:obs}. The dashed line represents how the fraction of 24\,$\mu$m bright/dark cores would change if the cores in the nearest bin were placed in the progressively larger distance bins. The uptick at large distances of 24\,$\mu$m-bright fraction is due to the particularly active IRDCs at these distances. }
\end{figure}

Our ability to detect 24\,$\mu$m counterparts worsens with increasing distance, which in turn will cause our estimates of the frequency of 24\,$\mu$m counterparts to be too low. Figure~\ref{fig:frac24} shows the fraction of cores which are 24\,$\mu$m-bright and 24\,$\mu$m-dark as a function of distance. We calculate this fraction only for the targets covered with MIPS observations. Because we become less sensitive to faint sources with increasing distance, we expect this to be an observational bias. 

The dashed line in Figure~\ref{fig:frac24} shows how we would expect the fraction of 24\,$\mu$m-bright to dark sources would fall off with increasing distance assuming distance was the only factor at play in hindering the detectability of counterparts. To make this estimate, we assume that the distribution of 24\,$\mu$m fluxes and also the fraction of 24\,$\mu$m-bright cores is representative of the sample. We then project this subset of cores to each distance bin and compute the fraction that would fall below the flux detection limit at that distance. This does not appear as a straight line because the fluxes are not smoothly distributed within the nearest bin.  We confirm that the general trend can be explained as a observational bias. The fact that we detect a rise in the detection frequency of counterparts is due to the very active regions (e.g. IRDC18454 near W43) where the we find a large fraction of 24\,$\mu$m-bright sources.

We cross-correlate our catalog with the young stellar object catalog by \citet{Robitaille2008} for the 27 IRDCs which are part of the GLIMPSE I and II datasets.  Within a matching radius of 10$''$, there are 40 objects in our sample that appear in the \citet{Robitaille2008} catalog, all but six of which are bonafide YSOs and not AGB contaminants (T. Robitaille, priv. communication).  All of the YSOs have 24\,$\mu$m counterparts and tend to have temperatures above the sample median (20\,K). 

\subsubsection{Other signposts of star formation}
\label{sss:ego}

Extended Green Objects (EGO) \citep{Cyganowski2008, Chambers2009} are areas of enhanced diffuse emission in the ``green'' band (4.5\,$\mu$m) of the {\em Spitzer}/IRAC camera. This is thought to be a result of shocked-H$_2$ emission line which falls into that band \citep{DeBuizer2010} or due to scattered continuum emission \citep{Takami2012}. In the catalog of \citet{Cyganowski2008}, five EGOs (in four IRDCs) are found within 10$''$ of a core in our catalog, and in the \citet{Chambers2009} catalog, six  ``green fuzzy'' (G.F.) regions in three IRDCs match with core positions. The matching cores and their EGO classification are given in Table~\ref{tab:egoassoc}. Ten of the eleven matching cores with EGO activity have a 24\,$\mu$m counterpart, and their temperature range from 13 to 24\,K.

\begin{table}
\caption{IRDC cores associations with EGOs \label{tab:egoassoc}}
\begin{tabular}{lcll}
Target & ID from &  &  \\
Name & Table~\ref{pointsrc} & Classificiation$^a$ & Reference \\
 \hline
IRDC310.39-0.30 & 2 & 4.5$\mu$m-possible & (1) \\ 
IRDC011.11-0.12 & 8 & MIR-possible & (1) \\ 
IRDC18182 &  4  & MIR-possible & (1) \\ 
		   & 11 & 4.5$\mu$m-likely & (1) \\ 
IRDC019.30+0.07 & 2 & G.F. & (2) \\ 
IRDC028.34+0.06 & 9 & G.F. & (2) \\ 
			     & 17 & G.F. & (2) \\ 
			     & 19 & G.F. & (2) \\ 
			     & 21 & G.F. & (2) \\ 
IRDC048.66-0.29 & 4 & 4.5$\mu$m-likely$^b$ & (1) \\ 
                            & 6 & G.F. & (2) \\ 
 \hline 
\end{tabular}

\tablefoottext{a}{Based on the scheme presented in \citet{Cyganowski2008} which used various criteria in the mid-infrared (MIR) and 4.5\,$\mu$m IRAC band, and also noted the level of certainty. We also not the positions in \citet{Chambers2009} that coincide with ``green fuzzy'' (G.F.) emission. }
\tablefoottext{b}{no 24~$\mu$m counterpart.}

\tablebib{
(1) \citet{Cyganowski2008};
(2) \citet{Chambers2009}
}
\end{table}

SiO emission is a well-known tracer of shocked gas, widely used to search for outflows originating from young stars.  The targets in our sample have not been uniformly surveyed for SiO emission, therefore we can not say with statistical certainty whether SiO is associated with just 24\,$\mu$m-bright cores or not.  Nonetheless, we compare our core catalog with the \citet{Beuther_IRDCoutflow} survey of SiO(2-1) emission overlaps with 12 of the IRDCs in the sample (see Table~\ref{tab:sample}). Further follow-up observations of SiO(2-1) with the IRAM 30-m telescope have been performed on 8 additional IRDCs in a recent survey by \citet{Vasyunina2011} and (H. Linz, in prep.). These single point observations do not have sufficient resolution (28$''$) to resolve the outflow structure, but we find that in 90\% of the cases, the nearest core to the peak position of the SiO emission coincides with a core with a 24\,$\mu$m counterpart.

Cores which are bright at 24\,$\mu$m appear to be (1) warmer, (2) consistent with YSO colors when detectable at mid-infrared bands, and (3) often in close spatial association with molecular outflow tracers. Unfortunately, a full sampling of all cores is not available or falls outside of the selection criteria of these supplementary surveys, so we can not quantify these trends. In any case, the catalog presented in this work provides an expanded range of conditions or evolutionary stages which can be further observationally explored.

\section{Discussion}
\label{s:disc}

\subsection{The nature of the cores}
\label{ss:nature}

We present a core (size $\sim$ 0.05 to 0.3\,pc) population for which we model the temperature, mass, and luminosity of the cold, outer shell of the core where the bulk of the mass resides. Despite not including the 24\,$\mu$m flux density in our SED fits, the shells surrounding 24\,$\mu$m-bright cores appear to be warmer than 24\,$\mu$m-dark cores. In addition, previous surveys of YSOs and outflows show that these independent indicators of embedded star formation tend to be coincident with 24\,$\mu$m-bright cores. While supplemental survey data are not uniformly available for the full EPoS sample, we are unable to show correlations with statistical significance. However, at face value, the warmer, 24\,$\mu$m-bright cores appear to be associated with more star formation activity than the colder, 24\,$\mu$m-dark cores, and hence the 24\,$\mu$m-bright cores could be in a more advanced evolutionary stage.

We find a tight relation between the far-infrared flux and the core luminosity (see Figure~\ref{lumfall}) which we would not expect if the flux was dominated by a stochastic external radiation field at PACS wavelengths. Still, at 24\,$\mu$m we find the largest scatter in the relationship where the influence of the external radiation field is stronger \citep{Pavlyuchenkov2012}. \citet{Wilcock2011} show that on larger scales (0.4 to 1~pc) temperature gradients due to enhanced interstellar radiation fields are important in modeling the SEDs of IRDC clumps. On the core scale, however, the impact of embedded sources will be greater and the influence of the external radiation field (at PACS wavelengths) will be more effectively shielded by the high column density material in which the cores are embedded.

The fits to the core temperatures are strongly peaked between 15 and 25\,K. Typical gas temperatures in IRDCs of the clouds on larger scales have been found to be between 8 to 17\,K \citep[e.g.][]{Sridharan:2005,Pillai_ammonia,Ragan2011a}, based on measurements of the NH$_3$(1,1) and (2,2) inversion transition lines. A similar range is found in the dust temperatures \citep[e.g.][]{Peretto2010}. Furthermore, high resolution measurements of temperature in IRDCs \citep{Devine2011,Ragan2011a} show small scale enhancements of gas temperature near regions of active star formation, i.e. when 24\,$\mu$m emission is present. Assuming the gas and dust are thermally coupled, which should be the case for the density regime of IRDCs \citep{goldsmith_tg}, the presence of protostars and gradual heating of the surroundings has already been observationally connected.

Every line of evidence presented above suggests that at least cores with a 24\,$\mu$m counterpart are protostellar. But what about the newly uncovered population of cores found only at 70\,$\mu$m and longward? Such sources comprise 34\% of our core catalog. Apart from the higher median temperature associated with 24\,$\mu$m-bright sources, there is little difference between the masses and luminosities of 24\,$\mu$m bright and dark sources. Are they objects at an earlier evolutionary stage, or are they at the same stage but we are somehow biased against detecting 24\,$\mu$m counterparts for some sources?

As we discussed above, the detection of 24\,$\mu$m counterpart is possible if our line of sight permits us to peer into a cavity formed by outflowing material to view emission from warm dust near the central source. Cores dark at 24\,$\mu$m can still contain protostars, but by chance our line of sight does not include the warm dust emission. Alternatively, they may be starless with some degree of external heating by the ISRF or a neighboring source. The 24\,$\mu$m data alone can not tell us which is the case for a given dark source. \citep{Arce_PPV} show that massive protostars can carve a large variety of cavities in their host core, so it is difficult to rule out either scenario based on the present statistics. Here, we instead turn the correlations seen in the cold {\em outer} core to inform our understanding of the 24\,$\mu$m-dark cores which comprise 34\% of our sample.

The warmer outer-core temperatures associated with 24\,$\mu$m-bright cores supports the interpretation that they are more evolved than 24\,$\mu$m-dark cores. Because of the observational issues mentioned above, there is unavoidable overlap such that protostellar cores are not detected at 24\,$\mu$m. The idea that 24\,$\mu$m-bright cores are in a more advance evolutionary stage than 24\,$\mu$m-dark has been suggested before in \citet{Chambers2009} when considering the correlation between 24\,$\mu$m-bright cores and outflow signatures. Further observations and modeling of evolving massive cores, and their indirect signatures, would be useful in determining the underlying paradigm of the SEDs we observe. 

An interesting population of cores is worthy of further study:  those cores which are detected at 100\,$\mu$m and 160\,$\mu$m, but not at 70\,$\mu$m or 24\,$\mu$m. We infer from their red colors (see Figure~\ref{fircolor}) that they are yet colder sources than the PACS cores we present here. However, since the SED only has these two fluxes and because the contamination with artifacts may be higher, we can not model them in the same way and do not present the catalog in this paper. Further studies at high-resolution at long wavelengths are needed to confirm whether these are indeed cold cores. Again, from our simple approximation (see Figure~\ref{fig:radtrans}) it is unlikely that such cores contain protostars, and thus they potentially represent the elusive prestellar stage of massive protostars.

\subsection{The most extreme cores}
\label{ss:highlum}
Within the sample there are 30 cores with luminosities greater than 10$^3 \lsun$ and 32 objects which have masses greater than 100\,$\msun$, with 25 objects overlapping from those subsets.  Only one core out of the 32 cores above 100\,$\msun$ lacks an IRAS or MIPS 24\,$\mu$m counterpart. All of the cores with L $> 10^3 \lsun$ have IRAS and/or MIPS 24\,$\mu$m counterparts.

\begin{figure}
\begin{center}
\includegraphics[scale=0.5]{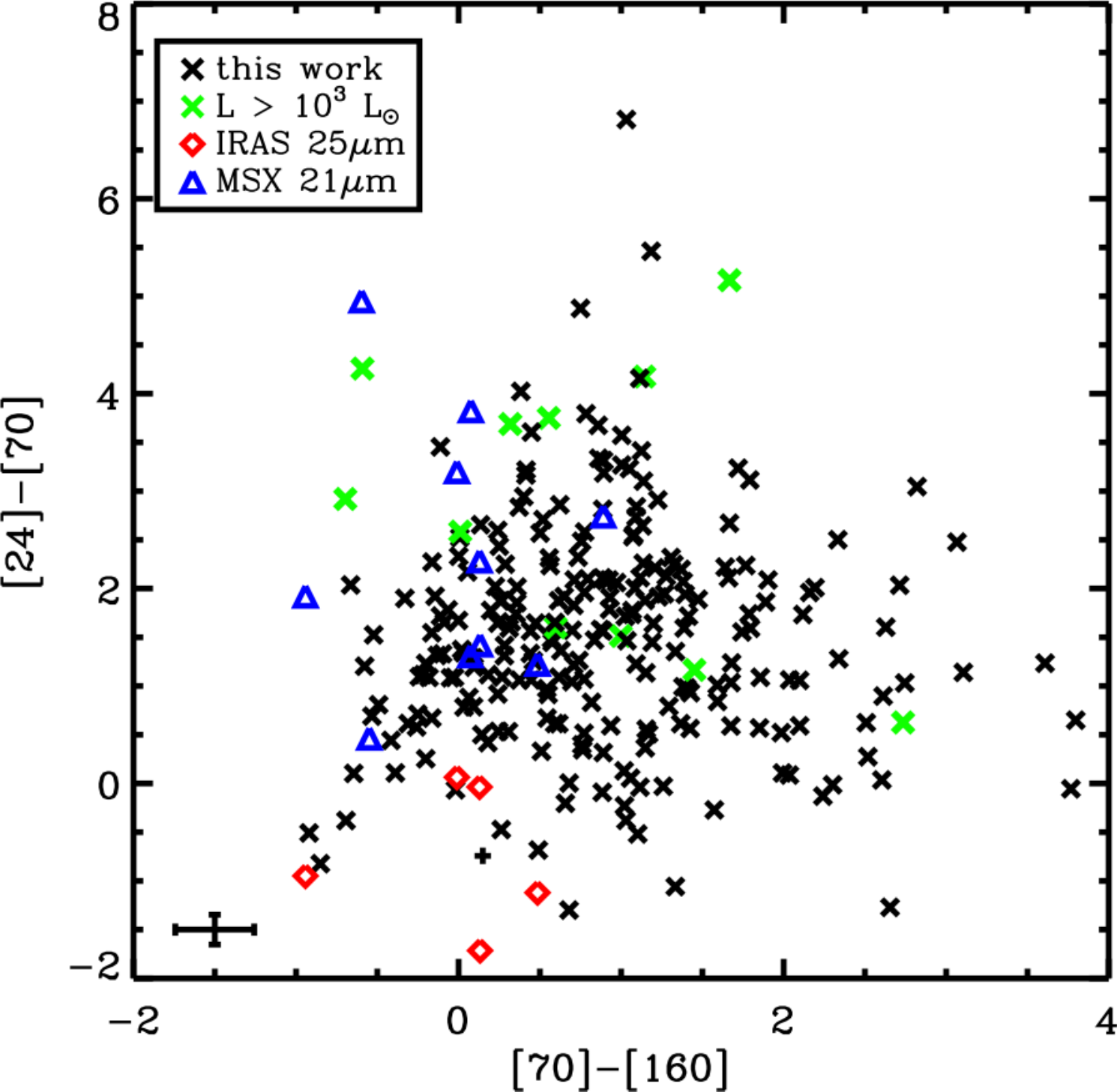}
\end{center}
\caption{[24] - [70] vs. [70] - [160] color-color diagram including only cores for which 24\,$\mu$m counterparts exist (black). The high-luminosity sources ($L > 10^3 \lsun$) are indicated in green. The typical error is plotted in the lower-left corner. Since very bright sources often saturated the MIPS 24\,$\mu$m detector, additional data points extracted from the IRAS point source catalog at 25\,$\mu$m (red diamonds) and the MSX point source catalog at 21\,$\mu$m (blue triangles) were used in lieu of a MIPS 24\,$\mu$m data point in this plot. Note that both the IRAS and MSX fluxes were integrated over significantly larger areas corresponding to the beam of the respective telescopes. \label{fircolor2}}
\end{figure}

The most luminous cores are not distinct in the PACS color-color diagram (Figure~\ref{fircolor}). In Figure~\ref{fircolor2} we show the [24] - [70] versus [70] - [160] color-color diagram, in which we highlight the  cores with L $> 10^3 \lsun$. Here we supplemented the 24\,$\mu$m data points with IRAS or MSX archival data at 25\,$\mu$m and 21\,$\mu$m, respectively, because often very luminous cores saturate the MIPS detector. We find that the luminous cores are also not distinct in this color space. In other words, while these cores span several orders of magnitude in luminosity, their spectral shape is fairly uniform.

\begin{figure*}
\begin{center}
\includegraphics[scale=0.5]{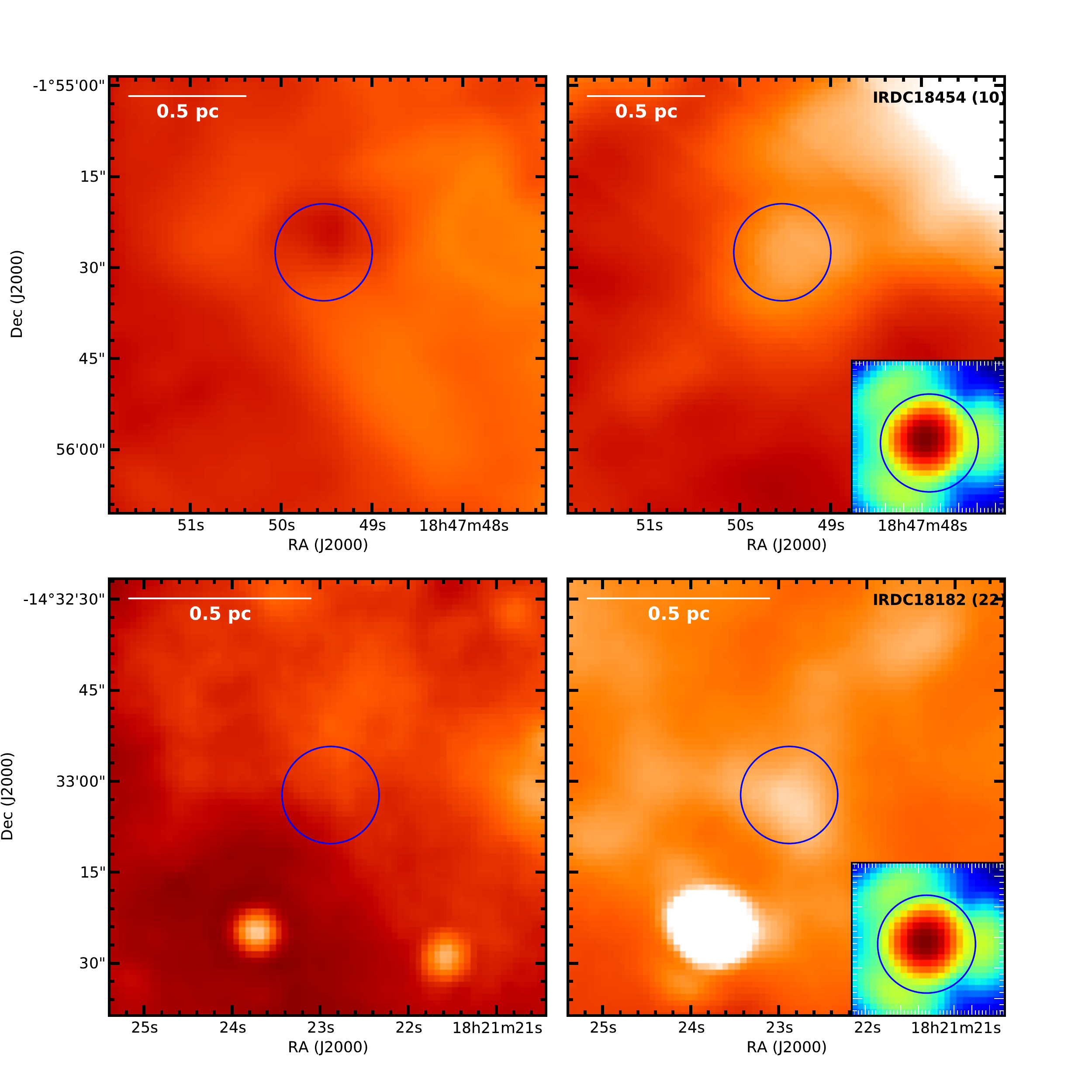}
\end{center}
\caption{Left panels show the MIPS 24\,$\mu$m image and the right panels show the PACS 100\,$\mu$m image of the most massive cores with no 24\,$\mu$m counterparts which are listed in Table~\ref{tab:extreme}. The top row is core 10 in IRDC18454 (estimated 138\,$\msun$), and the bottom row is core 22 in IRDC18182 (estimated 89\,$\msun$). The psf of {\em Herschel} at 100\,$\mu$m, rotated in the sense of the scan direction, is shown in the lower-right corner of the right-hand panels. The blue circles indicate the position of the massive core in both the MIPS 24\,$\mu$m and PACS 100\,$\mu$m image. \label{fig:darkcores1}}
\end{figure*}

\begin{figure*}
\begin{center}
\includegraphics[scale=0.5]{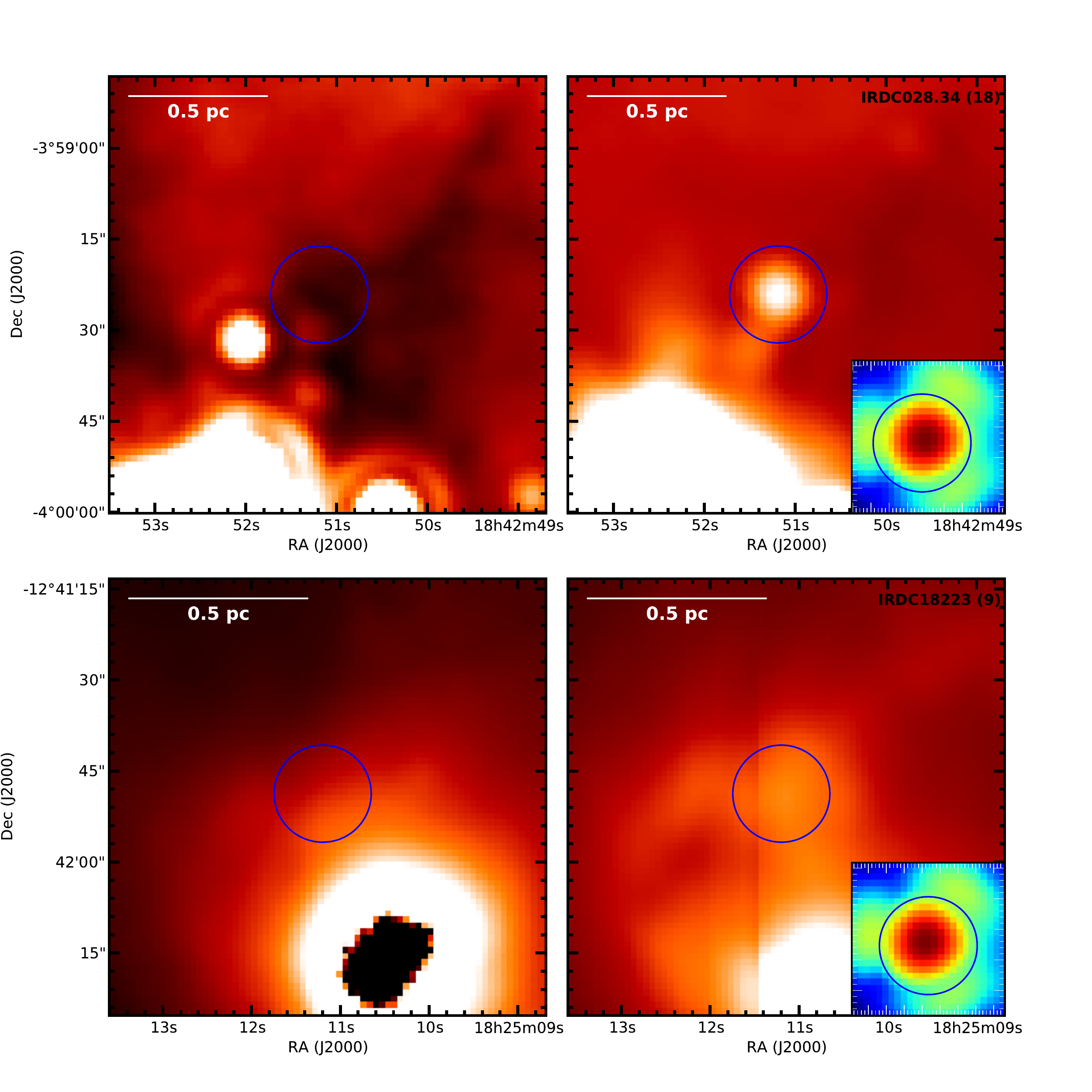}
\end{center}
\caption{Left panels show the MIPS 24\,$\mu$m image and the right panels show the PACS 100\,$\mu$m image of the most massive cores with no 24\,$\mu$m counterparts which are listed in Table~\ref{tab:extreme}. The top row is core 18 in IRDC028.34+0.06 (estimated 53\,$\msun$) , and the bottom row is core 9 in IRDC18223 (estimated 51\,$\msun$). The psf of {\em Herschel} at 100\,$\mu$m, rotated in the sense of the scan direction, is shown in the lower-right corner of the right panels.  The blue circles indicate the position of the massive core in both the MIPS 24\,$\mu$m and PACS 100\,$\mu$m image.\label{fig:darkcores2}}
\end{figure*}

\begin{figure*}
\begin{center}
   \begin{minipage}{0.45\linewidth}
    \raisebox{-5.6cm}{ \includegraphics[angle=90,width=\linewidth]{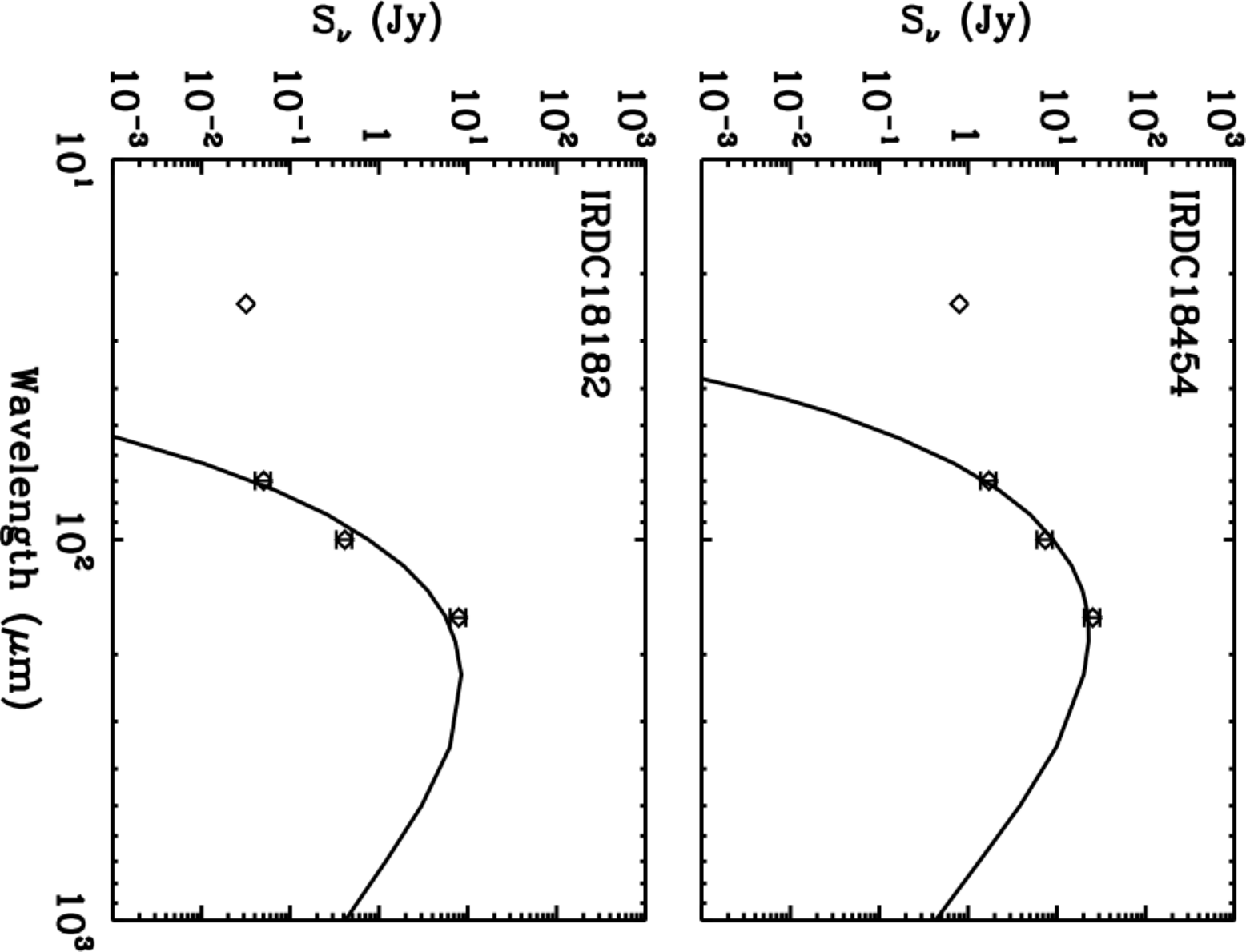}}
   \end{minipage} \hspace{-0.1cm}
   \begin{minipage}{0.45\linewidth}
    \raisebox{-5.6cm}{ \includegraphics[angle=90,width=\linewidth]{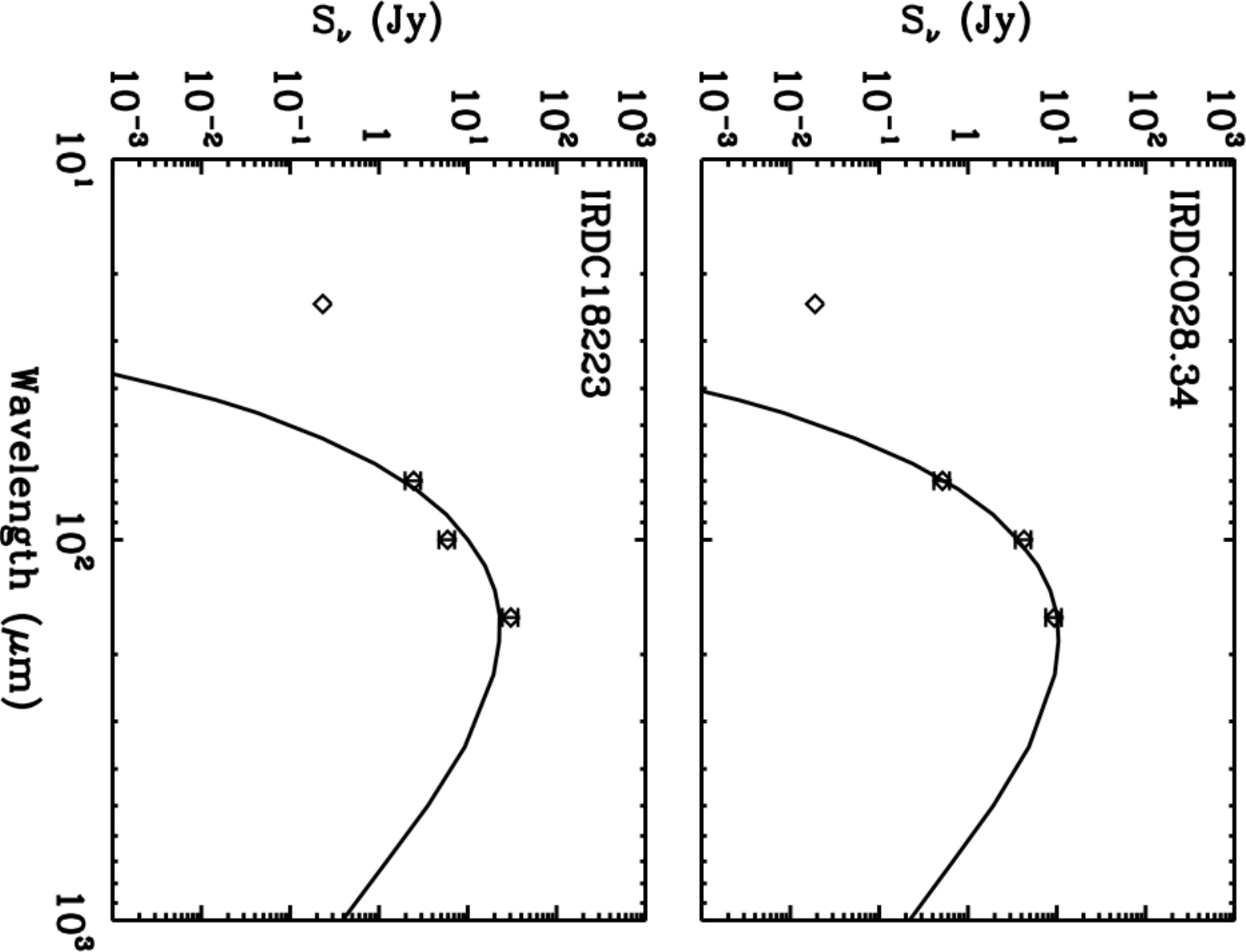}}
   \end{minipage}
 \end{center}
      \caption{\label{fig:darkcoresed} SEDs of massive cores from Figure \ref{fig:darkcores1} (left column) and Figure \ref{fig:darkcores2} (right column). The diamond plotted at the 24\,$\mu$m position represents an estimate of the local diffuse 24\,$\mu$m emission due to a nearby bright region. No 24\,$\mu$m counterpart is reported for any of these sources, but the diffuse emission hampers our ability to detect counterparts for these cores.}
\end{figure*}

In Table~\ref{tab:extreme}, we list the cores above 50\,$\msun$ that lack a 24\,$\mu$m counterpart. In Figures~\ref{fig:darkcores1} and \ref{fig:darkcores2}, we show the MIPS 24\,$\mu$m image and PACS 100\,$\mu$m image of these four 24\,$\mu$m-dark cores which are above 50\,$\msun$ based on our SED models. These sources have dust temperatures between 12 and 17\,K and luminosities from 37 to 354\,$\lsun$. In all cases, they are found in the immediate vicinity of cores with a bright 24\,$\mu$m emission source. In Figure~\ref{fig:darkcoresed} we show that the diffuse emission, likely due to the nearby source (in some cases HMPO) hampers our sensitivity to 24\,$\mu$m counterparts for the neighboring cores we feature in Figures~\ref{fig:darkcores1} and \ref{fig:darkcores2}.

\begin{table*}
\caption{Most extreme cores \label{tab:extreme}}
\begin{tabular}{lcccllll}
\hline \hline
Source & Index$^a$ & RA (J2000) & Dec (J2000) & Temperature & Luminosity & Mass & Remark \\
name & & ($^{h}$:$^{m}$:$^{s}$) & ($^{\circ}$:$^{'}$:$^{''}$) & (K) & ($\lsun$) & ($\msun$)  & \\
\hline
IRDC18454 & 10 & 18:47:49.5 &  -1:55:27 & 17 & 354 & 138 & \\
IRDC18182 & 22 & 18:21:22.9 & -14:33:02 & 12 & 37 & 89 &  \\
IRDC028.34 & 18 & 18:42:51.2 &  -3:59:24 & 16 & 110 & 53 &  \\
IRDC18223 & 9 & 18:25:11.2 & -12:41:48 & 17 & 157 &   51 & beside IRAS18223-1243\\
\hline 
\end{tabular}

\tablefoottext{a}{From Table~\ref{pointsrc}.}
\end{table*}

\subsection{Environment of cores}
\label{ss:environ}

For 44 of the 45 targets, there are sub-millimeter maps which we have used to estimate the total cloud mass from either the ATLASGAL survey, SCUBA archival maps, or published MAMBO 1.2\,mm maps (see Section~\ref{ss:atlasgal} and Table~\ref{tab:sample}). We compute total cloud masses ($M_{cloud}$) using these data assuming a single dust temperature of 20~\,K, as we do not yet have dust temperature maps for the large-scale emission in our sample in order to take into account temperature gradients.  We will revisit this added complexity in forthcoming work.

\begin{figure}
\begin{center}
\includegraphics[scale=0.55]{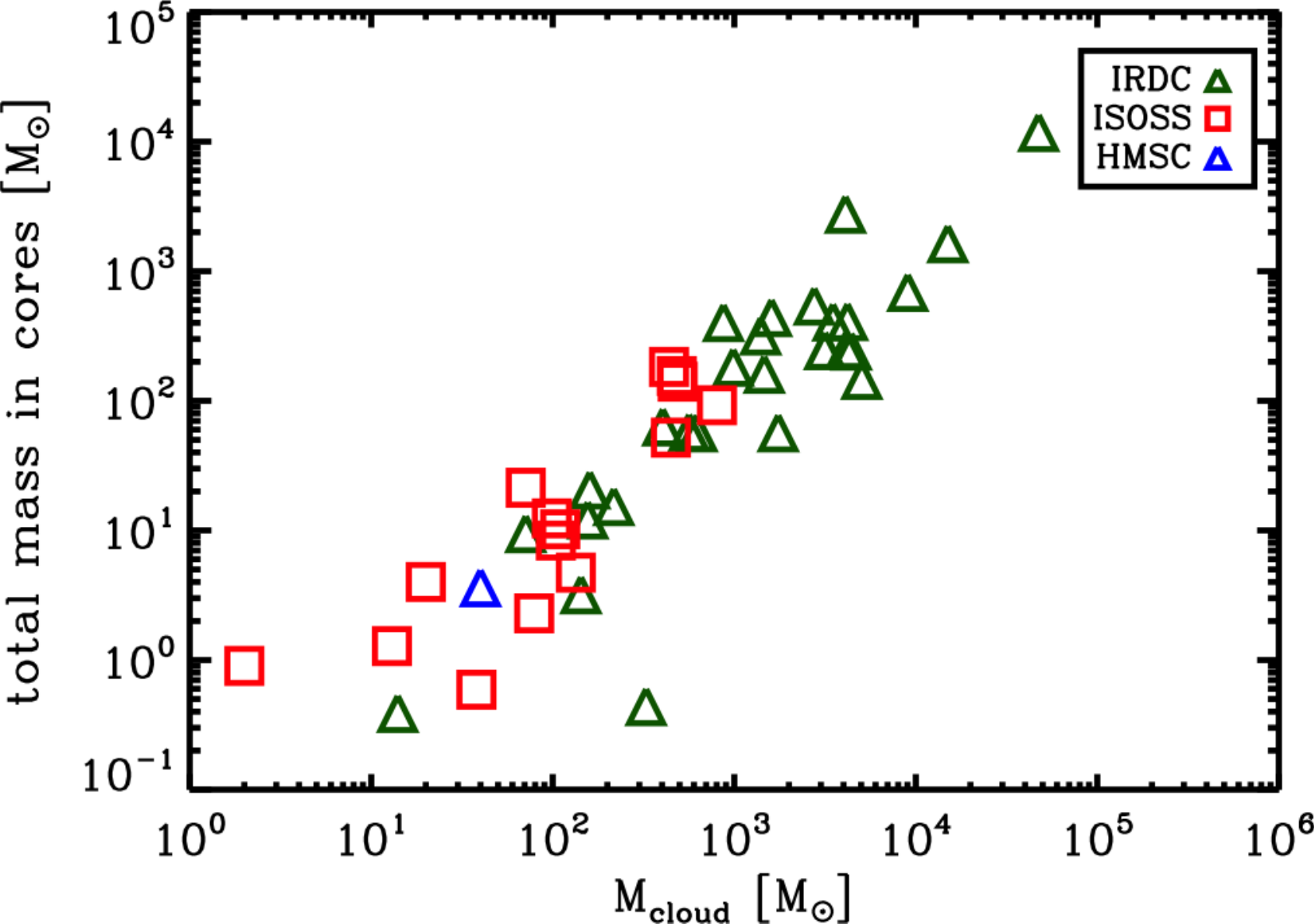}
\end{center}
\caption{Total mass found in cores versus the total mass of a given cloud, distinguishing between cores found in IRDCs (green) and ISOSS sources (red). \label{agfrac}}
\end{figure}

In Figure~\ref{agfrac} we show the mass found in cores as a function of total cloud mass. Clearly, in the clouds of higher mass, there is also more mass in cores. The ISOSS sources are more isolated and, based on the limited SCUBA observations, they have smaller gas reservoirs, which implies that they will likely form lower-mass clusters than the IRDCs. Interestingly, the median fraction of total core mass to total cloud mass is relatively constant at about 10\% for all 44 clouds considered.

The 496 cores we report here are not evenly distributed across the sample, with some clouds hosting just one core meeting our criteria, and others hosting over 30. We use a minimum spanning tree algorithm \citep[e.g.][]{Billot2011} to compute the mean projected core separation for each cloud. Average over the sample, the mean separation is roughly 0.5\,pc. We note, however, that because of the strict criteria of our point source extraction, this is certainly an underestimate of the ubiquity and proximity of cores of varying characteristics and evolutionary stages. 

\begin{figure}
\includegraphics[scale=0.55]{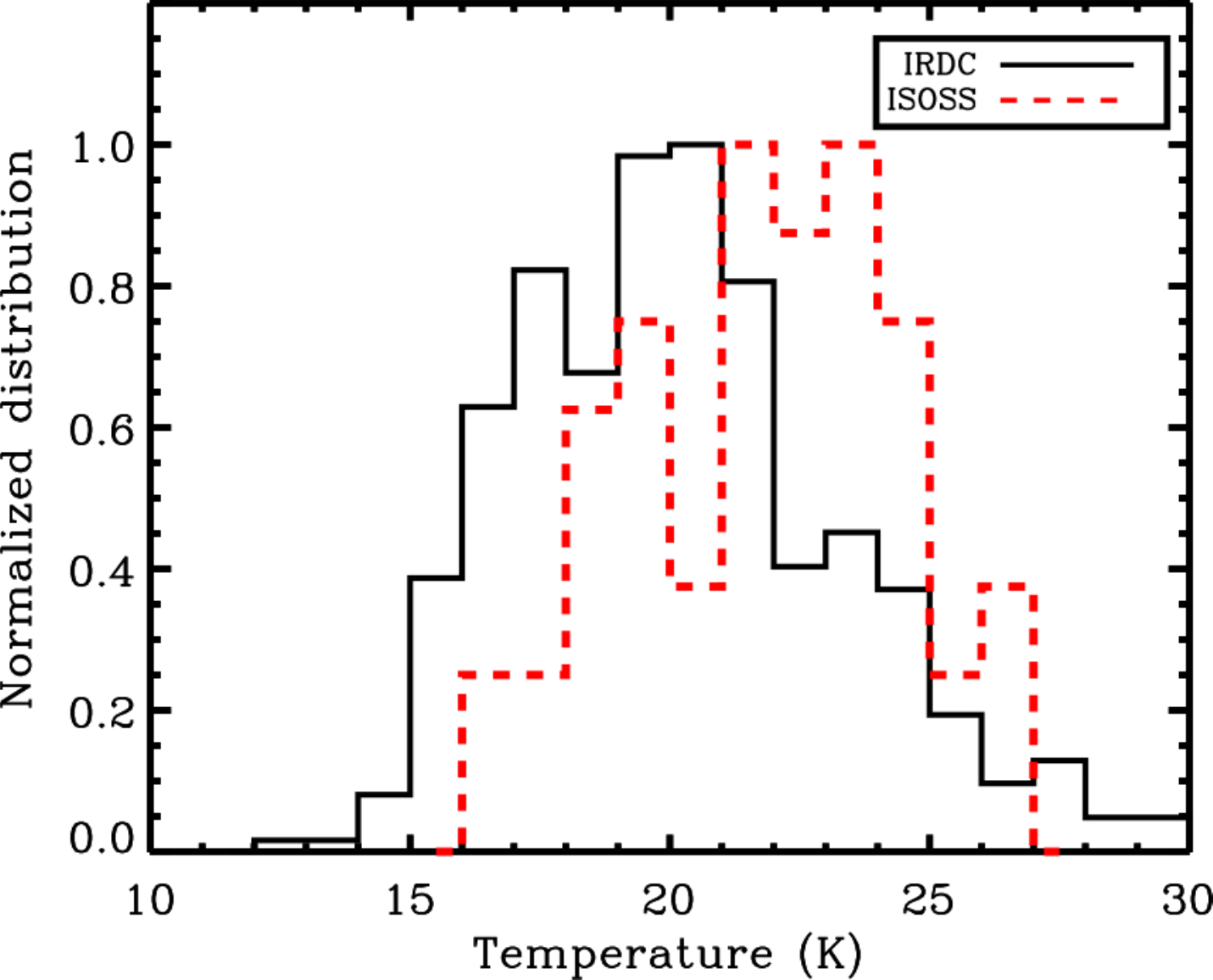}
\includegraphics[scale=0.55]{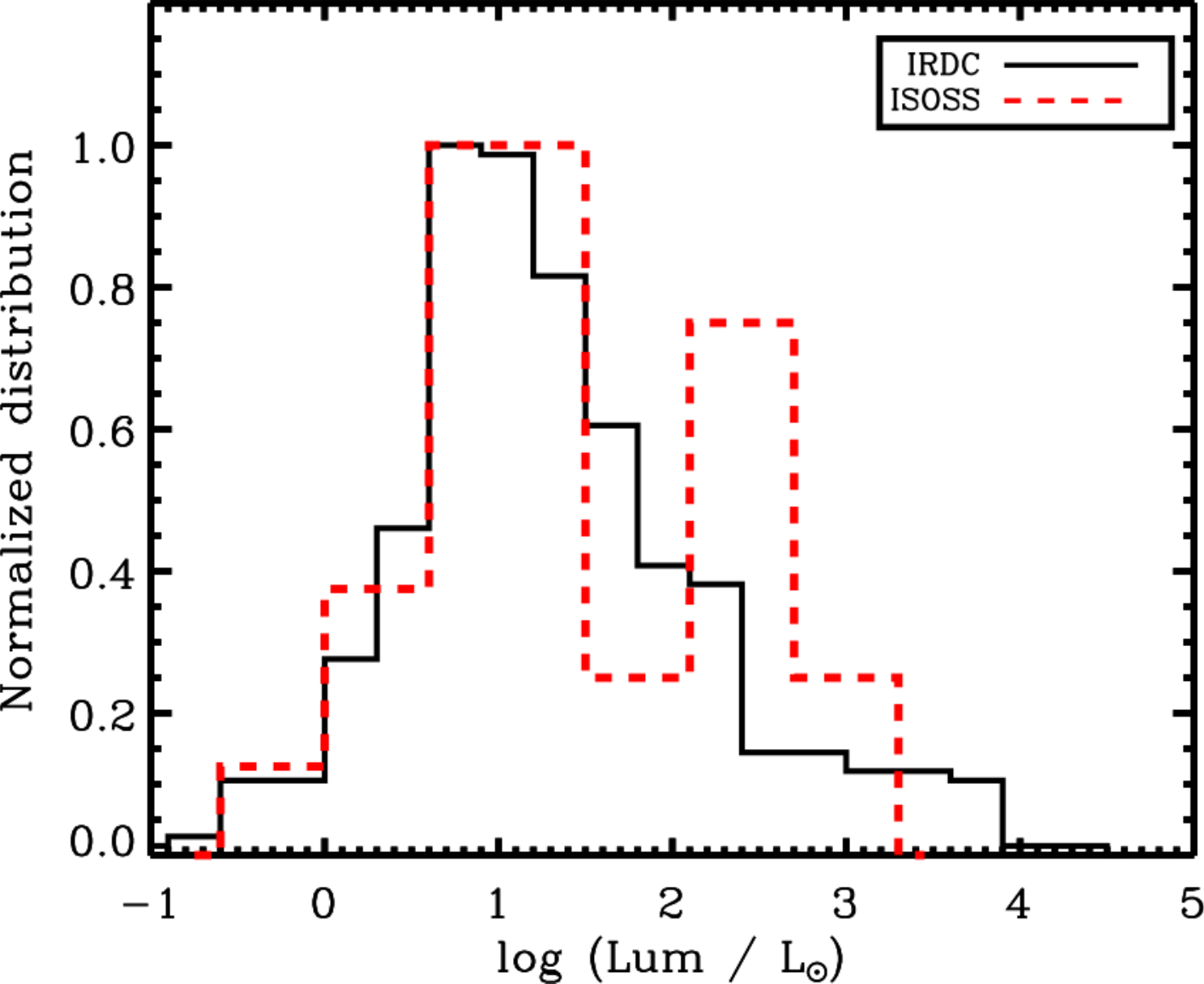}
\includegraphics[scale=0.55]{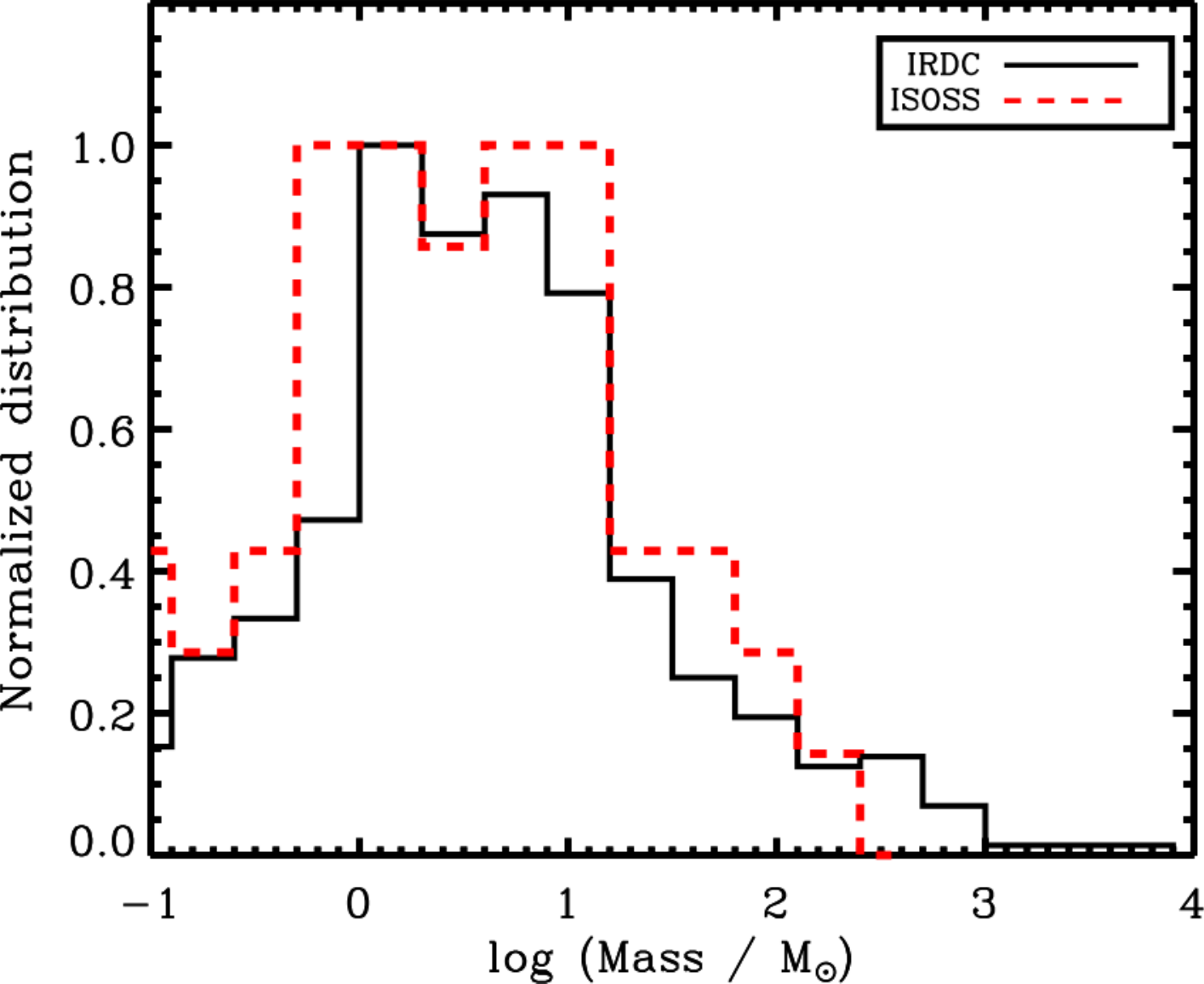}
\caption{Normalized histograms of the temperatures (top), luminosities (middle), and masses (bottom) of cores found in IRDCs (black) and those in ISOSS sources (red).\label{isossprop}}
\end{figure}

What is the difference between IRDCs and ISOSS sources in terms of their core properties? 
There are a total of 53 cores in the 16 ISOSS targets in our sample (average 3 per ISOSS target), while there are 448 cores in the 29 IRDCs (average 12 cores per IRDC).  In Figure~\ref{isossprop} we show the distributions of core properties for the IRDC cores and ISOSS cores. There is no major distinction between the populations, only that the median temperature of ISOSS cores is higher by $\sim$2\,K compared to just the IRDC cores, which is not surprising since most of the ISOSS cores are also 24\,$\mu$m-bright cores, which show a similar median shift compared to the 24\,$\mu$m-dark population (see Figure~\ref{propdistrib}).

\begin{figure*}
\begin{center}
\includegraphics[scale=0.8]{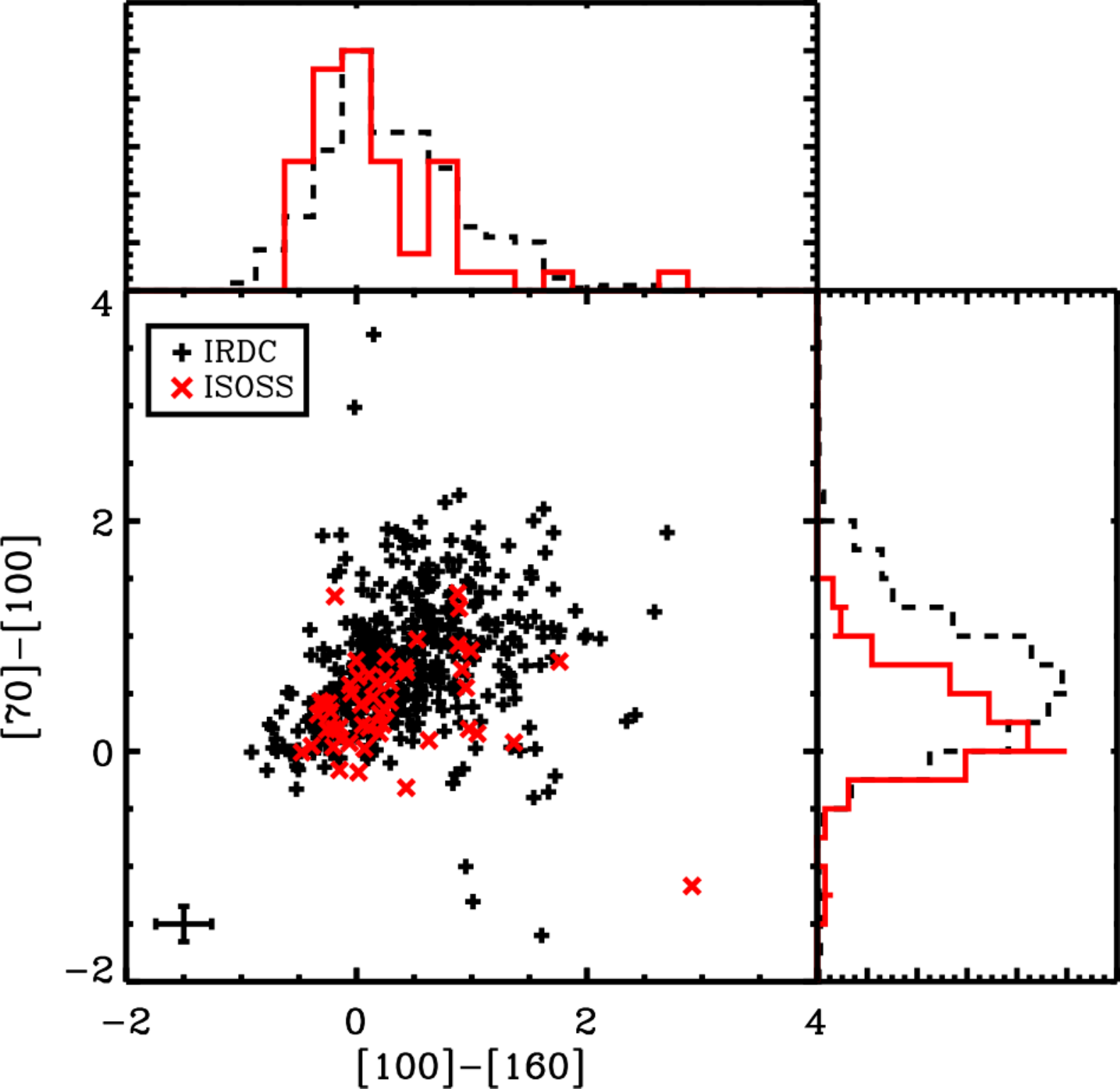}
\caption{PACS color plot of IRDC cores (black) and ISOSS cores (red). \label{isosscolor}}
\end{center}
\end{figure*}

As shown in Figure~\ref{fig:agmass_galcentricdist}, the ISOSS sources are our main probe of the outer Galaxy, where the ambient radiation field is lower than in the active Molecular Ring.  Still, the cores found in ISOSS targets have higher temperatures and luminosities than average, but essentially the same core masses. Figure~\ref{isosscolor} shows again the core colors, this time distinguishing between IRDC and ISOSS cores. Here the ISOSS cores are quite confined to blue colors. This implies that the ISOSS cores are not as heavily embedded objects, or that they are more evolved cores.

\section{Conclusions}
\label{s:conclusion}

We present an overview of the first results of the Earliest Phases of Star Formation survey with {\em Herschel}, focusing on the sample of 45 high-mass regions. The goal of the work presented here is to present the EPoS sample of IRDCs and ISOSS sources as a whole and profile the population of unresolved point sources, which we call cores, within them. We use PACS point source flux densities to construct and fit modified blackbodies to the spectral energy distributions of each core and use the fit to estimate its temperature, luminosity, and mass.  The main results of this work are as follows:

\begin{itemize}

\item We extract 496 point sources in the 45 IRDC structures in our sample.  Their sizes range from 0.05 to 0.3~pc, which are consistent with ``cores'' in the global context of star formation \citep{BerginTafalla_ARAA2007}. We model the SED of the cores based on the 70, 100, and 160\,$\mu$m point source fluxes. We find a wide range in core luminosities (0.1 to 10$^4 \lsun$, median 16\,$\lsun$) and masses (0.1 to a few 10$^3 \msun$, median 4\,$\msun$). The dust temperatures range from 13 to 30\,K (median ~20\,K).

\item The fluxes at 70, 100, and 160\,$\mu$m are good predictors of the core luminosity, with the tightest correlation at 160\,$\mu$m. We perform simple radiative transfer models which show that in cores housing protostars, emission at these wavelengths are determined mainly by the internal source properties. For starless cores, our models show that an amplified external radiation field can cause emission at these wavelengths to reach levels found in protostellar cores. Further work is needed to determine the effects of protostars with various parameters and that of anisotropic heating from neighboring sources. 

\item Most (66\%) of the cores have a counterpart at 24\,$\mu$m. Cores with 24\,$\mu$m counterparts tend to be marginally more massive and more luminous on average than their 24\,$\mu$m-dark brethren. To the extent which other surveys (e.g. {\em Spitzer} and molecular line observations) overlap with our sample, we find that the pre-existing evidence for star formation activity (e.g. YSO colors, outflow activity) almost always coincides with the presence of a 24\,$\mu$m counterpart, leading us to conclude that such cores contain protostars. Cores without a 24\,$\mu$m counterpart may also harbor protostars, but have not yet been probed for supporting evidence for embedded sources, or they may be starless cores with some level of external heating. Our radiative transfer models show that external heating is unlikely to account for the 24\,$\mu$m emission. In order to detect a 24\,$\mu$m counterpart, the inner core region containing the warm dust heated by an internal protostar must be exposed via a protostellar outflow clearing a cavity in the outer pare.  This leads us to conclude that when 24\,$\mu$m counterparts are detected, it is because a core is in a more evolved state.

\item Cores that have a counterpart at 24\,$\mu$m are warmer on average than cores without a 24\,$\mu$m counterpart. We conclude that while the mass and luminosities of these two populations are similar, the warmer cores have protostars that have heated a larger volume of dust within the core, which moderately increases the average core temperature traced by the PACS observations. We find that cores with larger fluxes at 24\,$\mu$m tend to have higher outer-core temperatures.  Warmer cores also correspond to ``bluer'' colors on the PACS color-color plot ([70] - [100] vs. [100] - [160]). 

\item In addition to the 496 cores cataloged in this paper, we find an additional 312 candidate cores which appear only at 100 and 160\,$\mu$m. We can not model these cores because the SED is too poorly constrained, but we find that their [100] - [160] color is consistent with yet colder dust temperatures, thus these may represent an evolutionary stage prior to protostellar formation.

\item The cores are warmer than typical temperatures of large-scale cloud structures in which they are embedded. We conclude that at PACS wavelengths, the core fluxes are due to the effects of internal heating sources and heating from the external radiation field does not play a significant role. We emphasize here that since we probe on {\em core scales} where the cores are embedded in high column density structures which shield them from the ambient radiation field. Further modelling of the dust temperature and column density structure on the ``clump'' and ``cloud'' scales will enable us to determine more precisely on what scales temperature gradients caused by external heating play an important role.

\item Cores within ISOSS sources tend to be 24\,$\mu$m-bright and less massive than cores in IRDCs. They serve as our probes of cores in the outer Galaxy, where we are biased against finding ``dark clouds'' in the same sense as IRDCs in the inner-Galaxy. As molecular gas is known to be more concentrated in the inner Molecular Ring of the Galaxy, these sources offer unique insight into the early phases of isolated ``massive'' star formation.

\end{itemize}

The legacy of the EPoS data will reach far beyond the overview presented here. In this paper, we focus on the core population, much of which was until now completely obscured by heavy extinction. These embedded cores will play a role in the further interpretation of the EPoS data, which will include a full treatment of the radiative transfer from the small to large scales. Future work will also address the detailed large-scale structure in dust temperature and column density, which together can better characterize the environments of the embedded cores.

\begin{acknowledgements}
This work was supported by the Deutsche Forschungsgemeinschaft priority program 1573 ("Physics of the Interstellar Medium").
PACS has been developed by a consortium of institutes led by MPE (Germany) and including UVIE (Austria); KU Leuven, CSL, IMEC (Belgium); CEA, LAM (France); MPIA (Germany); INAF-IFSI/OAA/OAP/OAT, LENS, SISSA (Italy); IAC (Spain). This development has been supported by the funding agencies BMVIT (Austria), ESA-PRODEX (Belgium), CEA/CNES (France), DLR (Germany), ASI/INAF (Italy), and CICYT/MCYT (Spain).
SPIRE has been developed by a consortium of institutes led by Cardiff University (UK) and including Univ. Lethbridge (Canada); NAOC (China); CEA, LAM (France); IFSI, Univ. Padua (Italy); IAC (Spain); Stockholm Observatory (Sweden); Imperial College London, RAL, UCL-MSSL, UKATC, Univ. Sussex (UK); and Caltech, JPL, NHSC, Univ. Colorado (USA). This development has been supported by national funding agencies: CSA (Canada); NAOC (China); CEA, CNES, CNRS (France); ASI (Italy); MCINN (Spain); SNSB (Sweden); STFC (UK); and NASA (USA).
This research has made use of NASAs Astrophysics Data System, as well as the SIMBAD database, operated at CDS, Strasbourg, France. This research made use of Tiny Tim/Spitzer, developed by John Krist for the Spitzer Science Center. The Center is managed by the California Institute of Technology under a contract with NASA.
\end{acknowledgements}


\begin{table*}[h]
\caption{IRDC sample \label{tab:sample}}
\begin{tabular}{lcccrrl}
\hline \hline
Source & RA (J2000) & Dec (J2000) & $v_{lsr}$ & Distance & $M_{cloud}^a$ & References \\
name & ($^{h}$:$^{m}$:$^{s}$) & ($^{\circ}$:$^{'}$:$^{''}$) & (km s$^{-1}$) & (kpc) & ($\msun$) &  \\
\hline
HMSC 07029$-$1215 & 07:05:10.8  & $-$12:19:02 & 12.0 & 0.63$^{+0.53}_{-0.51}$ & 40$^e$ & (1)(2)(3) \\
IRDC  310.39$-$0.30 & 13:56:04.9 & $-$62:13:42 & -53.0 & 5.02$^{+1.22}_{-1.22}$ & 1398 &  (4)$^d$ \\                                
IRDC  316.72$+$0.07 & 14:44:19.2 & $-$59:44:29 & -38.7 & 2.76$^{+0.46}_{-0.44}$ & 3165 &  (4)$^d$ \\                               
IRDC  320.27$+$0.29 & 15:07:45.0 & $-$57:54:16 & -31.0 & 2.25$^{+0.41}_{-0.42}$ & 156 &  (4)$^d$ \\                              
IRDC  321.73$+$0.05 & 15:18:13.1 & $-$57:21:52 & -32.7 & 2.32$^{+0.40}_{-0.41}$ & 564$^b$ &  (4)$^d$ \\                               
IRDC  004.36$-$0.06 & 17:55:45.6 & $-$25:13:51 & 11.4 & 3.28$^{+0.99}_{-1.60}$ & 327 &  (5)(6)$^d$\\
IRDC  009.86$-$0.04 & 18:07:37.4 & $-$20:26:20 & 17.8 & 2.66$^{+0.63}_{-0.80}$ & 143 & (5)(6)(7)$^d$(8)(9) \\
IRDC  010.70$-$0.13 & 18:09:45.9 & $-$19:42:12 & 29.0 & 3.46$^{+0.46}_{-0.55}$ & 4413 &  (10)(11)(12) \\
IRDC  011.11$-$0.12 & 18:10:27.7 & $-$19:20:59 & 29.2 & 3.41$^{+0.46}_{-0.54}$ & 5045 & (10)(11)(13)(14)(15)$^d$(16)(17)(18) \\
IRDC  18102 & 18:13:11.0 & $-$18:00:23 & 21.4 & 2.60$^{+0.54}_{-0.64}$ & 1464 &  (19)(20)(21)$^d$\\
IRDC  013.90$-$0.51 & 18:17:34.0 & $-$17:06:42 & 23.1 & 2.59$^{+0.50}_{-0.59}$ & 1747 &  (4)$^d$ \\
IRDC  015.05$+$0.09 & 18:17:40.3 & $-$15:49:10 & 29.2 & 2.95$^{+0.43}_{-0.49}$ & 160  & (6)$^d$(22)\\
IRDC  18151 & 18:17:50.5 & $-$12:07:55 & 29.8 & 2.65$^{+0.41}_{-0.46}$ & 872$^c$ & (19)(20)(21)$^d$\\
IRDC  18182 & 18:21:15.9 & $-$14:31:25 & 40.9 & 3.44$^{+0.35}_{-0.39}$ & 2760 &  (20)(21)$^d$\\
IRDC  18223 & 18:25:11.5 & $-$12:49:45 & 45.5 & 3.50$^{+0.33}_{-0.36}$ & 3501 &  (20)(21)$^d$(23) \\
IRDC  019.30$+$0.07 & 18:25:53.8 & $-$12:06:18 & 26.3 & 2.36$^{+0.43}_{-0.48}$ & 624 & (10)(11)(14)$^d$(22)(24)\\
IRDC  18306 & 18:33:33.4 & $-$08:31:55 & 54.8 & 3.64$^{+0.31}_{-0.32}$ & 1603 & (19)(20)(21)$^d$\\
IRDC  18308 & 18:33:32.3 & $-$08:37:11 & 73.7 & 4.43$^{+0.27}_{-0.28}$ & 3148 &  (19)(20)$^d$ \\
IRDC  18310 & 18:33:39.5 & $-$08:21:10 & 86.5 & 4.91$^{+0.26}_{-0.26}$ & 4098 &  (19)(20)$^d$ \\
IRDC  18337 & 18:36:24.1 & $-$07:41:33 & 56.2 & 3.66$^{+0.30}_{-0.32}$ & 4152 &  (19)(20)(21)$^d$ \\
IRDC  18385 & 18:41:17.3 & $-$05:10:04 & 47.0 & 3.13$^{+0.33}_{-0.35}$ & 987  & (19)(20)(21)$^d$ \\
IRDC  028.34$+$0.06 & 18:42:50.3 & $-$04:02:17 & 78.4 & 4.52$^{+0.30}_{-0.30}$ & 15011 &  (10)(11)(14)$^d$(22)(25)(26)\\
IRDC  18437 & 18:46:21.9 & $-$02:12:24 & 96.1 & 5.40$^{+0.40}_{-0.36}$ & 9035 & (19)(20)(21)$^d$ \\
IRDC  18454 & 18:47:55.0 & $-$01:55:32 & 94.6 & 5.35$^{+0.40}_{-0.36}$ & 47521 &  (19)(20)(21)$^d$ \\
IRDC  18530 & 18:55:29.0 & $+$02:17:43 & 75.9 & 4.60$^{+0.44}_{-0.40}$ & 4205 &  (19)(20)(21)$^d$ \\
IRDC  19175 & 19:19:51.3 & $+$14:01:37 & 7.7 & 0.78$^{+0.48}_{-0.48}$ & 71 &  (19)(20)(21)(27)$^d$\\
IRDC  048.66$-$0.29 & 19:21:44.4 & $+$13:49:24 & 34.0 & 2.63$^{+0.56}_{-0.52}$ & 217 &  (28)$^d$(22)(29) \\
IRDC  20081 & 20:10:15.8 & $+$27:28:50 & 5.7 & 1.07$^{+0.76}_{-0.76}$ & 14$^c$ &  (19)(20)(21)$^d$ \\
IRDC  079.31$+$0.36 & 20:31:58.0 & $+$40:18:20 & 0.1 & 1.56$^{+2.04}_{-1.56}$ & 803$^e$ &  (10)(11)(14)$^d$(17)(30) \\
\hline
ISOSS J04225$+$5150 & 04:22:32.2 & $+$51:50:31 & -37.7 & 4.24$^{+1.32}_{-1.11}$ & 404$^e$ & (12)(31)\\
ISOSS J06114$+$1726 & 06:11:24.5 & $+$17:26:26 & 22.7 & 5.91$^{+4.09}_{-2.61}$ & --  & (31) \\
ISOSS J06527$+$0140 & 06:52:45.6 & $+$01:40:14 & 45.0 & 4.80$^{+1.25}_{-1.07}$ & 437$^e$ &  (31) \\
ISOSS J18364$-$0221 & 18:39:10.8 & $-$02:18:48 & 33.1 & 2.35$^{+0.37}_{-0.40}$ & 99$^e$  &  (12)(32)(33)\\
ISOSS J19357$+$1950 & 19:35:45.9 & $+$19:50:58 & 48.7 & 4.75$^{+1.71}_{-1.71}$ & 134$^e$  &  (34)\\
ISOSS J19486$+$2556 & 19:48:36.8 & $+$25:56:55 & 20.3 & 2.75$^{+1.05}_{-1.05}$ & 71$^e$  &  (34)\\
ISOSS J19557$+$2825 & 19:55:47.0 & $+$28:25:29 & 27.8 & 3.52$^{+1.89}_{-1.89}$  & 110$^e$ & (35) \\
ISOSS J20093$+$2729 & 20:09:20.1 & $+$27:29:25 & 6.4 & 1.15$^{+0.77}_{-0.77}$ & 2$^e$   &  (31) \\
ISOSS J20153$+$3453 & 20:15:20.9 & $+$34:53:52 & 3.1 & 1.23$^{+1.14}_{-1.14}$ & 20$^e$  & (12)(34)\\
ISOSS J20298$+$3559 & 20:29:48.3 & $+$35:59:24 & 0.5 & 0.95$^{+3.50}_{-0.95}$ & 13$^e$  & (34)(36)\\
ISOSS J21311$+$5127 & 21:31:12.0 & $+$51:27:35 & -54.4 & 5.85$^{+0.65}_{-0.64}$ & 482$^e$ &  (35)\\
ISOSS J22164$+$6003 & 22:16:28.6 & $+$60:03:49 & -59.5 & 5.29$^{+0.66}_{-0.64}$ & 448$^e$ &  (37) (38)\\
ISOSS J22478$+$6357 & 22:47:54.1 & $+$63:57:11 & -40.1 & 3.23$^{+0.61}_{-0.60}$ & 104$^e$ &  (34)\\
ISOSS J23053$+$5953 & 23:05:23.1 & $+$59:53:53 & -52.5 & 4.31$^{+0.64}_{-0.62}$ & 488$^e$ &  (12)(31)(39)\\
ISOSS J23129$+$5944 & 23:12:56.6 & $+$59:44:30 & -38.6 & 3.05$^{+0.60}_{-0.60}$ & 38$^e$  &  (37)(40) \\
ISOSS J23287$+$6039 & 23:28:45.6 & $+$60:39:17 & -36.8 & 2.80$^{+0.60}_{-0.59}$ & 79$^e$ & (35) \\

\hline 
\end{tabular}

\tablefoottext{a}{Masses are computed with ATLASGAL 870~$\mu$m survey data unless otherwise indicated, using equation (1), assuming $T_{dust}$=20~K.}

\tablefoottext{b}{From \citet{Vasyunina2009} 1.2~mm MAMBO data assuming $T_d$=20~K and the recomputed distance.}

\tablefoottext{c}{From \citet{Beuther2002} 1.2~mm MAMBO data and the recomputed distance.}

\tablefoottext{d}{Indicates which publication the $v_{lsr}$ was taken and used for distance determination.}

\tablefoottext{e}{Computed with SCUBA 850~$\mu$m data.}

\tablebib{
(1) \citet{Forbrich2004};
(2) \citet{Forbrich2009};
(3) \citet{A&ASpecialIssue-Linz};
(4)  \citet{Vasyunina2011}; 
(5)  \citet{hennebelle_isogal};
(6) \citet{Teyssier2002};
(7) \citet{ragan_msxsurv};
(8) \citet{Ragan_spitzer}; 
(9) \citet{Ragan2011a,Ragan2012};
(10)  \citet{carey_msx}; 
(11)  \citet{carey_submmIRDC};
(12) \citet{Pitann2011};
(13) \citet{Johnstone_G11}; 
(14)  \citet{Pillai_ammonia}; 
(15)  \citet{Pillai_G11};
(16)  \citet{MIPSGAL}; 
(17)  \citet{Matthews2009}; 
(18)  \citet{A&ASpecialIssue-Henning}; 
(19)  \citet{sridharan_irdc};
(20)  \citet{Sridharan:2005};
(21) \citet{Beuther_IRDCoutflow}; 
(22) \citet{rathborne2006}; 
(23) \citet{A&ASpecialIssue-Beuther};
(24) \citet{Devine2011};
(25)  \citet{Wang_ammonia}; 
(26) \citet{Zhang2009}; 
(27) \citet{BeutherHenning2009};
(28) \citet{Ormel2005}; 
(29)  \citet{vanderwiel2008};
(30)  \citet{Redman2003};
(31)  \citet{Sunada2007};
(32) \citet{Birkmann2006};
(33) \citet{Hennemann2009};
(34) \citet{Hennemann2008};
(35) \citet{KrausePHD};
(36) \citet{Krause2003};
(37) \citet{Kerton2002};
(38) \citet{Kumar2006};
(39) \citet{Birkmann2007};
(40) \citet{KertonBrunt2003}
 }
\end{table*}

\begin{table*}[h]
\caption{Observation details \label{tab:obs}}
\begin{tabular}{lllcccl}
\hline \hline
Source & PACS & SPIRE & PACS & SPIRE & MIPS & remark \\
name & Obs ID & Obs ID & map size & map size & avail? (1) & (2)  \\
&  &  & ($' \times '$) & ($' \times '$) & &   \\
\hline
HMSC 07029$-$1215 &134218554(3-6) & 1342204850 & 6 $\times$ 6 & 9 $\times$ 9 & Y & MIPS:20635; SDP:1342186115  \\
IRDC 310.39$-$0.30 & 13421888(57-60) & 1342189513 &6 $\times$ 6 & 9 $\times$ 9 & MG & \\                                
IRDC 316.72$+$0.07 & 134220337(3-6) & 1342189514 & 7 $\times$ 7 & 9 $\times$ 9 & MG & \\                               
IRDC 320.27$+$0.29 & 134218939(0-3) & 1342189515 & 6 $\times$ 6 & 9 $\times$ 9 & MG & \\                              
IRDC 321.73$+$0.05 & 13421895(88-91) & 1342189516 & 7 $\times$ 7 & 9 $\times$ 9 & MG & \\                               
IRDC 004.36$-$0.06 & 13421935(18-21) & 1342192056 & 6 $\times$ 6 & 9 $\times$ 9 & MG & \\
IRDC 009.86$-$0.04 &  134220429(6-9) & 1342191190 & 6 $\times$ 6 & 9 $\times$ 9 & MG & \\
IRDC 010.70$-$0.13 &  134219180(1-4) & 1342192057 & 10 $\times$ 10 & 9 $\times$ 9 & MG & \\
IRDC 011.11$-$0.12 &  134218555(5-8) & 1342204952 & 6 $\times$ 12 & 9 $\times$ 12 & MG & SDP:1342186113 \\
IRDC 18102		 & 134221753(4-7) & 1342192061 & 6 $\times$ 6 & 9 $\times$ 9 & MG  & \\
IRDC 013.90$-$0.51 & 134219180(5-8) & 1342192062 & 10 $\times$ 10 & 9 $\times$ 9 & MG & \\
IRDC 015.05$+$0.09 & 13422070(57-60) & 1342192063 & 6 $\times$ 6 & 9 $\times$ 9 & MG & \\
IRDC 18151 		& 134219181(3-6) & 1342191188 & 6 $\times$ 6 & 9 $\times$ 9 & N & \\
IRDC 18182 		&  13421918(09-12) & 1342191189 & 7 $\times$ 7 & 9 $\times$ 9 & MG & \\
IRDC 18223 		&  13421855(59-62) & 1342204951 &  14 $\times$ 4 & 14 $\times$ 4 & MG & SDP:1342186112  \\
IRDC 019.30$+$0.07 & 13422177(67-70) & 1342192064 & 6 $\times$ 6 & 9 $\times$ 9 & MG & \\
IRDC 18306 		&  134221905(6-9)  & 1342192066 & 6 $\times$ 6 & 9 $\times$ 9 & MG & \\
IRDC 18308 		&  134221905(2-5)  & 1342192065 & 6 $\times$ 6 & 9 $\times$ 9 & MG  & \\
IRDC 18310 		& 134221906(0-3)  & 1342192067 & 6 $\times$ 6 & 9 $\times$ 9 & MG  & \\
IRDC 18337 		&  134221906(4-7)  & 1342192068 & 6 $\times$ 6 & 9 $\times$ 9 & MG & \\
IRDC 18385 		&  134221907(0-3)  & 1342192070 & 6 $\times$ 6 & 9 $\times$ 9 & MG & \\
IRDC 028.34$+$0.06 & 134221660(3-6) & 1342192071 & 7 $\times$ 7 & 9 $\times$ 9 & MG & \\
IRDC 18437 		& 134220903(3-6) & 1342192072 & 6 $\times$ 6 & 9 $\times$ 9 & MG & \\
IRDC 18454 		&  13421918(19-22) & 1342192073 & 6 $\times$ 6 & 9 $\times$ 9 & MG & \\
IRDC 18530 		&  13422090(29-32) & 1342192074 & 6 $\times$ 6 & 9 $\times$ 9 & MG & \\
IRDC 19175 		&  134219354(6-9) & 1342192075 & 6 $\times$ 6 & 9 $\times$ 9 & MG & \\
IRDC 048.66$-$0.29 & 134219349(2-5) & 1342192076 & 6 $\times$ 6 & 9 $\times$ 9 & MG & \\
IRDC 20081 		& 13421877(67-70) & 1342187653 & 6 $\times$ 6 & 9 $\times$ 9 & N & \\
IRDC 079.31$+$0.36 & 134218779(5-8) & 1342187718 & 6 $\times$ 6 & 9 $\times$ 9 & Y & MIPS:219,40184 \\
\hline
ISOSS J04225$+$5150 & 134219101(0-3) & 1342191185 & 6 $\times$ 6 & 9 $\times$ 9 & N & \\
ISOSS J06114$+$1726 &  134220423(2-5) & 1342192094 & 6 $\times$ 6 & 9 $\times$ 9 & N & \\
ISOSS J06527$+$0140 &  13422043(09-12) & 1342193795 & 6 $\times$ 6 & 9 $\times$ 9 & N & \\
ISOSS J18364$-$0221  & 134218556(3-5) & 1342204950 & 6 $\times$ 6 & 9 $\times$ 9 & MG & SDP:1342186111 \\
ISOSS J19357$+$1950 & 134219349(6-9) & 1342187526 & 6 $\times$ 6 & 9 $\times$ 9 & MG & \\
ISOSS J19486$+$2556 & 13421977(59-62) & 1342187652 & 6 $\times$ 6 & 9 $\times$ 9 & MG & \\
ISOSS J19557$+$2825 & 134218775(5-8) & 1342187527 & 6 $\times$ 6 & 9 $\times$ 9 & MG & \\
ISOSS J20093$+$2729 & 134218776(3-6) & 1342188092 & 6 $\times$ 6 & 9 $\times$ 9 & N & \\
ISOSS J20153$+$3453 & 134218775(1-4) & 1342188093 & 6 $\times$ 6 & 9 $\times$ 9 & N & \\
ISOSS J20298$+$3559 & 13421877(47-50) & 1342188094 & 6 $\times$ 6 & 9 $\times$ 9 & Y & MIPS:20444 \\
ISOSS J21311$+$5127 & 134218777(1-4) & 1342187654 & 6 $\times$ 6 & 9 $\times$ 9 & N & \\
ISOSS J22164$+$6003 & 134218700(3-6)& 1342186128 & 10 $\times$ 10 & 9 $\times$ 9 & Y & sdp; MIPS:50398 \\
ISOSS J22478$+$6357 & 13421878(09-12) & 1342187719 & 6 $\times$ 6 & 9 $\times$ 9 & N & \\
ISOSS J23053$+$5953 & 134218781(3-6) & 1342187720 & 6 $\times$ 6 & 9 $\times$ 9 & Y & MIPS:20444 \\
ISOSS J23129$+$5944 & 134218806(3-6) & 1342188688 & 6 $\times$ 6 & 9 $\times$ 9 & N & \\
ISOSS J23287$+$6039 & 134218810(2-5) & 1342187721 & 6 $\times$ 6 & 9 $\times$ 9 & Y & MIPS:30571\\
\hline 
\end{tabular}

\tablefoottext{1}{MG = MIPSGAL or MIPSGAL2 , Y= MIPS data from another project, N= no MIPS data available}

\tablefoottext{2}{SDP:xxxxxxxxxx = partial SPIRE maps done during the Science Demonstration Phase. MIPS:xxx = Spitzer program ID when MIPS 24\,$\mu$m was not covered by MIPSGAL. }
\end{table*}

\clearpage

\appendix
\section{Individual target descriptions}
IRDCs were discovered and subsequently named based on their appearance in absorption against the bright Galactic background at 8~$\mu$m and 15~$\mu$m based on Galactic plane surveys almost two decades ago. The follow-up studies focused on IRDCs which were largest and most easily identifiable in contrast: those lying in the inner Galaxy. Because most millimeter facilities at the time were located in the northern hemisphere, detailed work was done preferentially on IRDCs in the first quadrant of the Galaxy, which is reflected in the fact that much of our selection of ``well-studied'' sources are found in the first quadrant. For the EPoS sample, we selected IRDCs from these various surveys, favoring the targets that are relatively nearby (d $<$ 5\,kpc). They are denoted as IRDC LLL.LL-B.BB in our catalog, where ``L'' and ``B'' are the Galactic longitude and latitude.

Another strategy utilized in selecting IRDCs takes advantage of the fact that high-mass star formation is known to occur in a clustered mode, and there is ample evidence for triggering of several generations of high-mass star formation \citep[e.g.][]{Deharveng2010}, thus a portion of our sample is comprised of infrared-dark regions found in millimeter surveys near known high-mass protostellar objects \citep[HMPOs ][]{sridharan_irdc,  Beuther2002c, Beuther2002,  Beuther2002b, Beuther2002_masers}. \citet{Sridharan:2005} correlated the 1.2\,mm bolometer data from a well-studied HMPO sample with the corresponding MSX mid-infrared data. Within the 69 studied fields they found 56 gas clumps with associated mid-infrared shadows \citep{Sridharan:2005}, most of them with masses in excess of several 100\,M$_{\odot}$ and temperatures on the order of $\sim$15\,K \citep{Sridharan:2005}. Follow-up observations revealed outflow activity -- indicating already ongoing early star formation -- in $\sim$40\% of the sources, and dense gas abundances (e.g., CH$_3$OH and CH$_3$CN) are similar to those of low-mass starless cores \citep{Beuther_IRDCoutflow}. 

Using the 170~$\mu$m ISOPHOT \citep{Lemke1996} Serendipity Survey \citep[ISOSS][]{Bogun1996}, we selected 16 candidates for massive ($M > 100~\msun$), cold ($T_{dust} < 18\,K$) objects based on preliminary studies \citep{Krause2003, Krause2004, Birkmann2007}. For these sources, the naming convention derives from the FK5 (J2000) coordinates, $r$ and$d$, represented as ISOSSJrrrrr$\pm$dddd.  As shown in Figure~\ref{fig:gal_distrib2}, these objects represent the bulk of the objects from the outer Galaxy, which will have characteristically lower infrared background emission. The ISOSS sample has a large range in distances, but the median distance, 3.52\,kpc, is only slightly higher than the median distance of the sample, 3.23\,kpc.

\noindent {\it  IRDC\, 004.36-0.06:}
\citet{hennebelle_isogal} first identified this source as a peak in the extinction from ISOGAL observations. \citet{Teyssier2002} detect $^{13}$CO and strong C$^{18}$O emission associated with this sources. 
\\

\noindent {\it  IRDC\, 009.86-0.04:}  
Another inner-Galaxy object from \citet{hennebelle_isogal} that exhibits strong CO emission in \citet{Teyssier2002}. \citet{ragan_msxsurv} conducted further molecular studies on this IRDC, detecting N$_2$H$^+$ (1-0) and widespread CS(2-1) emission at the same velocity as \citet{Teyssier2002}.  \citet{ragan_msxsurv} find a cloud mass of 1500\,$\msun$ based on the integrated map of N$_2$H$^+$ emission.  \citet{Ragan_spitzer} present {\em Spitzer} observations of this cloud.  There are 31 YSOs in the vicinity of this cloud, two of which are classified in the earliest embedded phase. 
\\

\noindent {\it  IRDC\, 010.70-0.13:}
\citet{carey_submmIRDC} conduct continuum observations of this target and find a bright, compact source.  \citet{Peretto2009} find nine fragments within this IRDC. 
\\

\noindent {\it  IRDC\, 011.11-0.12:}
This was one of the first IRDCs subject to intense study in the literature, first in the infrared \citep{carey_msx}, the millimeter \citep{carey_submmIRDC, Johnstone_G11}, and in molecular line emission \citep{Pillai_ammonia, Pillai_G11}.  The filament is about 10~pc long consisting of over 1800\,$\msun$ (from 8~$\mu$m extinction measurements), and the bulk of the gas has an average kinetic temperature of 12\,K. 

This object was observed as part of the Herschel science demonstration phase, and \citet{A&ASpecialIssue-Henning} present the continuum properties of the pre- and proto-stellar cores uncovered at PACS wavelengths.  
\\

\noindent {\it  IRDC\, 013.90-0.51:}
\citet{Peretto2009} identify this object as a {\em Spitzer} dark core, and \citet{Vasyunina2009} perform 1.2\,mm continuum observations showing it breaks into four main peaks. The southern-most source (P3) showed double-peaked molecular line profiles in \citet{Vasyunina2011} indicating either the presence of two components superimposed along the line of sight or internal motion within the cloud.
\\

\noindent {\it  IRDC\, 015.05+0.09:} 
\citet{hennebelle_isogal} and \citet{Teyssier2002} identify this object as one of the most opaque objects, which results in depletion of the CO isotopomers onto the dust grains.
\citet{rathborne2006} find that this IRDC, with a total mass of 158\,$\msun$, resolves into five 1.2\,mm flux peaks with masses of 105, 83, 22, 43, and 29\,$\msun$, assuming 15\,K dust. 
\citet{Wang2006} report water maser associated with MM1 \citep[of][]{rathborne2006}.\\

\noindent {\it  IRDC\, 019.30+0.07:}
This source was targeted in \citet{carey_msx, carey_submmIRDC} detected in H$_2$CO and SCUBA 450/850~$\mu$m surveys, indicating two main spatial peaks in emission. 
\citet{Pillai_ammonia} confirm this morphology in NH$_3$ emission, find gas temperatures of 18.4 and 14.3\,K, and masses of 893 and 823\,$\msun$ for these regions.  \citet{rathborne2006} find a total mass of 3168\,$\msun$ for the entire complex, though much lower masses (113 and 114\,$\msun$) for the two respective peaks.  Finally, \citet{Devine2011} find a total mass of 1130\,$\msun$, concentrated in four NH$_3$ clumps.
\\

\noindent {\it  IRDC\, 028.34+0.06:} 
Of all the sources in \citet{carey_submmIRDC}, IRDC028.34+0.06 shows the strongest submillimeter emission (in two primary peaks at 450 and 850$\mu$m) and, thus, has the greatest estimated mass (up to several hundred $\msun$ each depending on the choice of dust emissivity index, $\beta$).  \citet{Pillai_ammonia} examine this source, here resolving into five subregions in NH$_3$ emission, with temperatures ranging from 13.2 to 16.6\,K and total mass of 4263\,$\msun$ (from ammonia measurements).  Observations by \citet{rathborne2006} at 1.2\, mm find a total cloud mass of 15895\,$\msun$ (the most massive in their sample) and resolve this target into 18 peaks.  In three of those peaks (MM4, MM6, and MM9), \citet{Wang2006} detect water masers.
\citet{Wang_ammonia} perform high resolution observations in NH$_3$, linking the presence of embedded protostars to the enhancement of linewidth and an increase in rotation temperature.
\citet{Zhang2009} obtain SMA continuum and spectral line observations, further resolving the dominant cores into multiple substructures.
\\

\noindent {\it  IRDC\, 048.66-0.29:}
Situated close to the GMC complex W51, this IRDC was targeted for SCUBA 450 and 850\,$\mu$m and HCO$^+$(3-2) observations by \citet{Ormel2005}.  They found that (at the resolution of the JCMT) this IRDC broke into three cores (P1, P2, and EP) near 100\,$\msun$ each, and the line data show that (intermediate mass) star formation is ongoing within the cores. \citet{rathborne2006} surveyed this IRDC with 1.2\,mm continuum observations, reporting a total cloud mass of 917\,$\msun$ for the cloud and 52 and 39\,$\msun$ masses for the dark cores reported in \citet{Ormel2005}, though the core sizes in \citet{rathborne2006} are considerably smaller, which could account for the difference. The star forming properties were studied in detail with {\em Spitzer} in \citet{vanderwiel2008}, in which 13 YSOs were identified near the cloud. The {\em Spitzer} observations also resolved P1 (from \citet{Ormel2005}) into two peaks, but found the star formation properties to be consistent.\\

\noindent {\it  IRDC\, 079.31+0.36:}
This IRDC is among the nearest in our sample and was first observed by \citet{carey_submmIRDC}. 
\citet{Redman2003} characterize the ongoing star formation in this IRDC which may have been triggered.
\citet{Pillai_ammonia} find high column densities ($> 10^{23}$ cm$^{-3}$) and gas tempertatures between 12 and 15\,K in this region. This object was included in the SCUBPOL legacy survey \citet{Matthews2009}.
\\

\noindent {\it  IRDC\, 310.39-0.30:}
\citet{Vasyunina2009} find the millimeter continuum emission peak coincides with the absorption peak in the mid-infrared, from which they estimate the mass is 1320\,$\msun$. The continuum peak is stronger in the MIR dark source than for the nearby MIR-bright source, indicating that this IRDC could contain a range of evolutionary stages. The MIR-bright sources are included in the \citet{Cyganowski2008} EGO catalog, and \citet{Vasyunina2011} detect weak SiO emission in this IRDC.

\noindent {\it  IRDC\, 316.72+0.07:}
This object hosts a small cluster of embedded infrared sources and has a mass between 400 and 950\,$\msun$ derived from 8\,$\mu$m extinction \citep{Linz2007}, or 561\,$\msun$ based on 1.2\,mm emission.\\

\noindent {\it  IRDC\, 320.27+0.29:}      
This cloud is comprised of a compact eastern component and an elongated western cloud, which are 50 and 70\,$\msun$, respectively, based on 1.2\,mm continuum observations \citep{Vasyunina2009}. Molecular line emission observed by \citet{Vasyunina2011} is weak. \\
                   
\noindent {\it  IRDC\, 321.73+0.05:}
Two main clumps of approximately equal mass \citep[about 100\,$\msun$ ][]{Vasyunina2009} comprise this IRDC. The eastern source, with a column density of 3.2 $\times$ 10$^{22}$ cm$^{-2}$, shows a weak source at 24\,$\mu$m, yet \citet{Vasyunina2011} detect SiO and asymmetric HCO$^+$ emission, indicating the presence of a shock and inflowing gas, both indicators of star formation activity at an early stage. The western source also exhibits multiple peaks in the molecular line emission attributed to complex internal motions throughout the cloud.\\

\noindent {\it  HMSC\, 07029:}
The clump HMSC07029 (a.k.a. UYSO 1) had been recognised as a distinct peak in
a (sub-)millimeter survey of massive protocluster-cores in the outer galaxy by
\citet{Klein2005}. It is a distinct intermediate-mass core clearly
offset from the neighboring IRAS source 07029-1215 which in the pre-{\em Spitzer}
era was just detectable at wavelengths $\ge 450 \mu$m
citep{Forbrich2004}. However, strong CO outflow activity is also
associated with this core. \citet{Forbrich2009} collected more data on
this source and revealed an impressive system of crossed H$_s$ shocked
emission jets apparently emanating from the core which in {\em Spitzer} 70\,$\mu$m
images was 10$''$ offset from the interferometric position in the millimeter.
Our EPoS {\em Herschel} data could eventually disentangle the 70\,$\mu$m peak
position which is coinciding with the interferometric position, from the
surrounding extended PDR emission. More details about this enigmatic region
are given in our A\&A special issue paper \citep{A&ASpecialIssue-Linz}.\\

\noindent {\it IRDC\,18102:}
Although originally selected as a high-mass protostellar object (HMPO), the singular 1.2\,mm continuum peak \citep{Beuther2002} in this regions turns out to be mid-infrared dark in MSX and offset by less than $1'$ from an infrared-bright and centimeter-bright ultracompact H{\sc ii} region. SiO emission, known to be associated with molecular outflows, is reported in \citet{Beuther_IRDCoutflow} associate with the millimeter continuum peak. Since the infrared-dark and bright
regions are in so close proximity, interaction between the different
evolutionary stages is expected.\\

\noindent {\it IRDC\,18151:}
The name-giving IRAS\,18151-1208 is a well-known
high-mass outflow-disk region \citep{Davis2004,Fallscheer2011}. 
In the vicinity of this active site, one finds two similarly massive gas-clumps 
of which the one is 8\,$\mu$m-dark and hosts a H$_2$O maser \citep{Beuther2002_masers}.\\

\noindent {\it  IRDC\,18182:}
The central IRAS\,18182-1433 is a well-known
``hot core''-type high-mass star-forming region \citep{Zapata2006,Beuther2006}. 
The infrared-dark gas-clumps are found
approximately $2'$ to the southeast and northeast.\\

\noindent {\it  IRDC\,18223:}
This filamentary dark cloud is located directly south
of the luminous IRAS\,18223-1243. The region contains regions is
several evolutionary stages, from IRDCs with already embedded ongoing
star formation \citep[e.g. IRDC 18223-3][]{Beuther2005, Beuther_protostars_IRDC, Fallscheer2009} to potentially real
high-mass starless cores that are even dark at 70\,$\mu$m \citep{A&ASpecialIssue-Beuther}.\\

\noindent {\it  IRDC\,18306:}
About $2'$ offset to the northeast from the central IRAS\,18306-0835 we
find filamentary infrared-dark structures that are associated with millimeter
continuum emission, broken into three main 1.2\,mm peaks.\\

\noindent {\it  IRDC\,18308:}
Again we find filamentary, mid-infrared dark millimeter continuum structures
extending from IRAS\,18308-0841 toward the north
and containing five main 1.2\,mm peaks. \\

\noindent {\it  IRDC\,18310:}
This region hosts H{\sc ii} region, IRAS\,18310-0825, a compact mid-infrared-dark filament, totalling five distinct millimeter continuum peaks.\\

\noindent {\it  IRDC\,18337:}
This region contains three strong 1.2\,mm continuum
sources that appear to be laid out in a linear structure. While the
IRAS\,18337-0743 has strong mid-infrared emission, the second source
is still -- although weaker -- mid-infrared bright, and the third
source is mid-infrared dark. This region is likely a particularly
good candidate to investigate sequential star formation.\\

\noindent {\it  IRDC\,18385:}
A $2'$  millimeter continuum filament stems from IRAS\,18385-0512, and an isolated infrared-dark globule lies to the south east.\\

\noindent {\it  IRDC\,18437:}
This region contains a chain of millimeter continuum sources, including IRAS\,18437-0216, in approximate north-south direction with different degrees of associated mid-infrared emission. Polarized sub-millimeter continuum emission is observed by \citet{Matthews2009}.\\

\noindent {\it  IRDC\,18454:}
This complex of several millimeter continuum sources is located
at the north-eastern edge of the prominent Galactic mini-starburst W43
\citep{Motte2003, BallyW43, Nguyen2011} at the same systemic velocity, thus, 
at presumably the same distance.  The millimeter continuum sources
exhibit several velocity components. \citet{Beuther_18454} explore whether
these are due to chance projections or physical interactions. Furthermore, it
is in general interesting to find infrared dark massive gas clumps in
so close proximity to one of the most active star-forming regions in
our Galaxy.\\

\noindent {\it  IRDC\,18530:}
Extending about $3'$ south of IRAS\,18530+0215 one finds
a filamentary infrared-dark structure that is associated with millimeter
continuum emission.\\

\noindent {\it  IRDC\,19175:}
This regions hosts a spiral-like filamentary structure
extending several arc-minutes from IRAS\,19175+1357 distributed in seven 1.2\,mm clumps. Two of the mid-infrared dark gas clumps have already been studied by
\citet{BeutherHenning2009} who found many smaller sub-structures within
them.\\

\noindent {\it  IRDC\,20081:} 
IRAS\,20081+2720 lies at the center of a compact infrared-dark filament with two compact 1.2\,mm peaks to the north and south.\\

\noindent {\it ISOSS sources}

The ISOSS portion of the sample lies largely in the outer Galaxy, thus there are fewer detailed studies in the literature to date. In many cases a nearby (possibly associated) IRAS source was included in a maser survey. These resulted in non-detections in most cases, but we list the exceptions below along with the descriptions of the few detailed studies.\\

\noindent {\it  ISOSSJ04225+5150:}
Birkmann (2006, PhD) found three compact cores in this region. The
first core (SMM1) has a mass of 275~M$_\odot$ and a temperature of
21.4\,K. The most massive core (SMM3, 510~M$_\odot$, 17\,K) show
indications for shocked outflows and an associated PDR \citep{Pitann2011}


\noindent {\it  ISOSSJ06527+0140:}
\citet{Migenes1999} and \citet{Sunada2007} detect a water maser at the position of IRAS source IRAS 06501+0143.\\

\noindent {\it  ISOSSJ18364-0221:}
\citet{Birkmann2006} found that ISOSS J18364-0221 broke into two compact cores (SMM1 and SMM2). SMM1 is a protostellar core (16.5\,K, 75\,$\msun$) with energetic outflows. The second core (SMM2) is a more massive (280\,$\msun$), pre-stellar object of lower temperature (12\,K). \citet{Hennemann2009} performed follow-up millimeter continuum and molecular line observations of SMM1 and found it was actually comprised of two sub-clumps (SMM1 North and South).  While SMM1 South is detected at 24 and 70~$\mu$m and SMM1 North is undetected in the infrared, both are associated with molecular outflows reaching up to 4.6~pc from the core.\\

\noindent {\it  ISOSSJ19357+1950:}   
SCUBA observations \citep{Hennemann2008} reveal three resolved clumps at 450 $\mu$m: SMM1 North (0.37 pc, 70 M$_\odot$) and South (0.31 pc, 50 M$_\odot$), and SMM2 in the South-West (0.39 pc, 120 M$_\odot$). SMM2 hosts a deeply embedded protostellar object that may evolve to a massive YSO.\\

\noindent {\it  ISOSSJ19486+2556:}
\citet{Hennemann2008} find three submillimeter clumps (SMM1-3) in this region, located along a chain from North-East to South-West.  SMM2 and SMM3 are associated with bright 24 and 70 $\mu$m sources and have masses of 40 and 60 M$_\odot$, whereas SMM1 (100 M$_\odot$) is not detected with \emph{Spitzer}.\\

\noindent {\it  ISOSSJ20153+3453:}
A single core with a gas mass of $\sim120$~M$_\odot$ and a dust
temperature of 17\,K was identified in this region by \citet{Hennemann2008}.
Two neighboring mid-infrared sources are detected towards SMM1, the one towards the submillimeter peak is interpreted as embedded YSO that is actively accreting as suggested by the presence of ``fuzzy'' emission prominent at 4.5\,$\mu$m. 
Futher {\em Spitzer} observations indicate the presence of a PDR.
Excited gas suggests shocked outflows in this region \citep{Pitann2011}.\\

\noindent {\it  ISOSSJ20298+3559:}
This region harbors four submillimeter clumps SMM1-4 and has been studied in detail in \citet{Krause2003} except for SMM4 which is offset to the west.  SMM1 (0.14 pc, 10 $\msun$) and SMM2 (unresolved: <0.07 pc, 3 $\msun$) are joined by extended submillimeter emission.
SMM4 is more extended (0.34 pc, 60 $\msun$) than the other clumps.
SMM1 hosts an embedded intermediate-mass YSO, and a Class II-like intermediate-mass YSO is associated with SMM4 \citep{Hennemann2008}.\\

\noindent {\it  ISOSSJ22478+6357:}
An elongated clump (SMM1) is resolved in both an Eastern (0.14 pc, 60 M$_\odot$) and a Western component (0.24 pc, 120 M$_\odot$) at 450 $\mu$m. Towards SMM1 East, a Class II-like YSO of intermediate central mass has been identified among several 24 $\mu$m sources \citep{Hennemann2008}.\\

\noindent {\it  ISOSSJ23053+5953:}
\citet{Birkmann2007} identified two cores in this region with $\sim
200$~M$_\odot$ each (at 19.5\,K and 17.3\,K). HCO$^+$ kinematics indicate
in-falling gas\ in both cores. Several outflow features were found in
this region by interferometric, spectroscopic and NIR narrow-band
observations \citep[see][]{Birkmann2007,Pitann2011}.\\

 
 



\section{Images}
\begin{figure*}[h]
\includegraphics[width=0.95\textwidth]{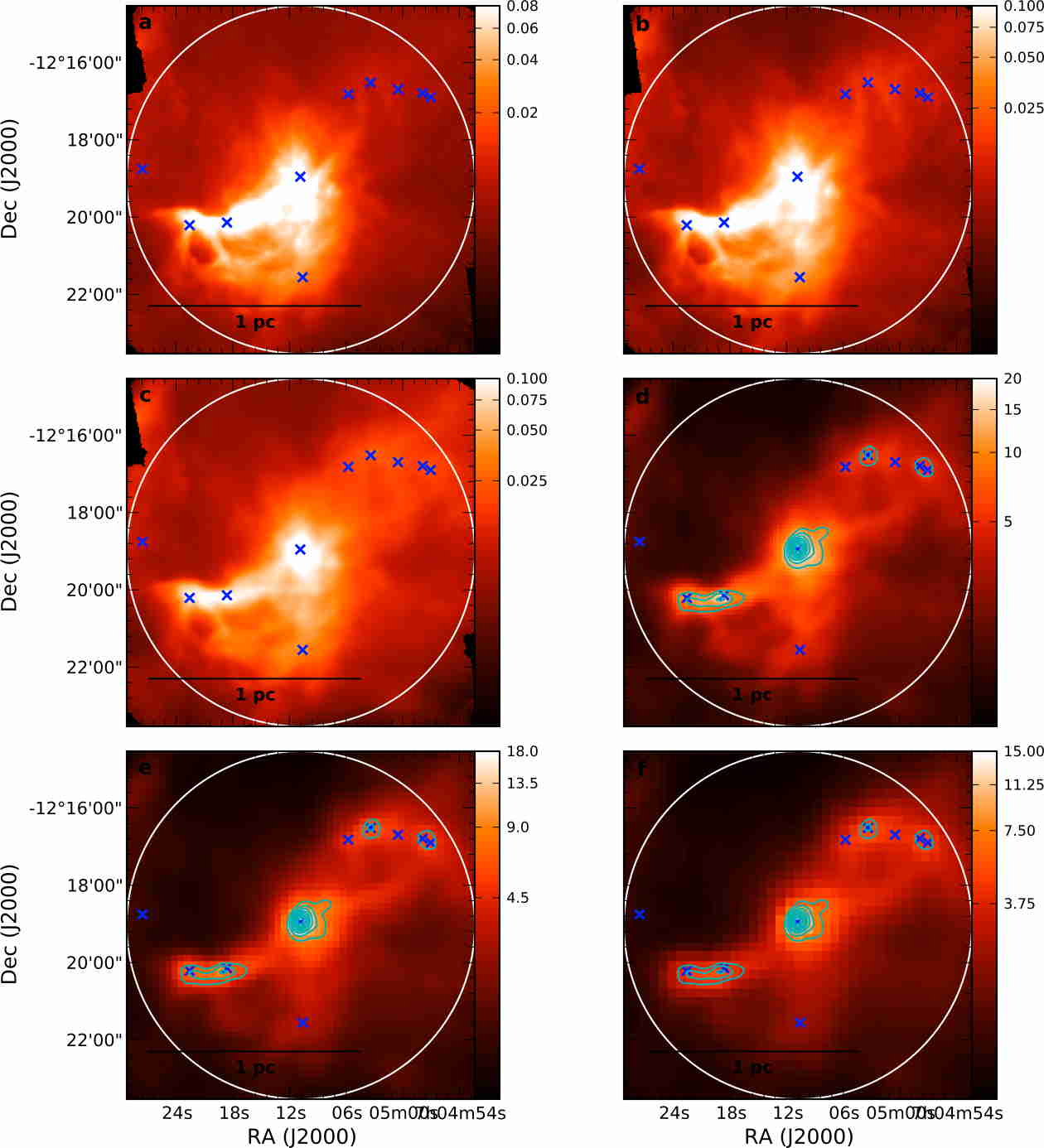}
\caption{\label{fig:hmsc} The {\em Herschel} (a) PACS 70\,$\mu$m , (b) PACS 100\,$\mu$m, (c) PACS 160\,$\mu$m, (d) SPIRE 250\,$\mu$m, (e) SPIRE 350\,$\mu$m, and (f) SPIRE 500\,$\mu$m images of HMSC 07029. The flux scales on the right edge of each PACS panel are in Jy pixel$^{-1}$ and Jy beam$^{-1}$ for the SPIRE panels. Point sources are marked with a blue $\times$, and the white solid line marks the boundary defined in Section~\ref{ss:atlasgal}. Cyan contours on the SPIRE panels show SCUBA 850\,$\mu$m emission from 20\% of peak flux in increments of 10\%: 0.32, 0.49, 0.65, 0.81, 0.97, 1.13, 1.30, 1.46, 1.62 Jy beam$^{-1}$.}
\end{figure*}

\clearpage

\begin{figure*}[h]
\includegraphics[width=0.95\textwidth]{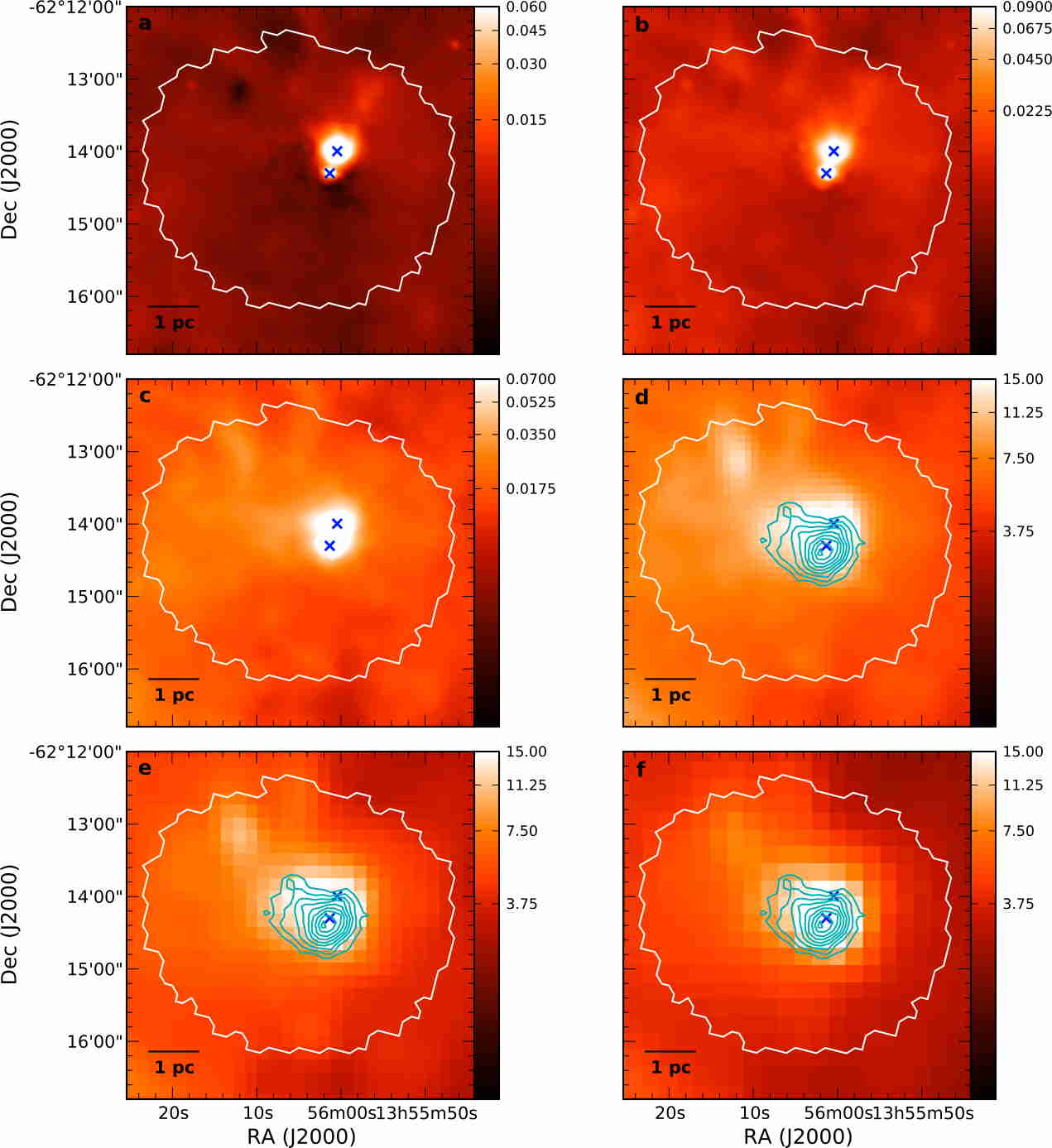}  
\caption{Same image layout as Figure~\ref{fig:hmsc} but for IRDC 310.39-0.30. Cyan contours on the SPIRE panels (d, e, and f) show ATLASGAL 870\,$\mu$m: 0.3, 0.45, 0.6, 0.75, 0.9, 1.05, 1.2, 1.35, 1.5 Jy beam$^{-1}$.}
\end{figure*}

\clearpage

\begin{figure*}[h]  
\includegraphics[width=0.95\textwidth]{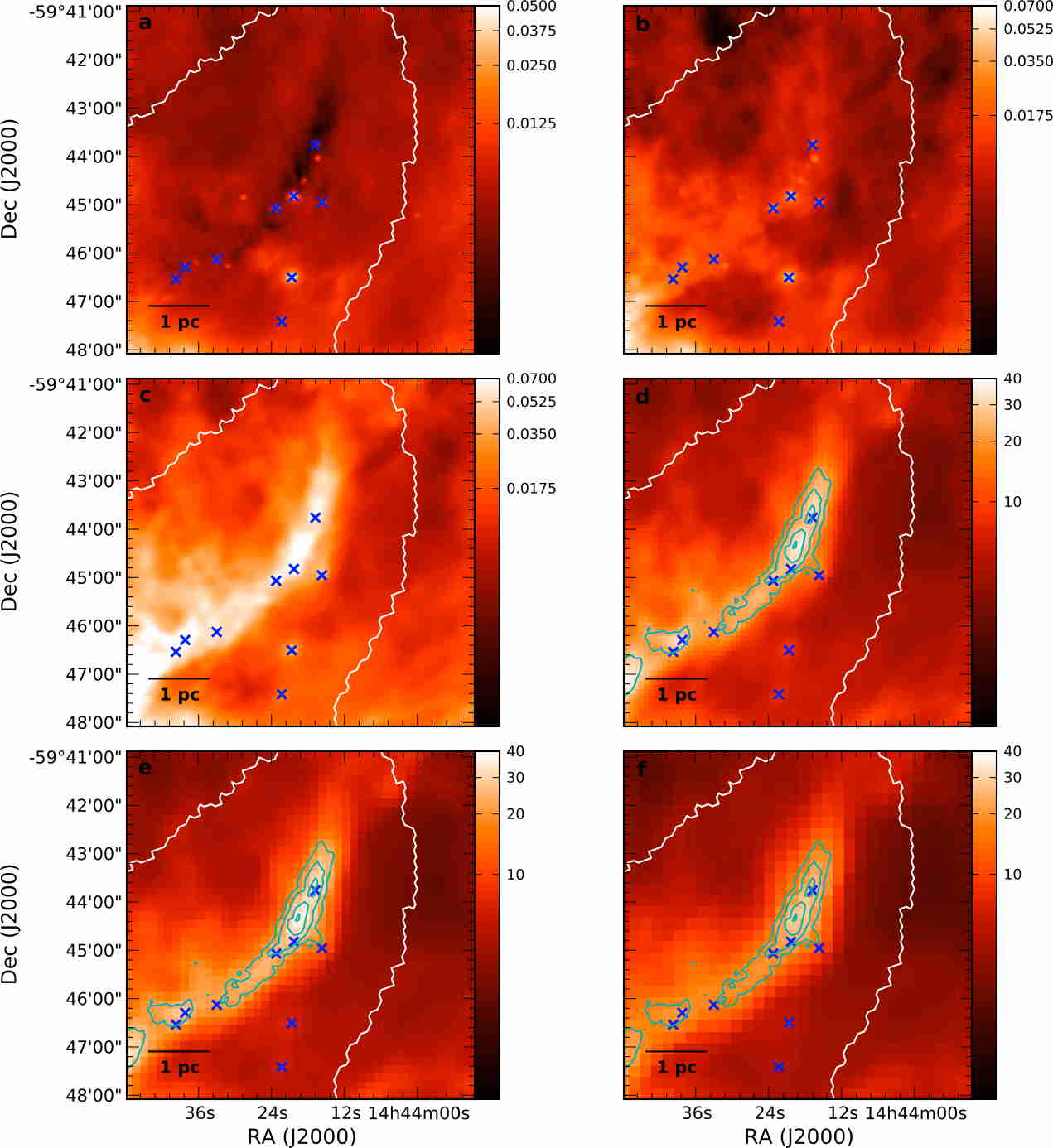}   
\caption{Same image layout as Figure~\ref{fig:hmsc} but for IRDC 316.72+0.07. Cyan contours on the SPIRE panels (d, e, and f) show ATLASGAL 870\,$\mu$m: 0.4, 0.6, 0.8, 1.0, 1.2, 1.4, 1.6, 1.8, 2.0 Jy beam$^{-1}$.}  
\end{figure*}

\clearpage

\begin{figure*}[h]  
\includegraphics[width=0.95\textwidth]{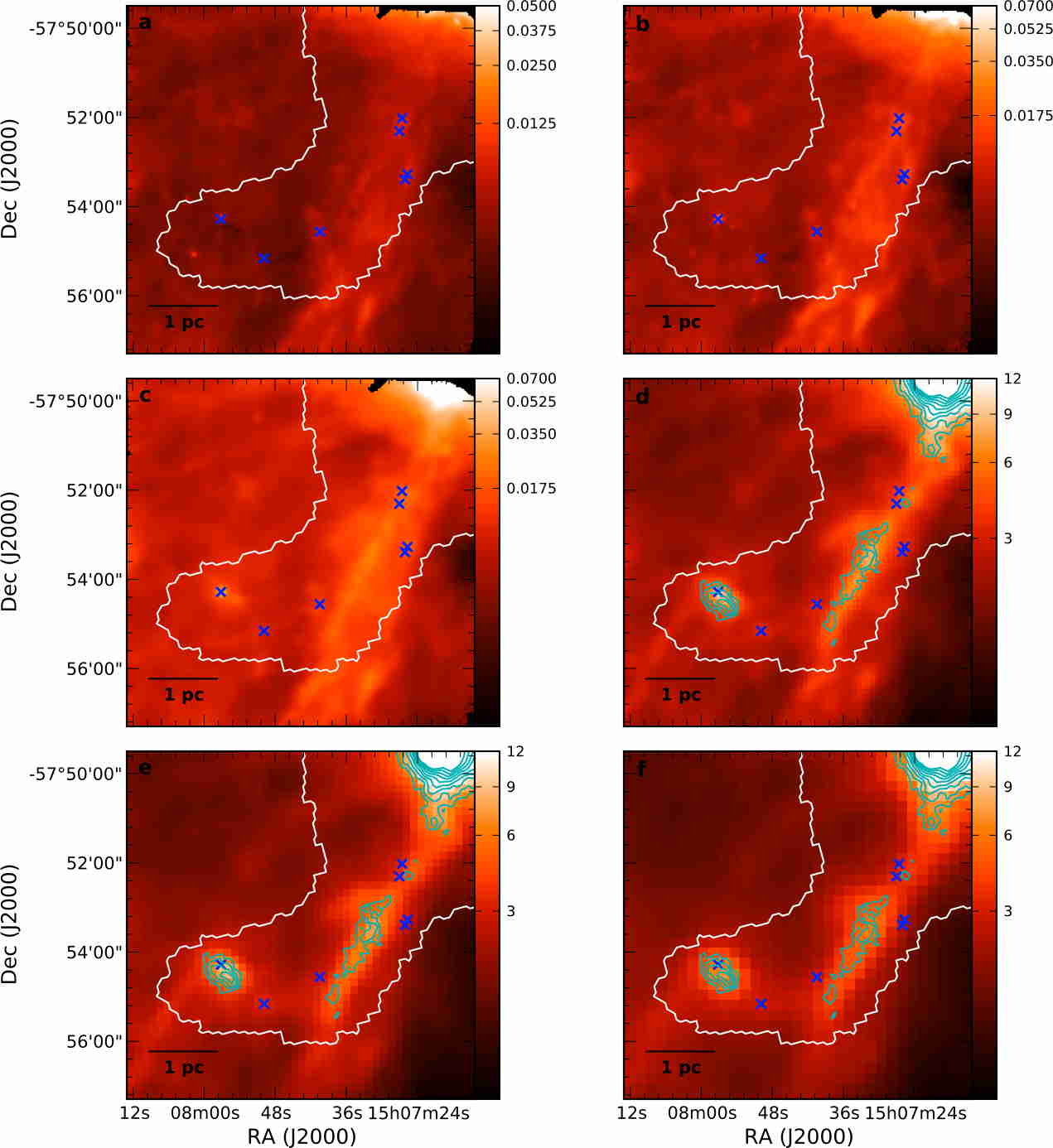}  
\caption{Same image layout as Figure~\ref{fig:hmsc} but for IRDC 320.27+0.29. Cyan contours on the SPIRE panels (d, e, and f) show ATLASGAL 870\,$\mu$m: 0.2, 0.3, 0.4, 0.5, 0.6, 0.7, 0.8, 0.9, 1.0 Jy beam$^{-1}$.}  
\end{figure*}

\clearpage

\begin{figure*}[h]
\includegraphics[width=0.95\textwidth]{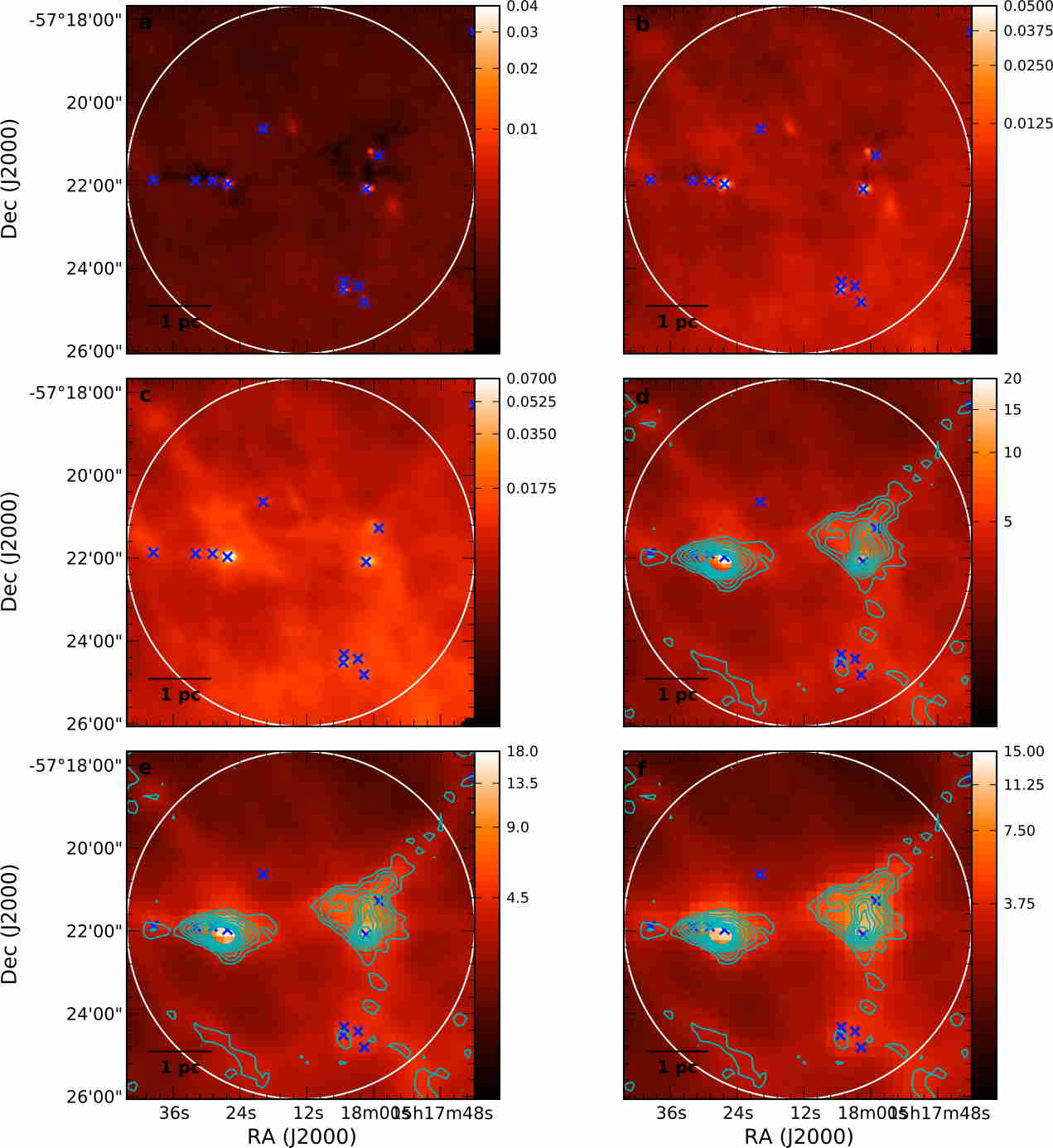}  
\caption{Same image layout as Figure~\ref{fig:hmsc} but for IRDC 321.73+0.05. Cyan contours on the SPIRE panels (d, e, and f) show MAMBO 1.2\,mm: 0.06, 0.09, 0.12, 0.18, 0.21, 0.24, 0.27, 0.3 Jy beam$^{-1}$. }  
\end{figure*}

\clearpage

\begin{figure*}[h]  
\includegraphics[width=0.95\textwidth]{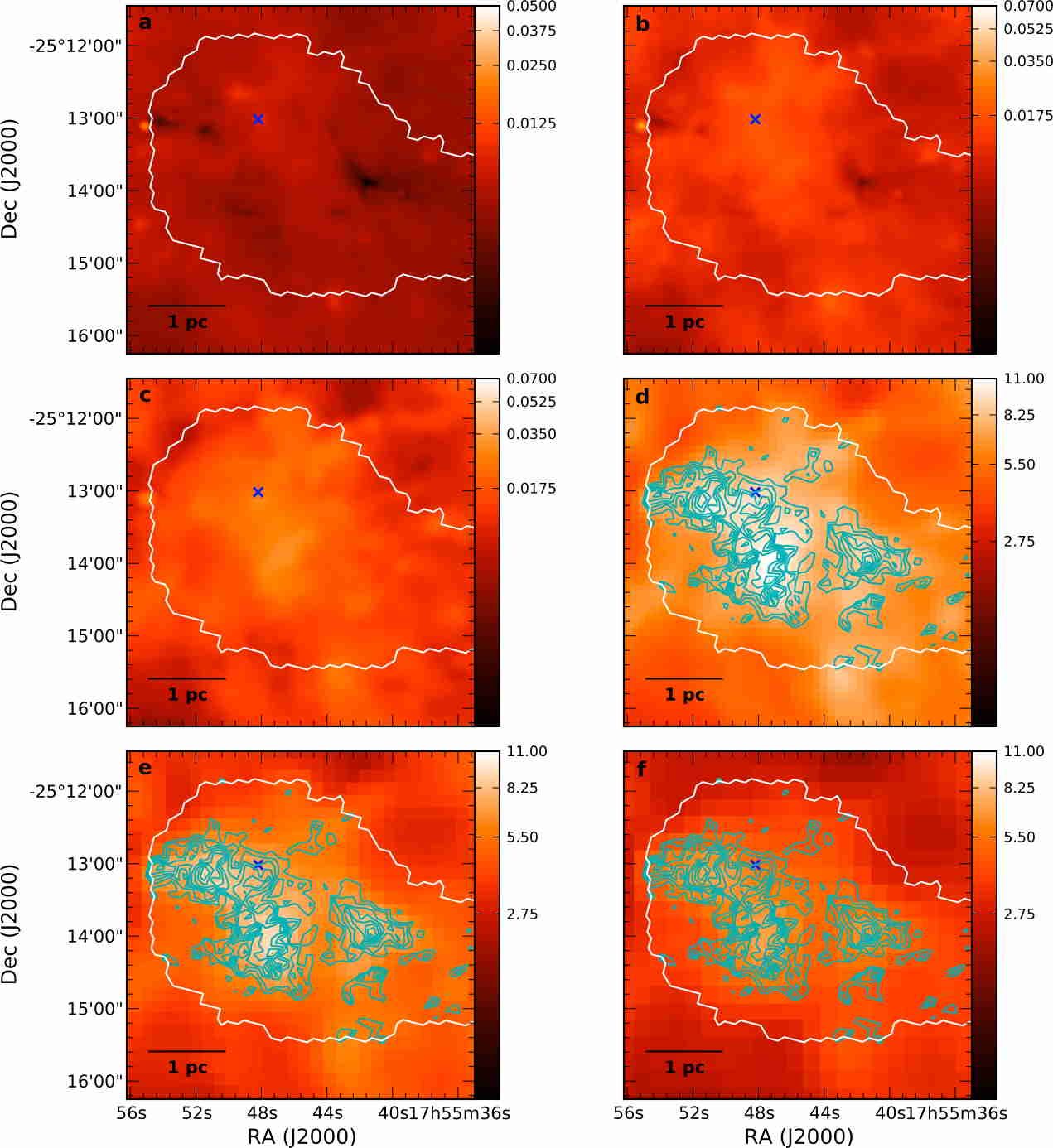}  
\caption{Same image layout as Figure~\ref{fig:hmsc} but for IRDC 004.36-0.06. Cyan contours on the SPIRE panels (d, e, and f) show ATLASGAL 870\,$\mu$m: 0.07, 0.11, 0.14, 0.18, 0.21, 0.25, 0.28, 0.32, 0.35 Jy beam$^{-1}$.}  
\end{figure*}

\clearpage

\begin{figure*}[h]
\includegraphics[width=0.95\textwidth]{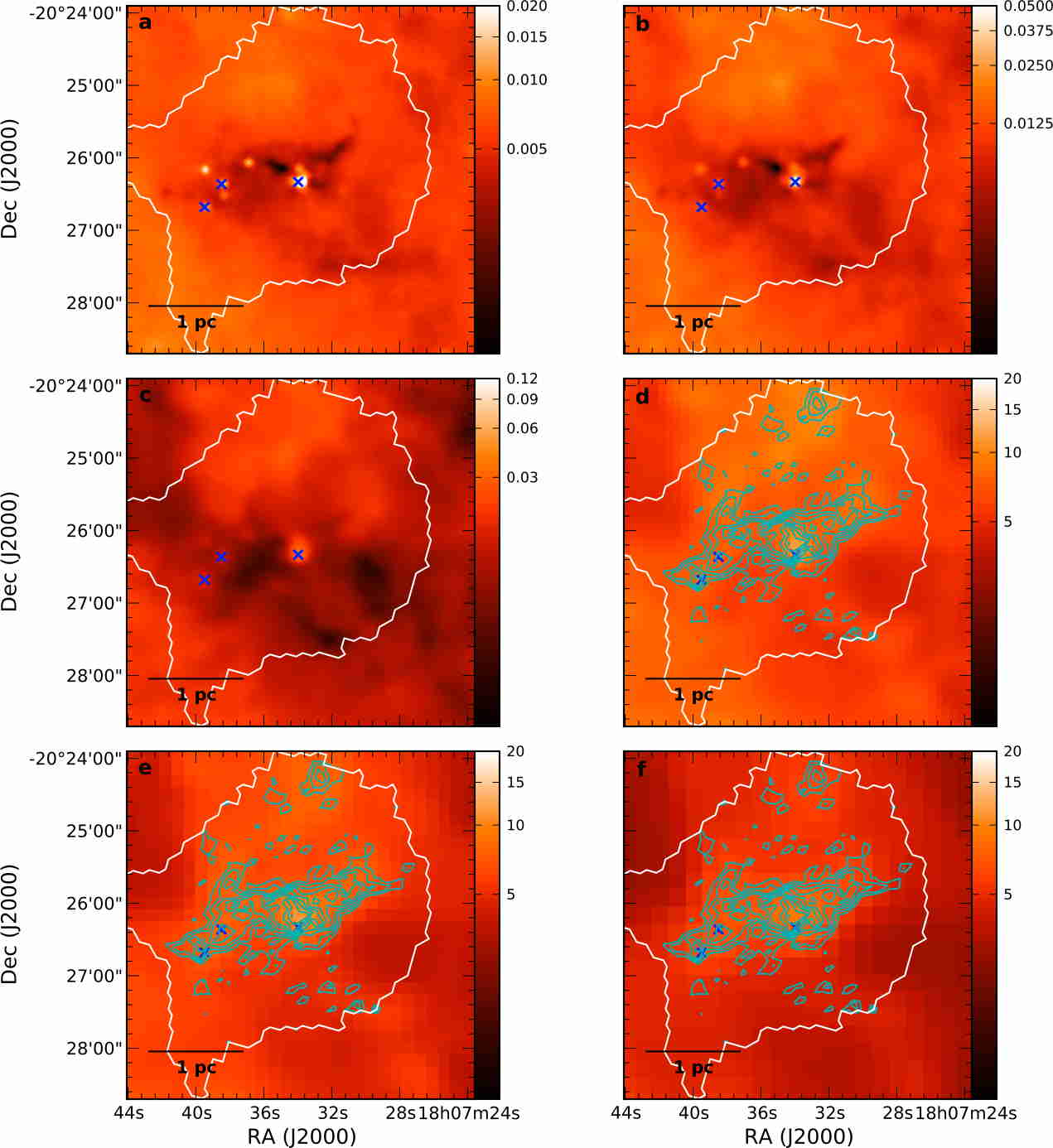}  
\caption{Same image layout as Figure~\ref{fig:hmsc} but for IRDC 009.86-0.04. Cyan contours on the SPIRE panels (d, e, and f) show ATLASGAL 870\,$\mu$m: 0.08, 0.13, 0.168, 0.21, 0.25, 0.29, 0.34, 0.38, 0.42 Jy beam$^{-1}$.}  
\end{figure*}

\clearpage

\begin{figure*}[h]  
\includegraphics[width=0.95\textwidth]{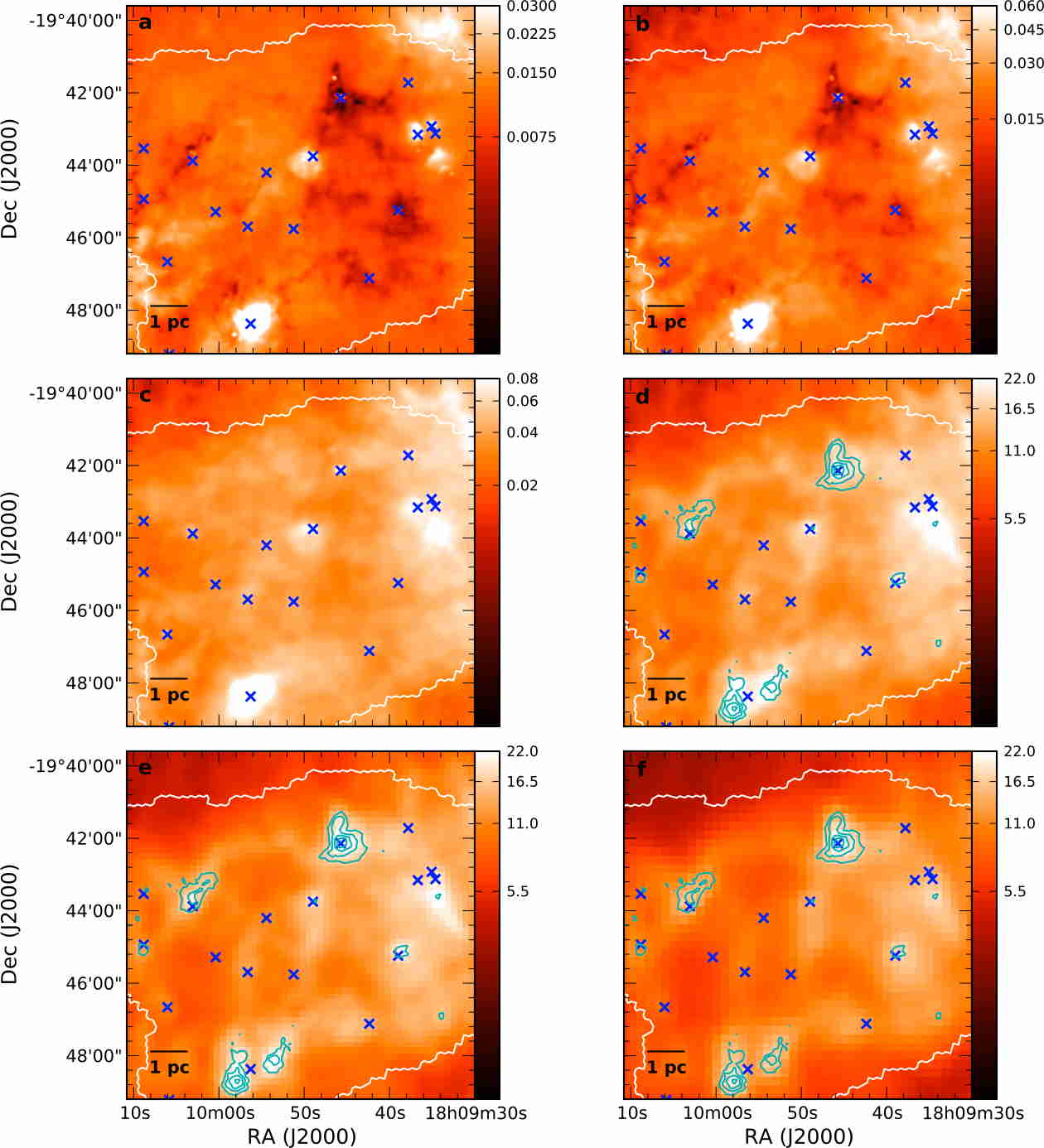}  
\caption{Same image layout as Figure~\ref{fig:hmsc} but for IRDC 010.70-0.13. Cyan contours on SPIRE panels (d, e, and f) show ATLASGAL 870\,$\mu$m: 0.39, 0.59, 0.78, 0.98, 1.17 , 1.37, 1.56, 1.76, 1.95 Jy beam$^{-1}$.}  
\end{figure*}

\clearpage

\begin{figure*}[h]  
\includegraphics[width=0.95\textwidth]{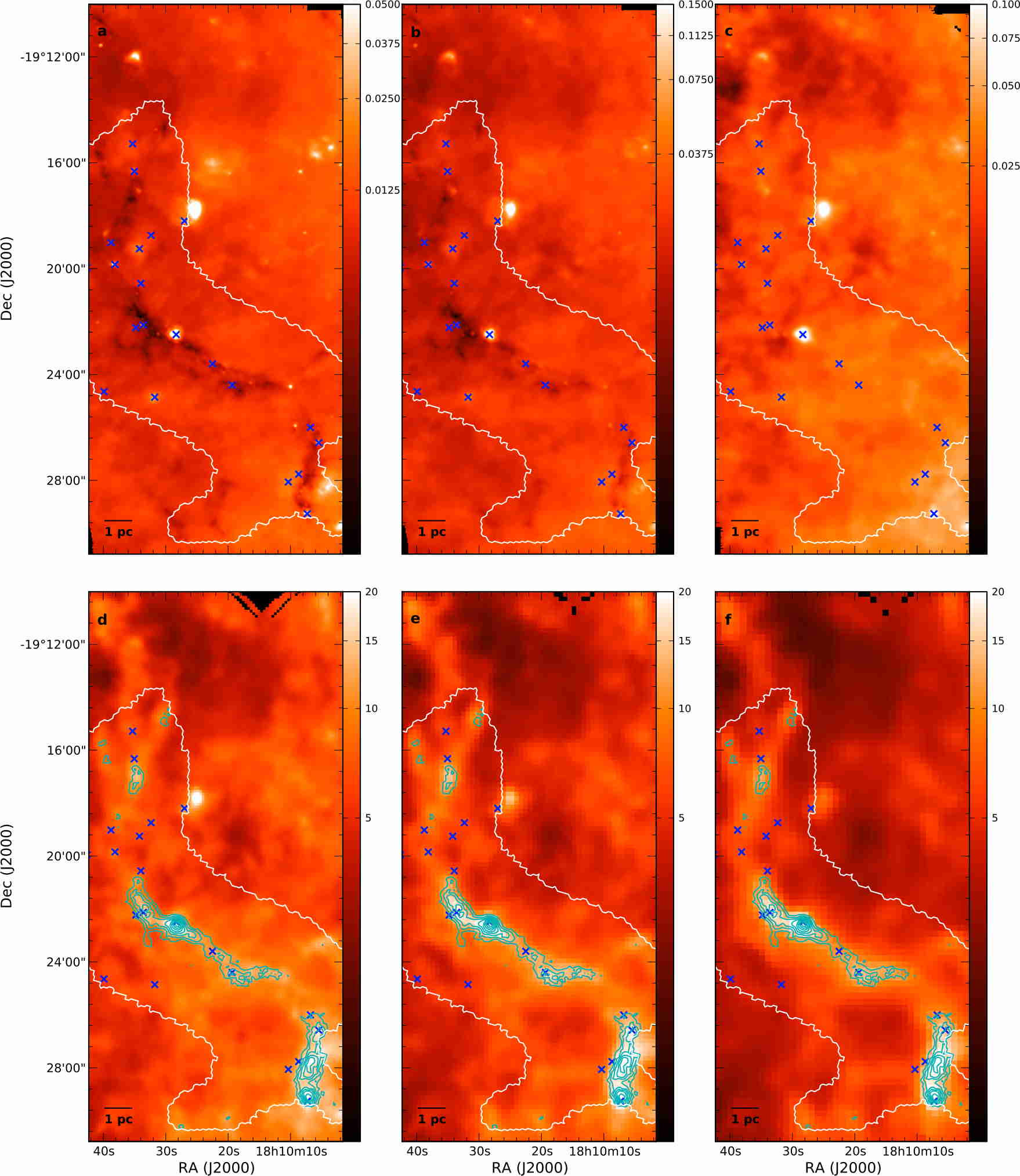}  
\caption{Same image layout as Figure~\ref{fig:hmsc} but for IRDC 011.11-0.12. Cyan contours on the SPIRE panels (d, e, and f) show ATLASGAL 870\,$\mu$m: 0.34, 0.51, 0.68, 0.85, 1.02, 1.19, 1.36, 1.53, 1.7 Jy beam$^{-1}$.}  
\end{figure*}

\clearpage

\begin{figure*}[h]  
\includegraphics[width=0.95\textwidth]{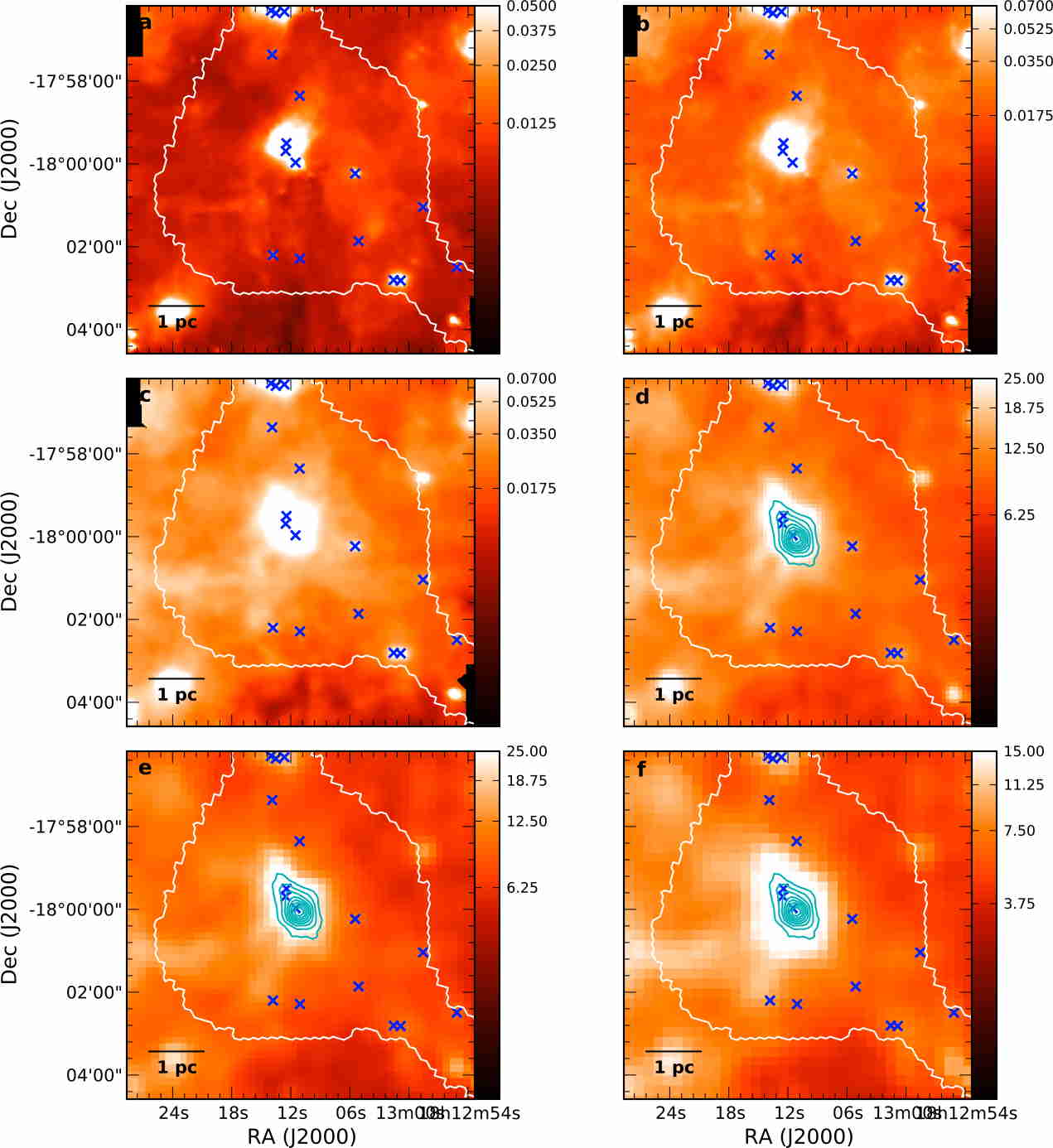}  
\caption{Same image layout as Figure~\ref{fig:hmsc} but for IRDC 18102. Cyan contours on the SPIRE panels (d, e, and f) show ATLASGAL 870\,$\mu$m: 0.68, 1.03, 1.37, 1.71, 2.05, 2.40, 2.74, 3.08 Jy beam$^{-1}$.}  
\end{figure*}

\clearpage

\begin{figure*}[h]  
\includegraphics[width=0.95\textwidth]{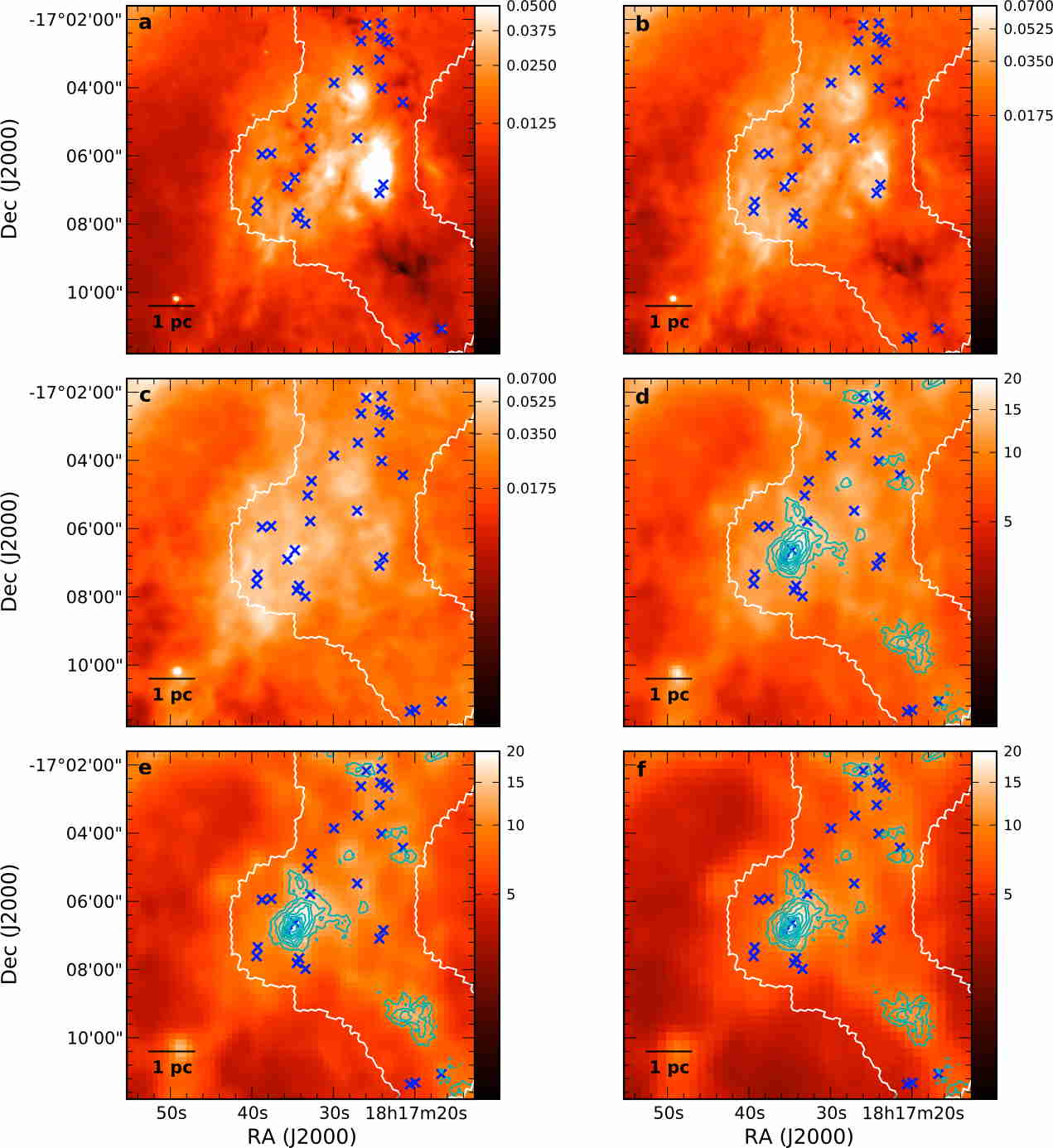}  
\caption{Same image layout as Figure~\ref{fig:hmsc} but for IRDC 013.90-0.51. Cyan contours on the SPIRE panels (d, e, and f) show ATLASGAL 870\,$\mu$m: 0.25, 0.37, 0.50, 0.62, 0.74, 0.87, 0.99, 1.12, 1.24 Jy beam$^{-1}$.}  
\end{figure*}

\clearpage

\begin{figure*}[h]  
\includegraphics[width=0.95\textwidth]{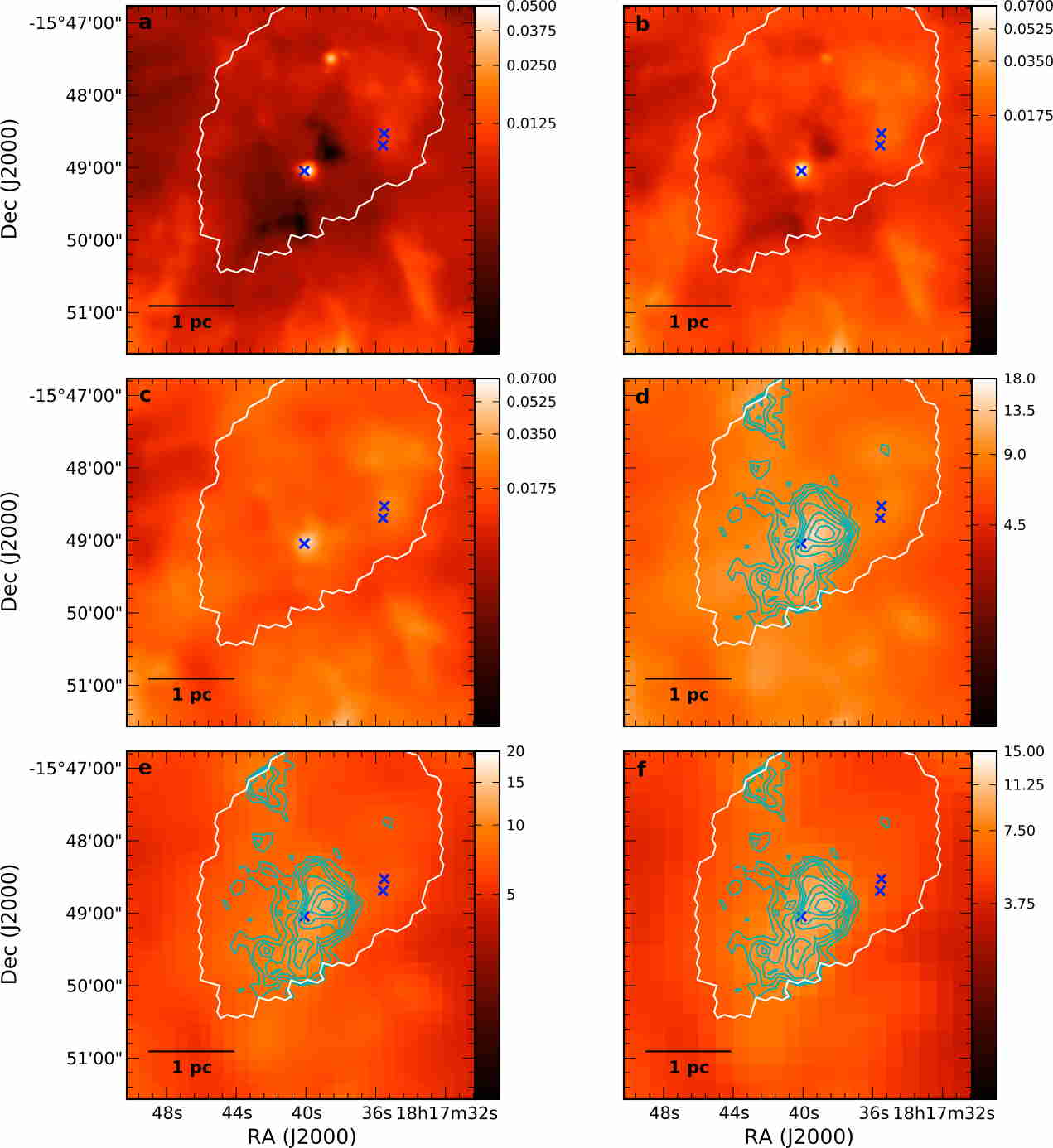}  
\caption{Same image layout as Figure~\ref{fig:hmsc} but for IRDC 015.05+0.09. Cyan contours on the SPIRE panels (d, e, and f) show ATLASGAL 870\,$\mu$m: 0.08, 0.13, 0.17, 0.21, 0.25, 0.29, 0.34, 0.38, 0.42 Jy beam$^{-1}$.}  
\end{figure*}

\clearpage

\begin{figure*}[h]  
\includegraphics[width=0.95\textwidth]{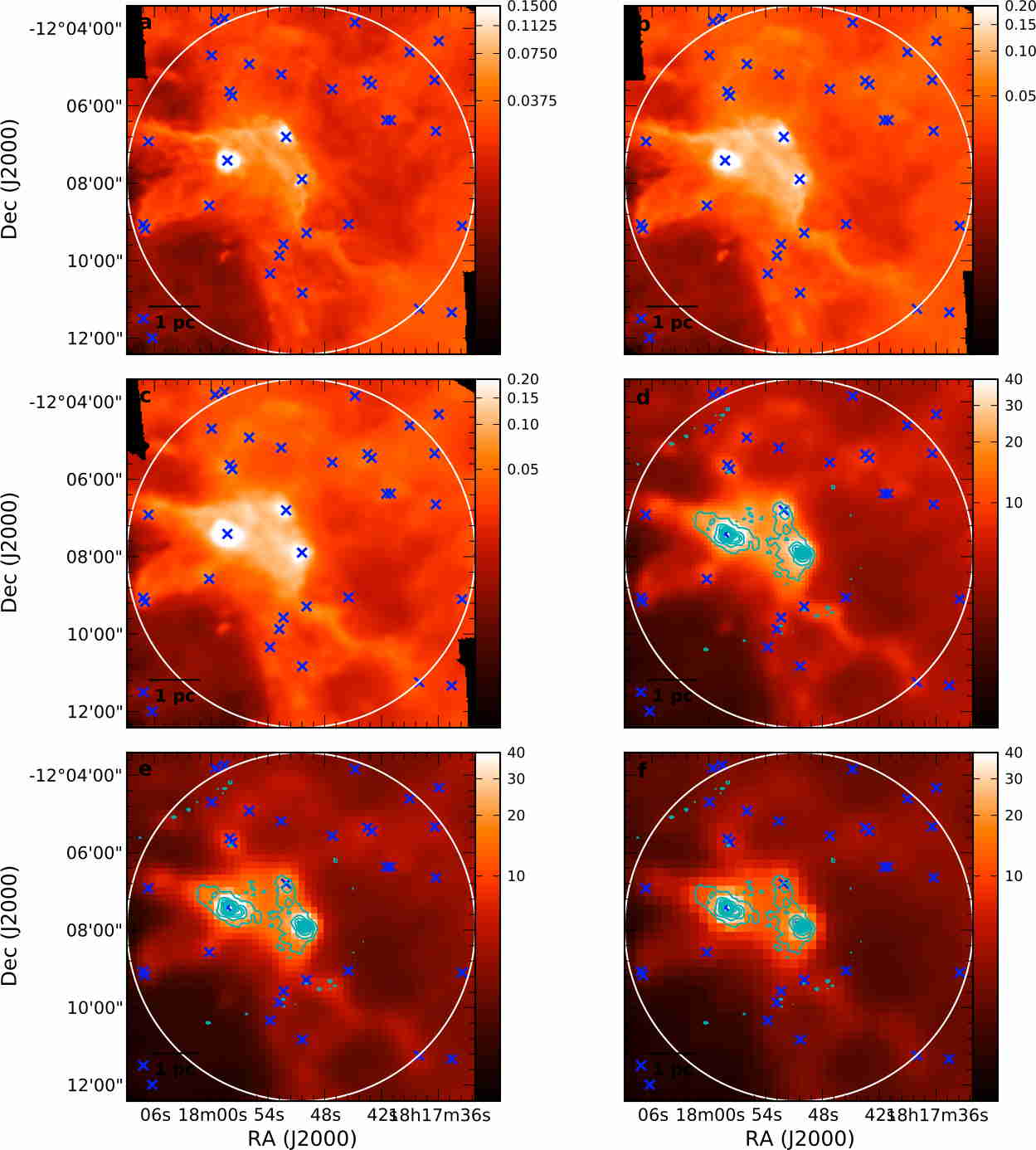}  
\caption{Same image layout as Figure~\ref{fig:hmsc} but for IRDC 18151. Cyan contours on the SPIRE panels (d, e, and f) show MAMBO 1.2\,mm: 0.13, 0.20, 0.27, 0.35, 0.40, 0.47, 0.54, 0.61, 0.67 Jy beam$^{-1}$. }  
\end{figure*}

\clearpage

\begin{figure*}[h]  
\includegraphics[width=0.95\textwidth]{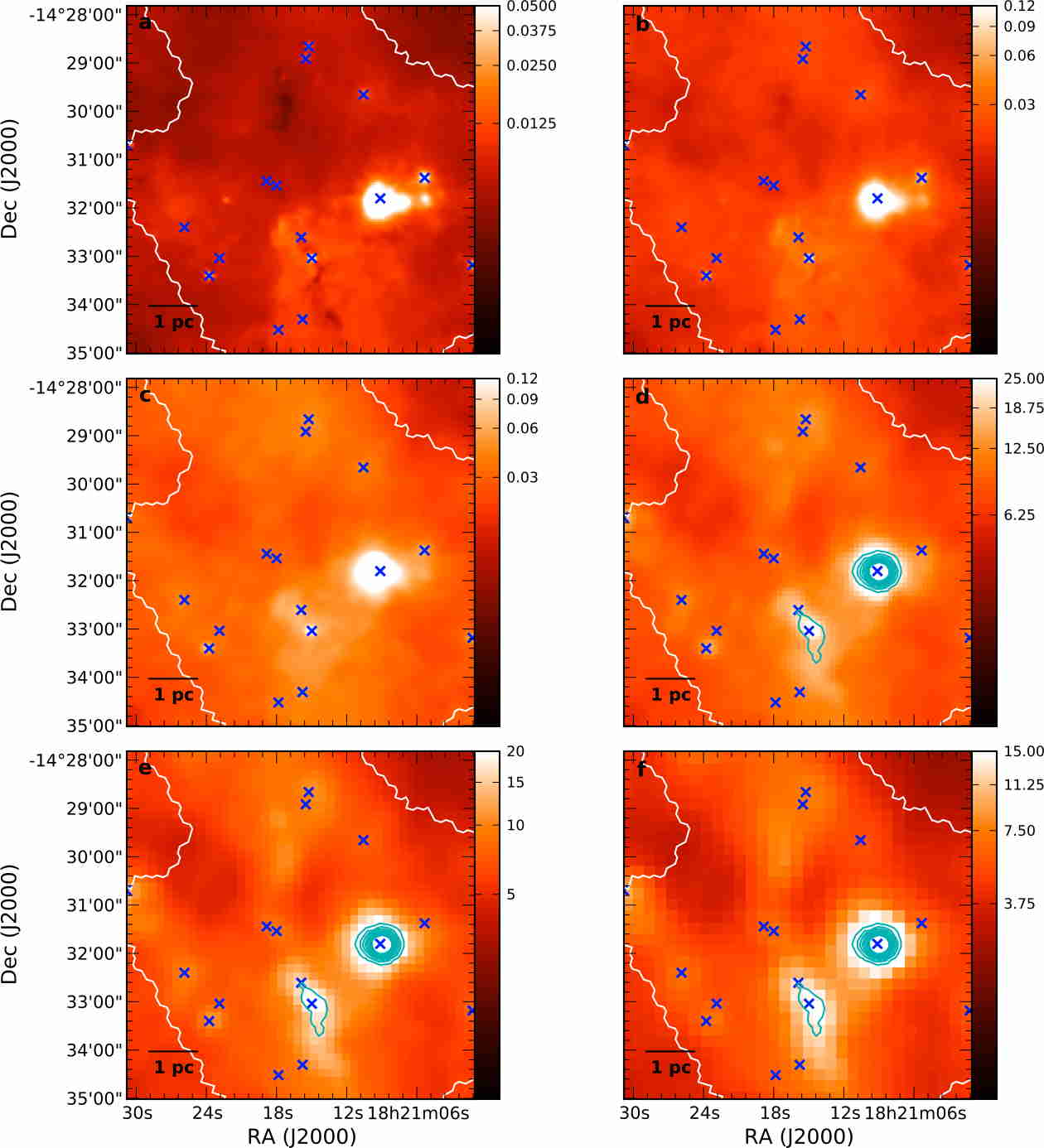}  
\caption{Same image layout as Figure~\ref{fig:hmsc} but for IRDC 18182.Cyan contours on the SPIRE panels (d, e, and f) show ATLASGAL 870\,$\mu$m: 0.6, 0.9, 1.2, 1.5, 1.8, 2.1, 2.4, 2.7, 3.0 Jy beam$^{-1}$.}  
\end{figure*}

\clearpage

\begin{figure*}[h]  
\includegraphics[width=0.95\textwidth]{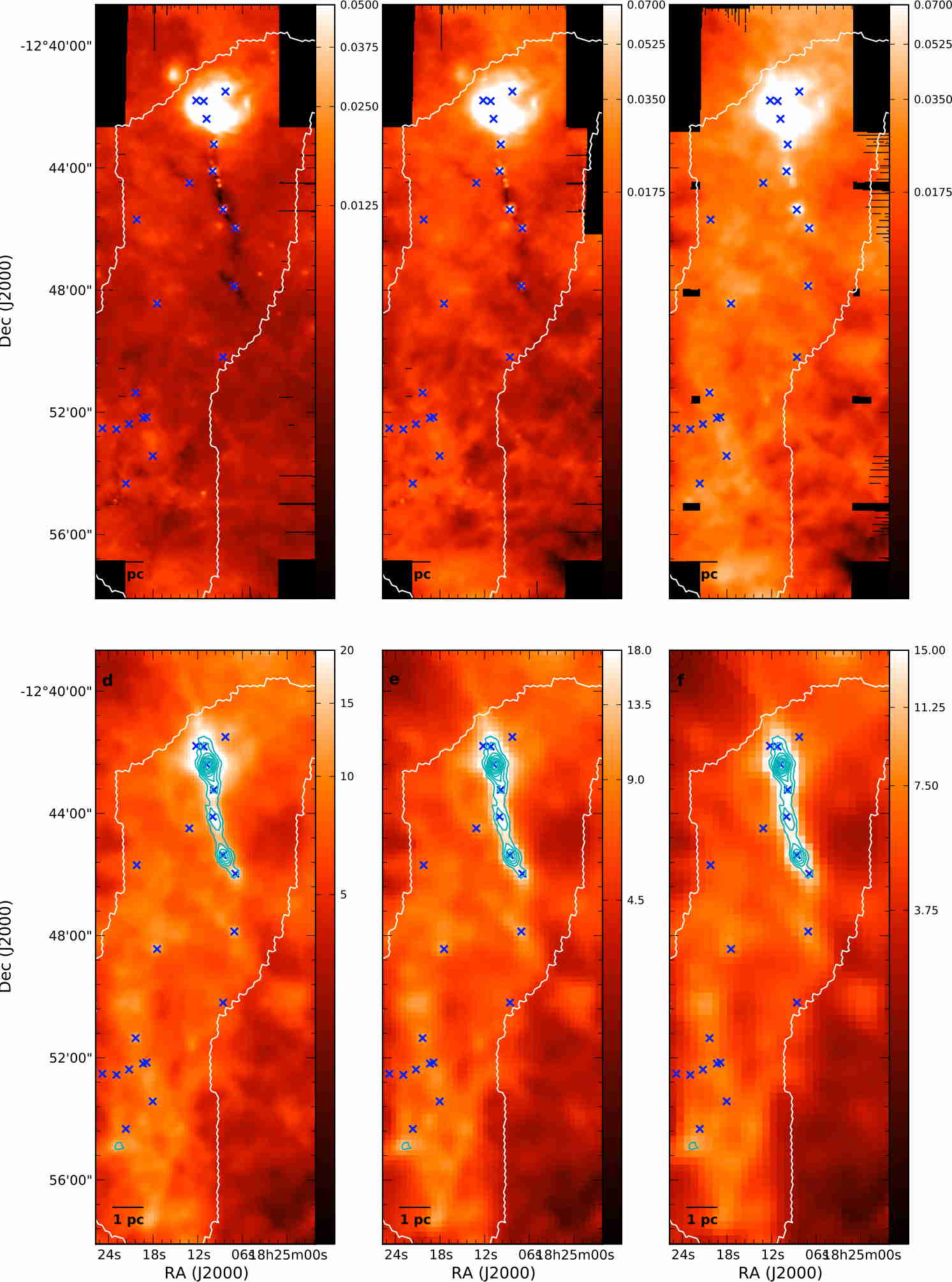}  
\caption{Same image layout as Figure~\ref{fig:hmsc} but for IRDC 18223. Cyan contours on the SPIRE panels (d, e, and f) show ATLASGAL 870\,$\mu$m: 0.49, 0.74, 0.99, 1.24, 1.48, 1.73, 1.98, 2.22, 2.47 Jy beam$^{-1}$.}  
\end{figure*}

\clearpage

\begin{figure*}[h]  
\includegraphics[width=0.95\textwidth]{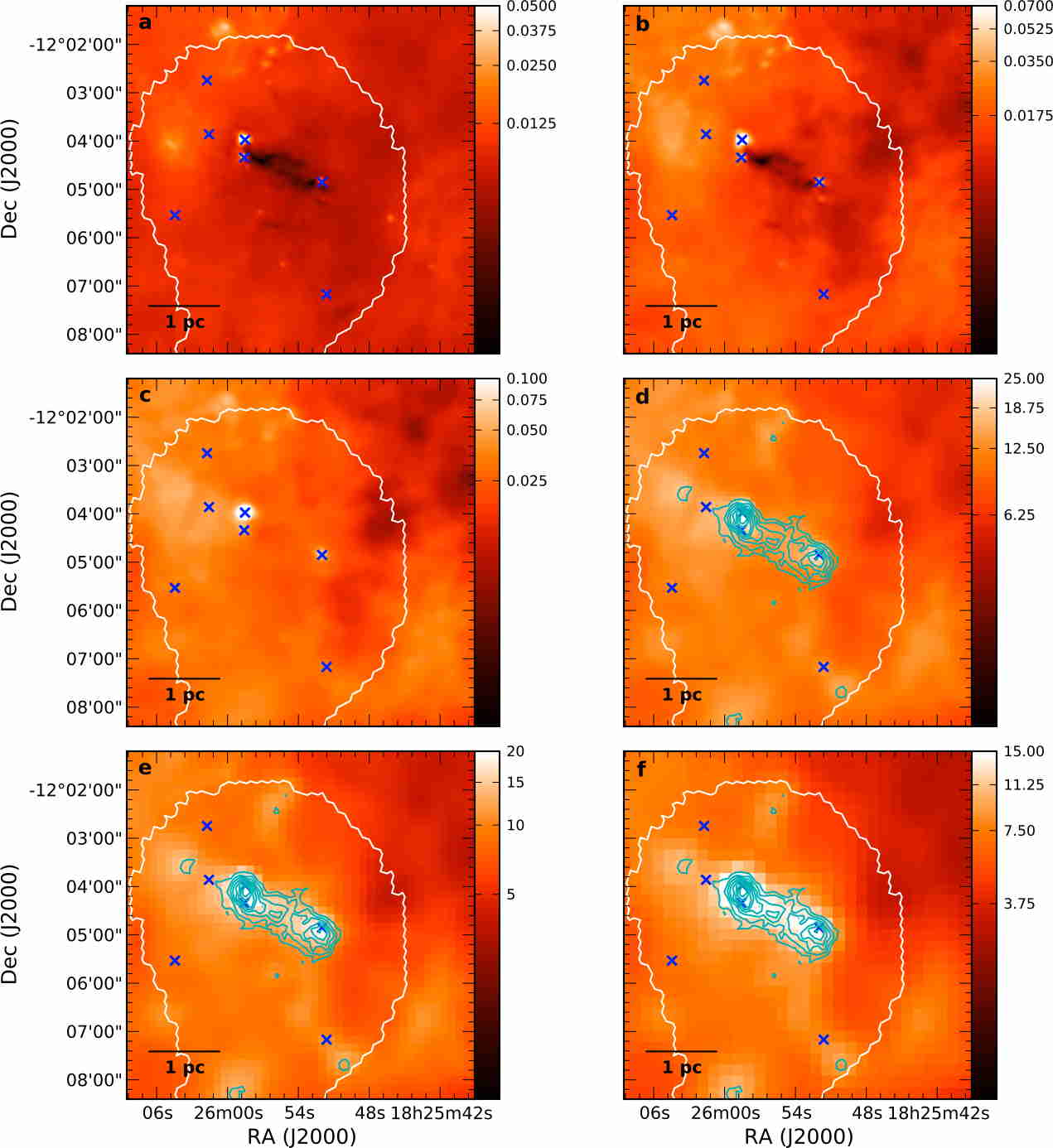}  
\caption{Same image layout as Figure~\ref{fig:hmsc} but for IRDC 19.30+0.07. Cyan contours on the SPIRE panels (d, e, and f) show ATLASGAL 870\,$\mu$m: 0.25, 0.37, 0.50, 0.62, 0.74, 0.87, 0.99, 1.12, 1.24 Jy beam$^{-1}$.}  
\end{figure*}

\clearpage

\begin{figure*}[h]  
\includegraphics[width=0.95\textwidth]{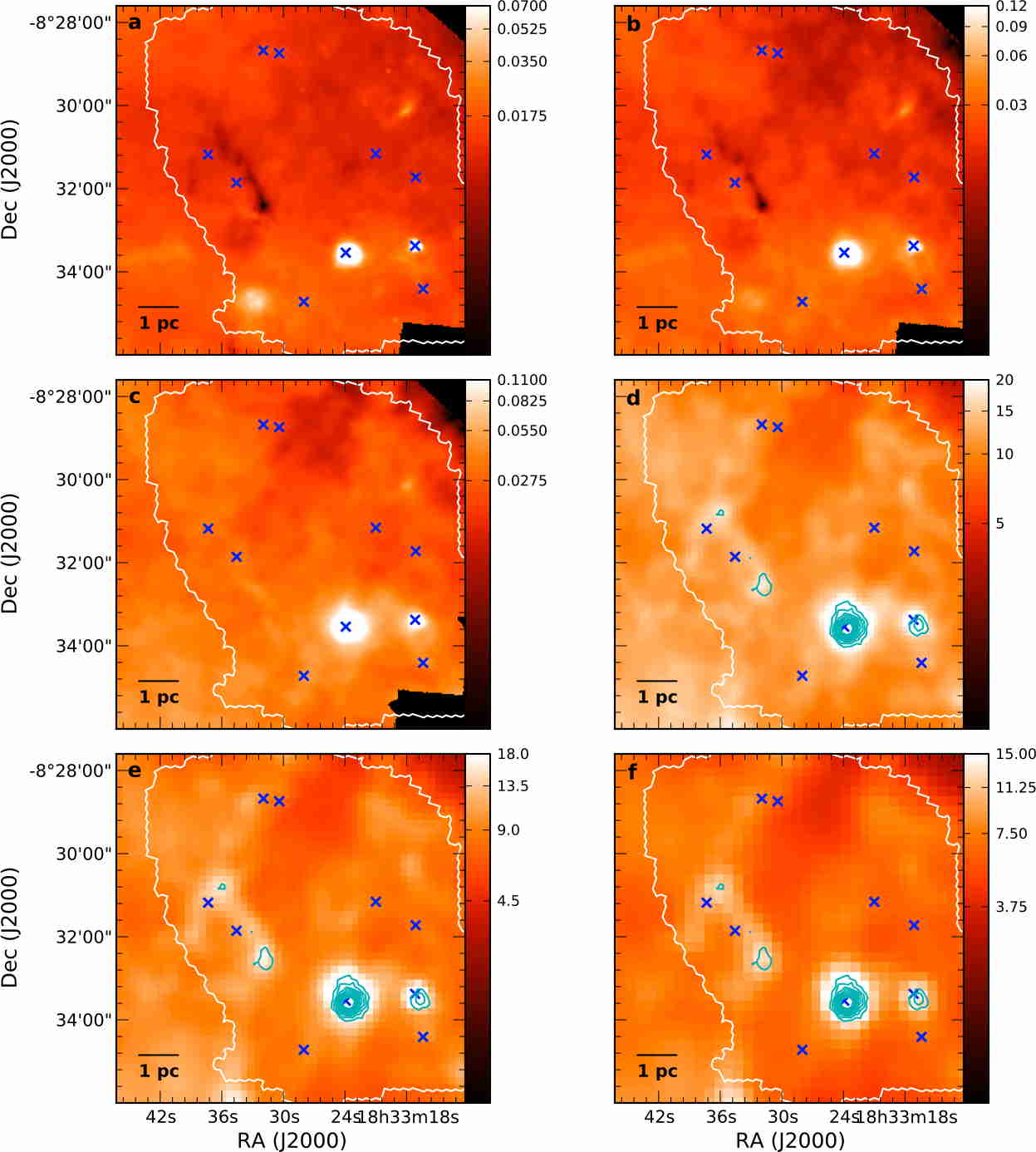}  
\caption{Same image layout as Figure~\ref{fig:hmsc} but for IRDC 18306. Cyan contours on the SPIRE panels (d, e, and f) show ATLASGAL 870\,$\mu$m: 0.4, 0.6, 0.8, 1.0, 1.2, 1.4, 1.6 ,1.8, 2.0 Jy beam$^{-1}$.}  
\end{figure*}

\clearpage

\begin{figure*}[h]  
\includegraphics[width=0.95\textwidth]{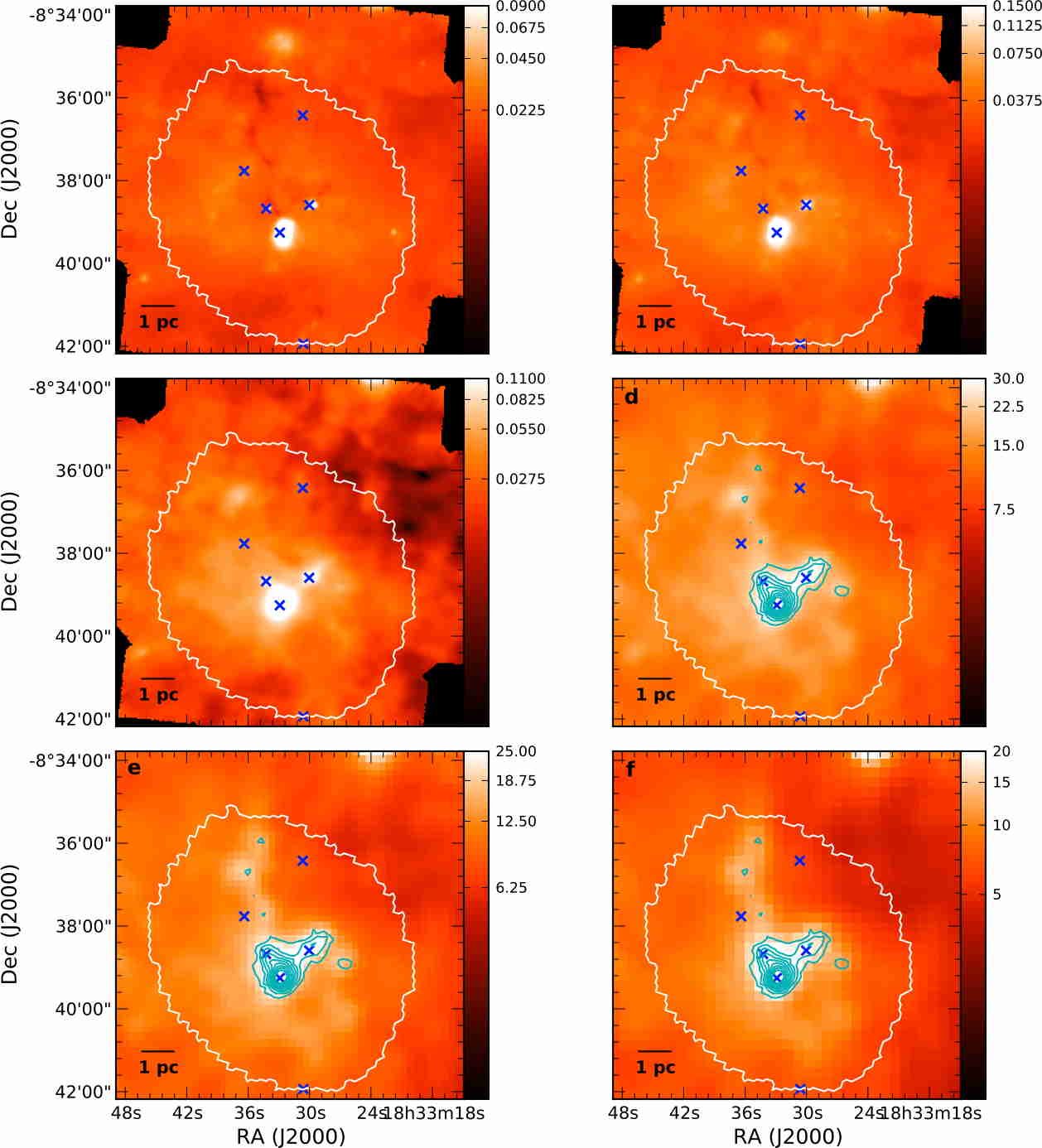}  
\caption{Same image layout as Figure~\ref{fig:hmsc} but for IRDC 18308. Cyan contours on the SPIRE panels (d, e, and f) show ATLASGAL 870\,$\mu$m: 0.4, 0.6, 0.8, 1.0, 1.2, 1.4, 1.6 ,1.8, 2.0 Jy beam$^{-1}$.}  
\end{figure*}

\clearpage

\begin{figure*}[h]  
\includegraphics[width=0.95\textwidth]{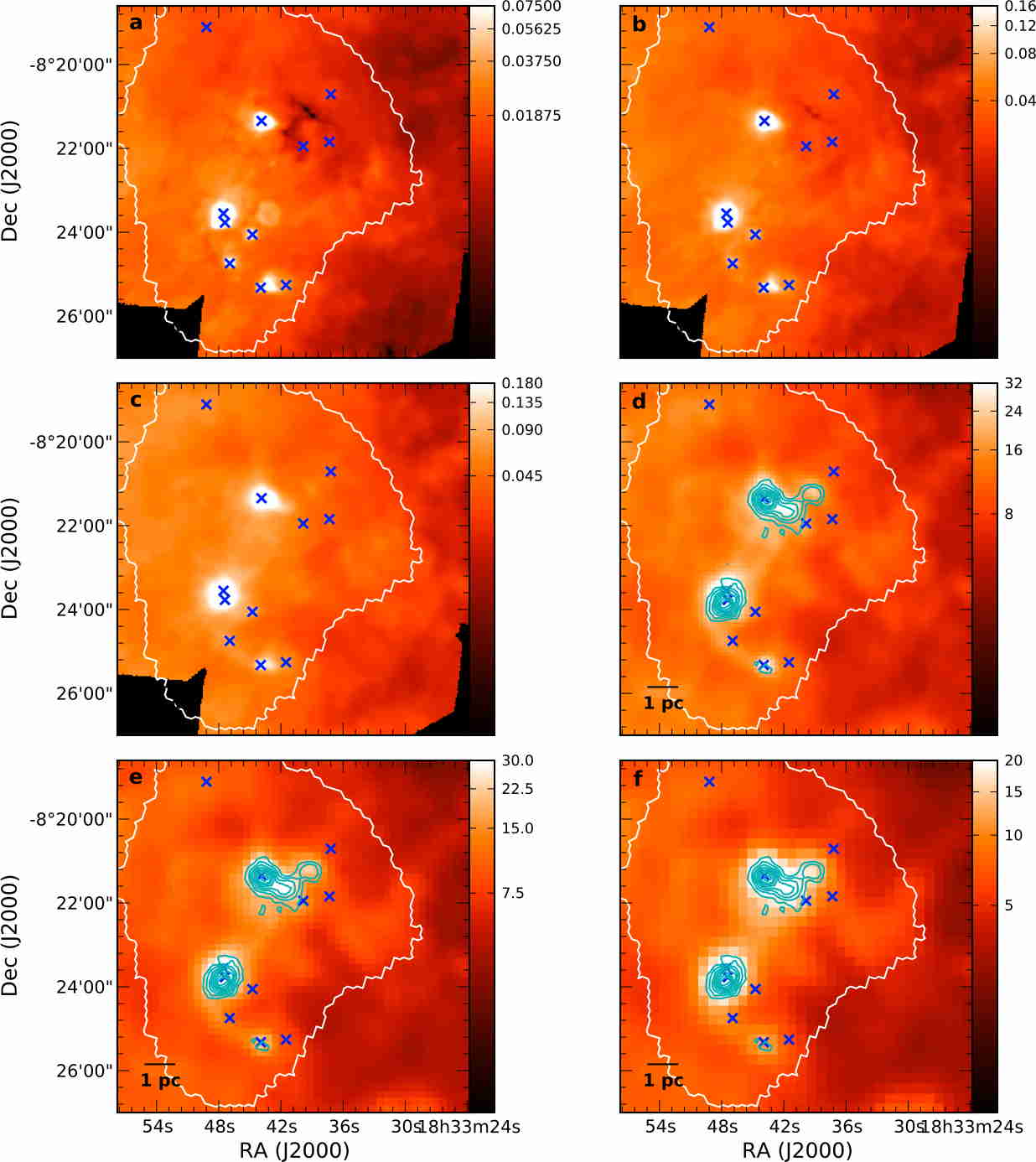}  
\caption{Same image layout as Figure~\ref{fig:hmsc} but for IRDC 18310. Cyan contours on the SPIRE panels (d, e, and f) show ATLASGAL 870\,$\mu$m: 0.4, 0.6, 0.8, 1.0, 1.2, 1.4, 1.6 ,1.8, 2.0  Jy beam$^{-1}$.}  
\end{figure*}

\clearpage

\begin{figure*}[h]  
\includegraphics[width=0.95\textwidth]{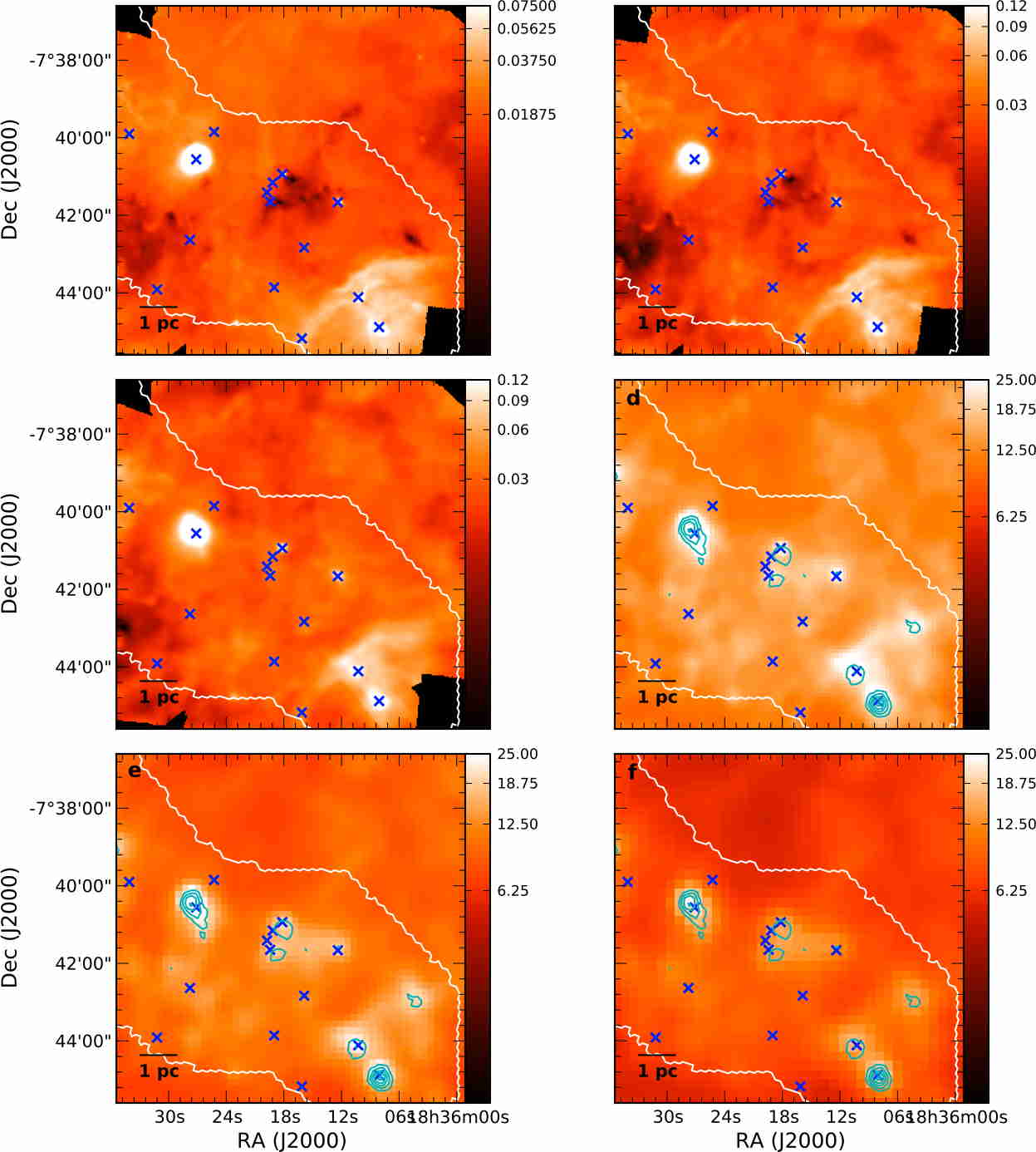}  
\caption{Same image layout as Figure~\ref{fig:hmsc} but for IRDC 18337. Cyan contours on the SPIRE panels (d, e, and f) show ATLASGAL 870\,$\mu$m: 0.4, 0.6, 0.8, 1.0, 1.2, 1.4, 1.6 ,1.8, 2.0 Jy beam$^{-1}$.}  
\end{figure*}

\clearpage

\begin{figure*}[h]  
\includegraphics[width=0.95\textwidth]{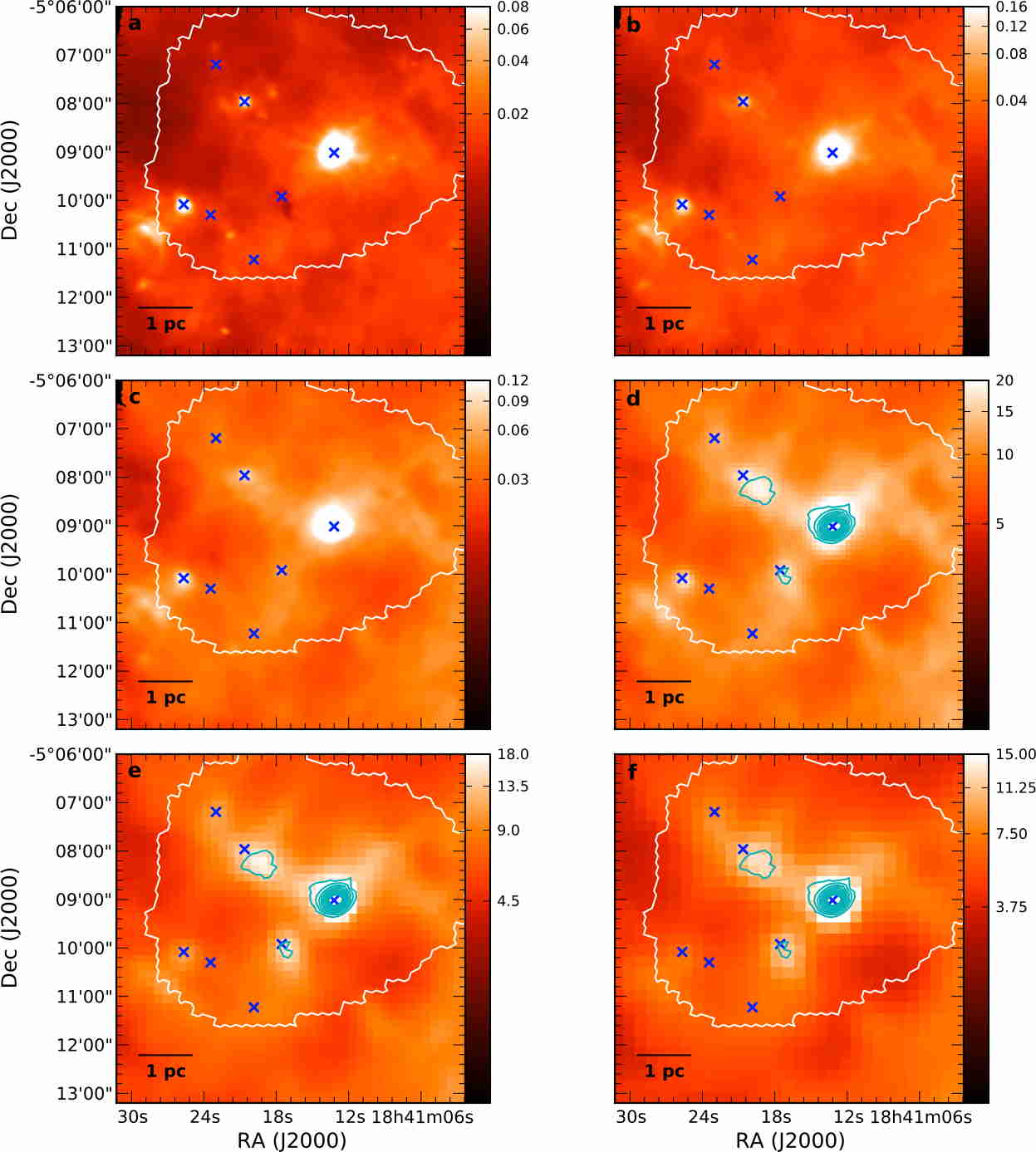}  
\caption{Same image layout as Figure~\ref{fig:hmsc} but for IRDC 18385. Cyan contours on the SPIRE panels (d, e, and f) show ATLASGAL 870\,$\mu$m: 0.4, 0.6, 0.8, 1.0, 1.2, 1.4, 1.6 ,1.8, 2.0 Jy beam$^{-1}$.}  
\end{figure*}

\clearpage

\begin{figure*}[h]  
\includegraphics[width=0.95\textwidth]{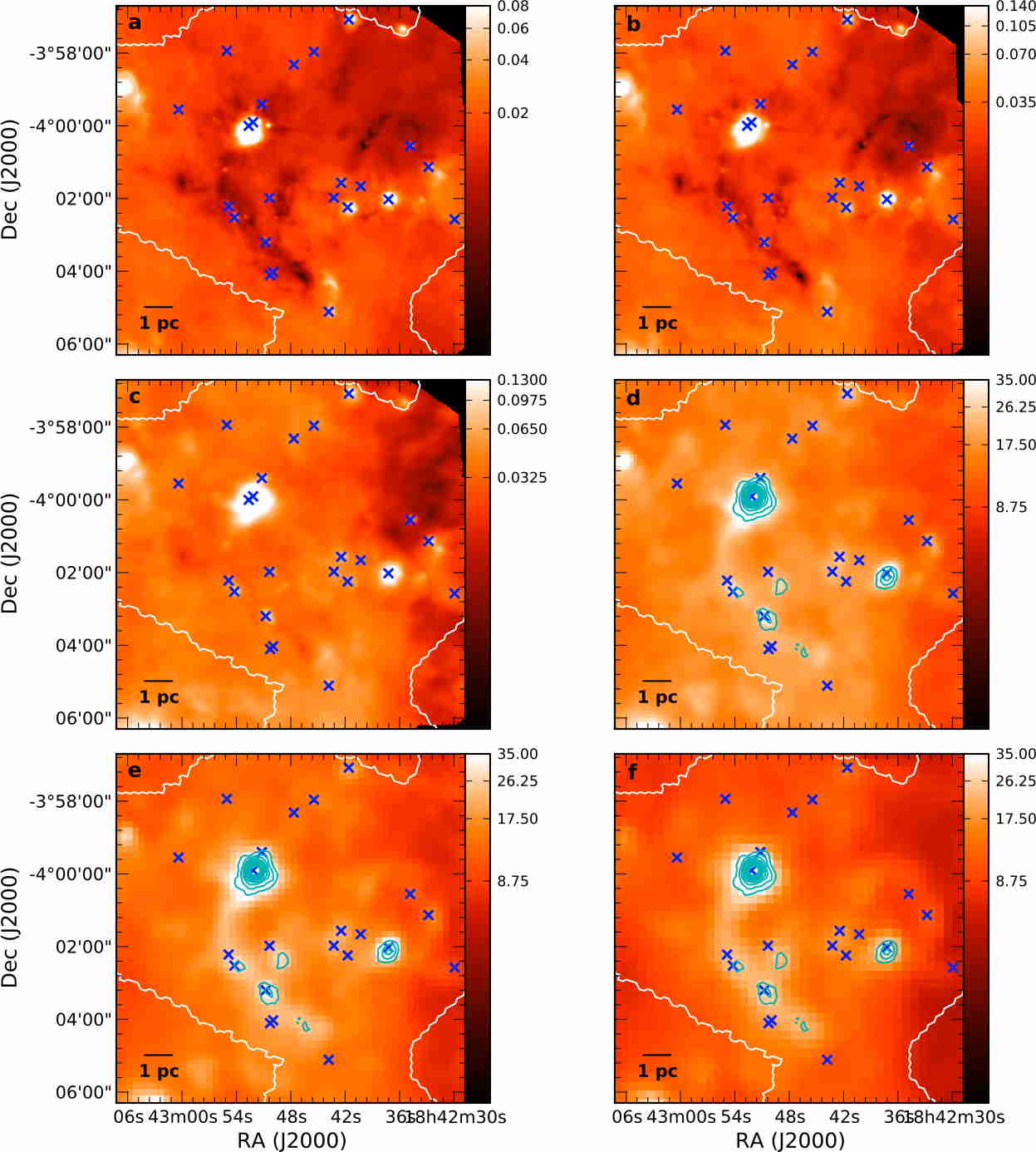}  
\caption{Same image layout as Figure~\ref{fig:hmsc} but for IRDC 028.34+0.06. Cyan contours on the SPIRE panels (d, e, and f) show ATLASGAL 870\,$\mu$m: 0.8, 1.2, 1.6, 2.0, 2.4, 2.8, 3.2, 3.6, 4.0 Jy beam$^{-1}$.} 
\end{figure*}

\clearpage

\begin{figure*}[h]  
\includegraphics[width=0.95\textwidth]{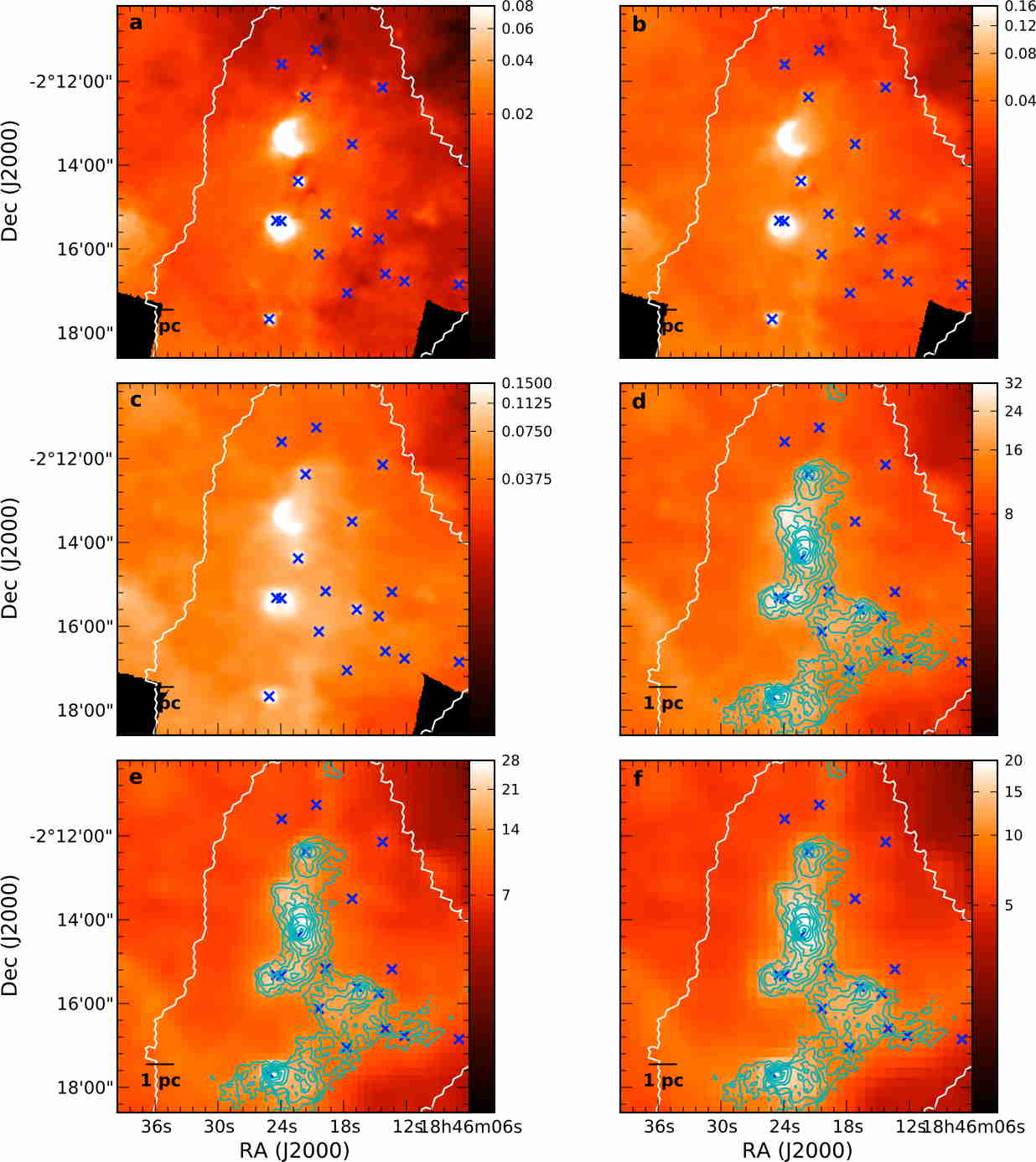}  
\caption{Same image layout as Figure~\ref{fig:hmsc} but for IRDC 18437. Cyan contours on the SPIRE panels (d, e, and f) show ATLASGAL 870\,$\mu$m: 0.2, 0.3, 0.4, 0.5, 0.6, 0.7, 0.8, 0.9, 1.0 Jy beam$^{-1}$.}  
\end{figure*}

\clearpage

\begin{figure*}[h]  
\includegraphics[width=0.95\textwidth]{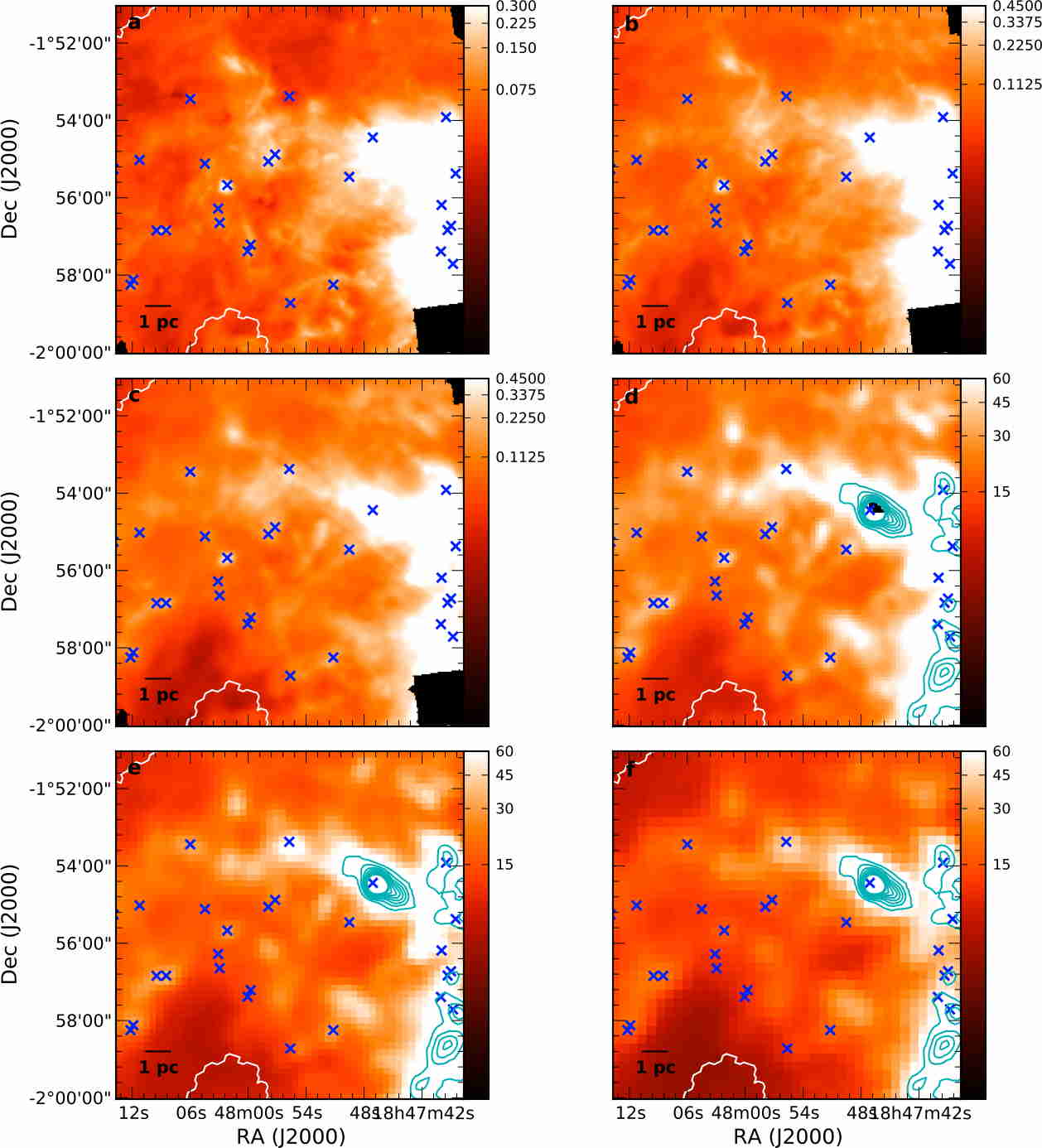}  
\caption{Same image layout as Figure~\ref{fig:hmsc} but for IRDC 18454. Cyan contours on the SPIRE panels (d, e, and f) show ATLASGAL 870\,$\mu$m: 2.0, 3.0, 4.0, 5.0, 6.0, 7.0, 8.0, 9.0, 10.0 Jy beam$^{-1}$.}  
\end{figure*}

\clearpage

\begin{figure*}[h]  
\includegraphics[width=0.95\textwidth]{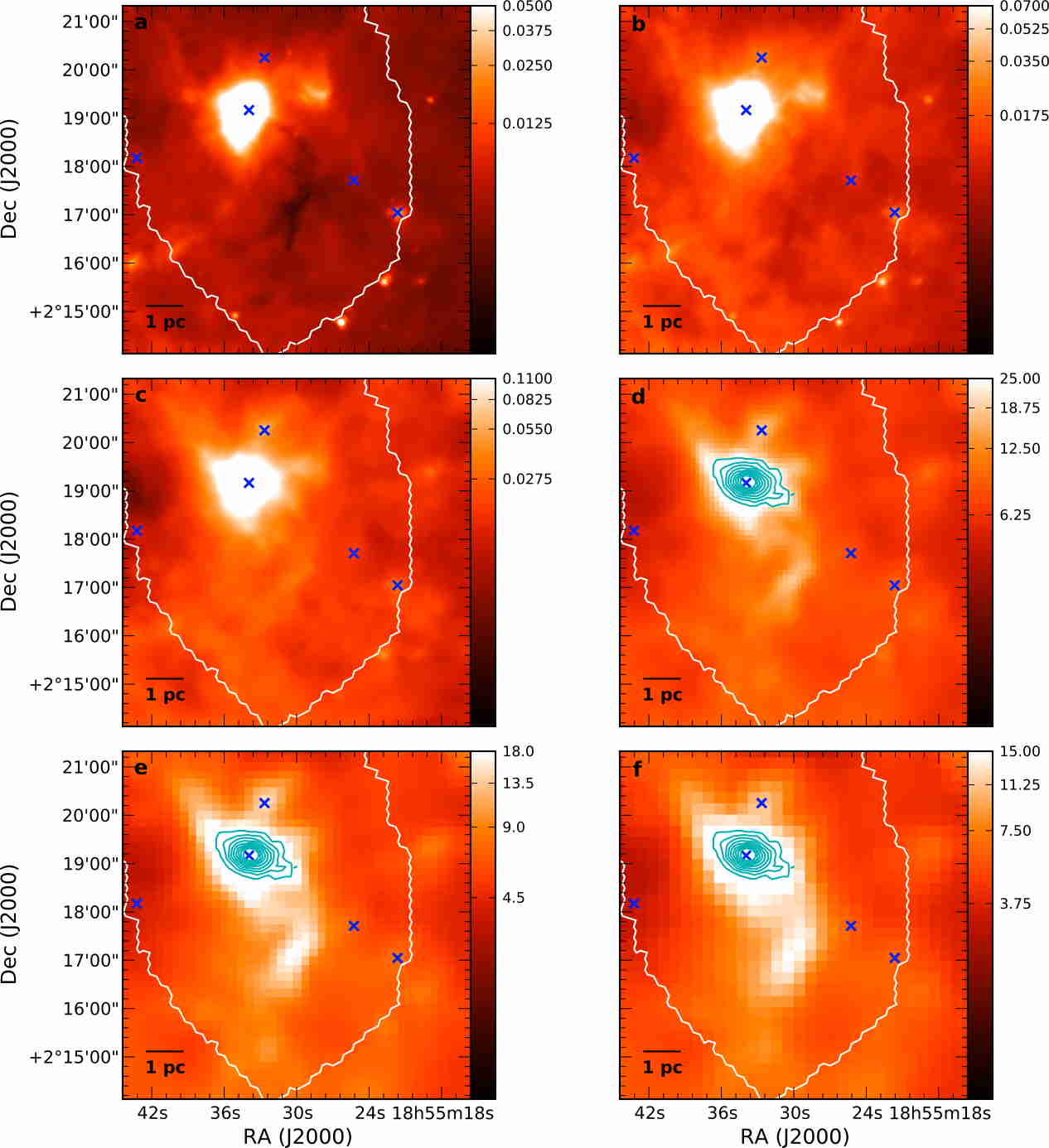}  
\caption{Same image layout as Figure~\ref{fig:hmsc} but for IRDC 18530. Cyan contours on the SPIRE panels (d, e, and f) show ATLASGAL 870\,$\mu$m: 0.6, 0.9, 1.2, 1.5, 1.8, 2.1, 2.4, 2.7, 3.0 Jy beam$^{-1}$.}  
\end{figure*}

\clearpage

\begin{figure*}[h]  
\includegraphics[width=0.95\textwidth]{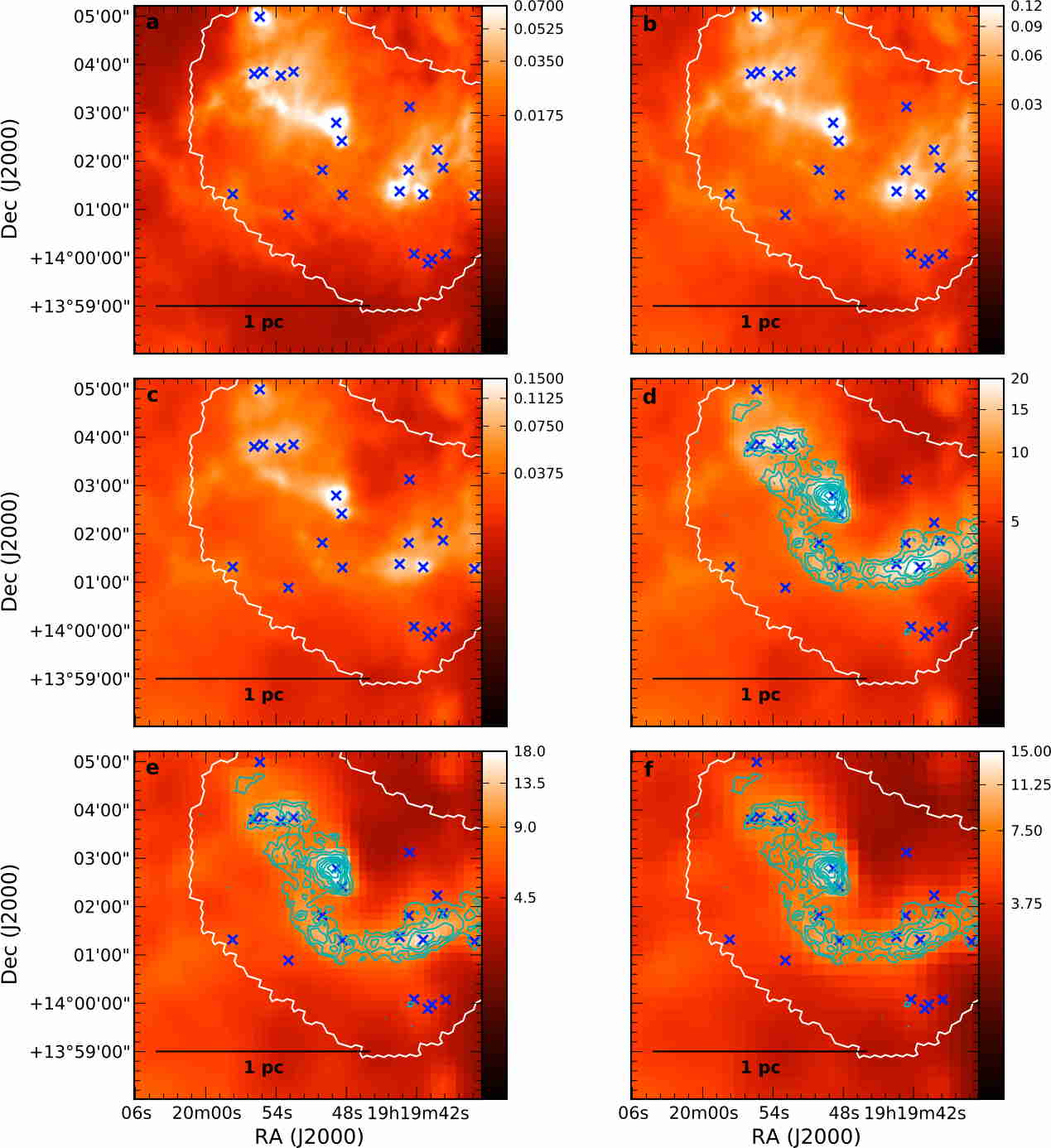}  
\caption{Same image layout as Figure~\ref{fig:hmsc} but for IRDC 19175. Cyan contours on the SPIRE panels (d, e, and f) show ATLASGAL 870\,$\mu$m: 0.19, 0.29, 0.38, 0.48, 0.57, 0.67, 0.76, 0.86, 0.95 Jy beam$^{-1}$.}  
\end{figure*}

\clearpage

\begin{figure*}[h]  
\includegraphics[width=0.95\textwidth]{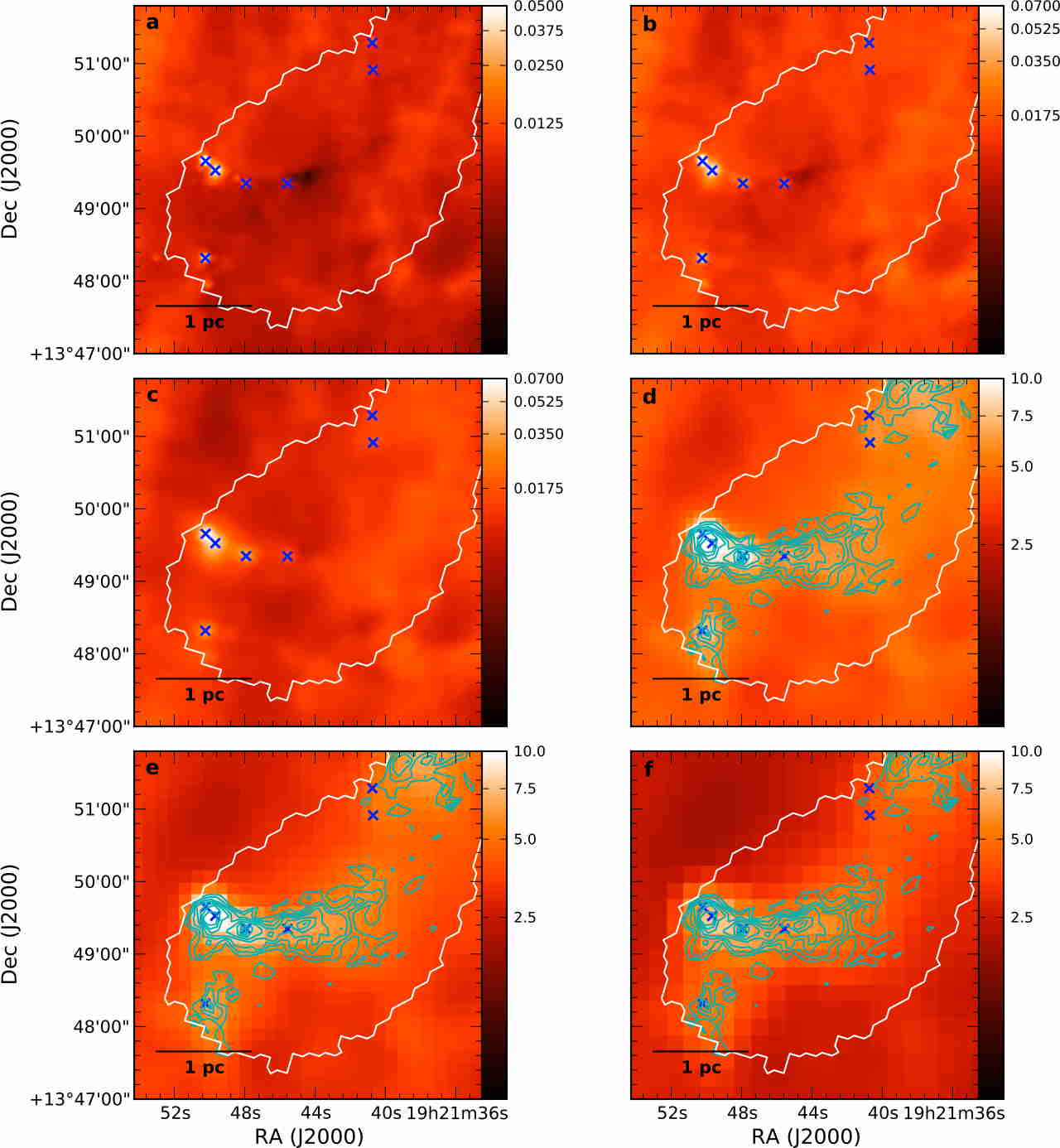}  
\caption{Same image layout as Figure~\ref{fig:hmsc} but for IRDC 048.66-0.29. Cyan contours on the SPIRE panels (d, e, and f) show ATLASGAL 870\,$\mu$m: 0.106, 0.159, 0.21, 0.265, 0.318, 0.371, 0.424, 0.477, 0.53 Jy beam$^{-1}$.}  
\end{figure*}

\clearpage

\begin{figure*}[h]  
\includegraphics[width=0.95\textwidth]{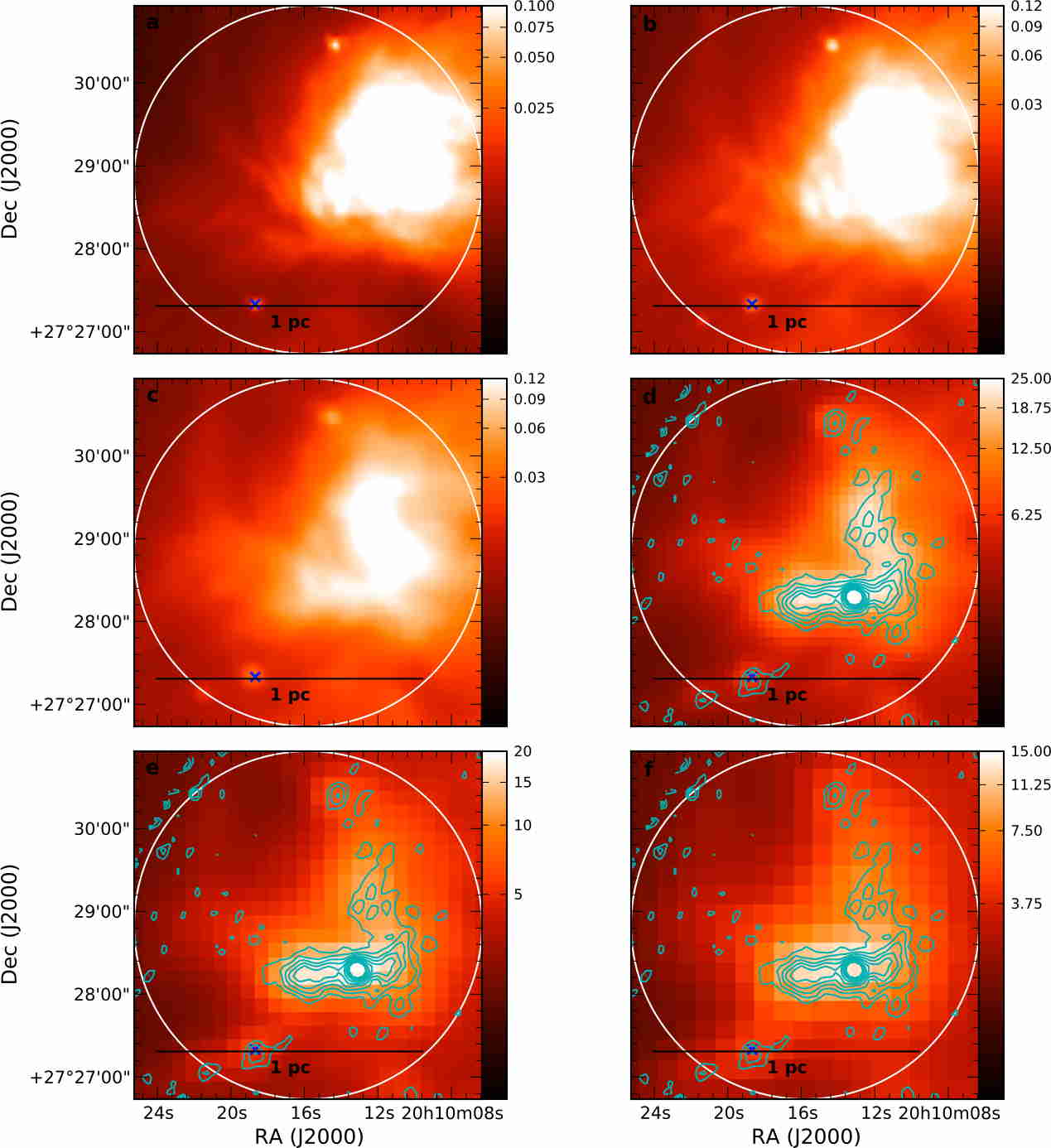}  
\caption{Same image layout as Figure~\ref{fig:hmsc} but for IRDC 20081. Cyan contours on the SPIRE panels (d, e, and f) show MAMBO 1.2\,mm: 0.04, 0.07, 0.09, 0.11, 0.13, 0.14, 0.16, 0.18 Jy beam$^{-1}$.}  
\end{figure*}

\clearpage

\begin{figure*}[h]  
\includegraphics[width=0.95\textwidth]{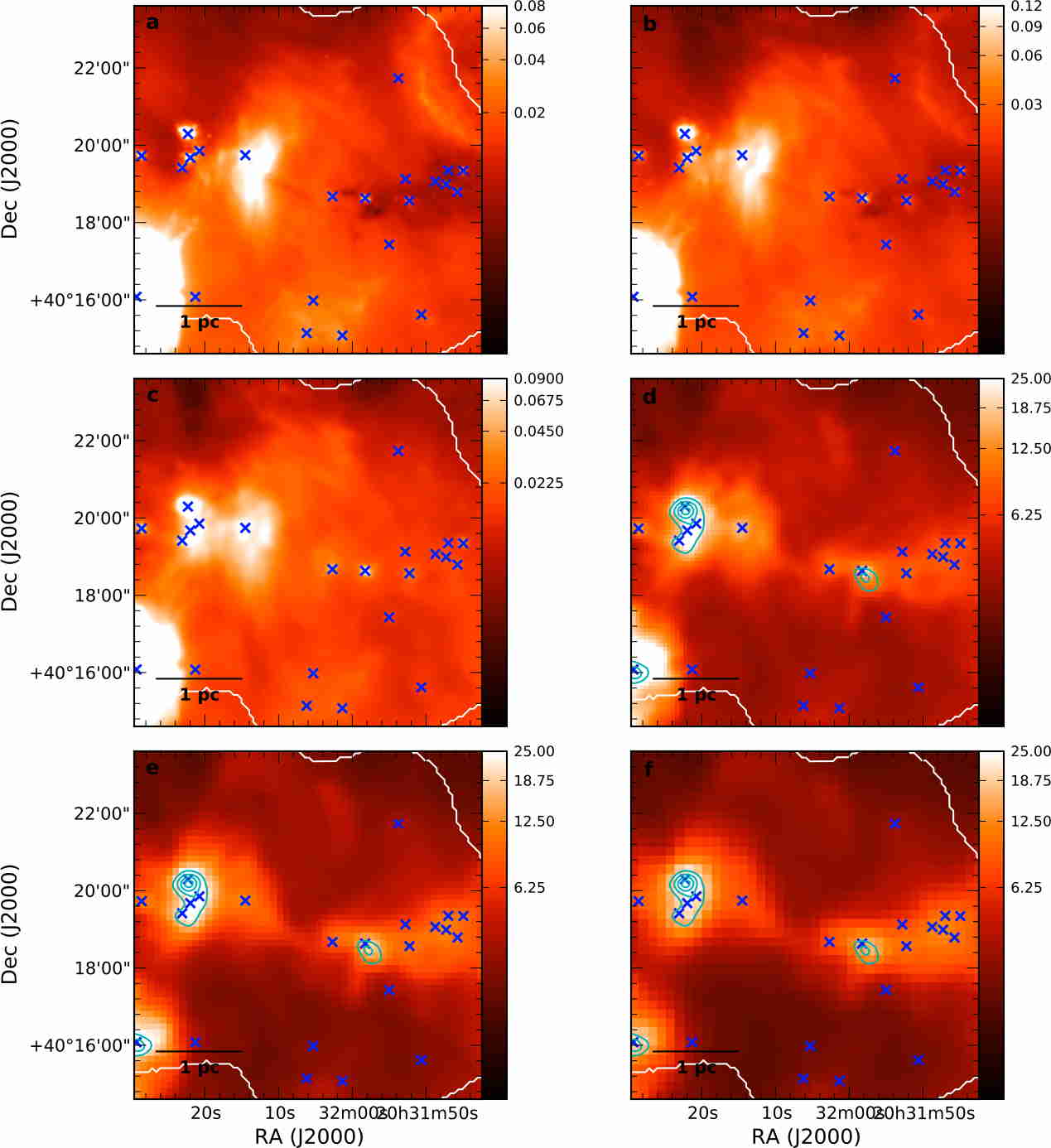}  
\caption{Same image layout as Figure~\ref{fig:hmsc} but for IRDC 079.31+0.36. Cyan contours on the SPIRE panels (d, e, and f) show SCUBA 850\,$\mu$m: 1.0, 1.5, 2.0, 2.5, 3.0, 3.5, 4.0, 4.5, 5.0 Jy beam$^{-1}$.}  
\end{figure*}

\clearpage

\begin{figure*}[h]  
\includegraphics[width=0.95\textwidth]{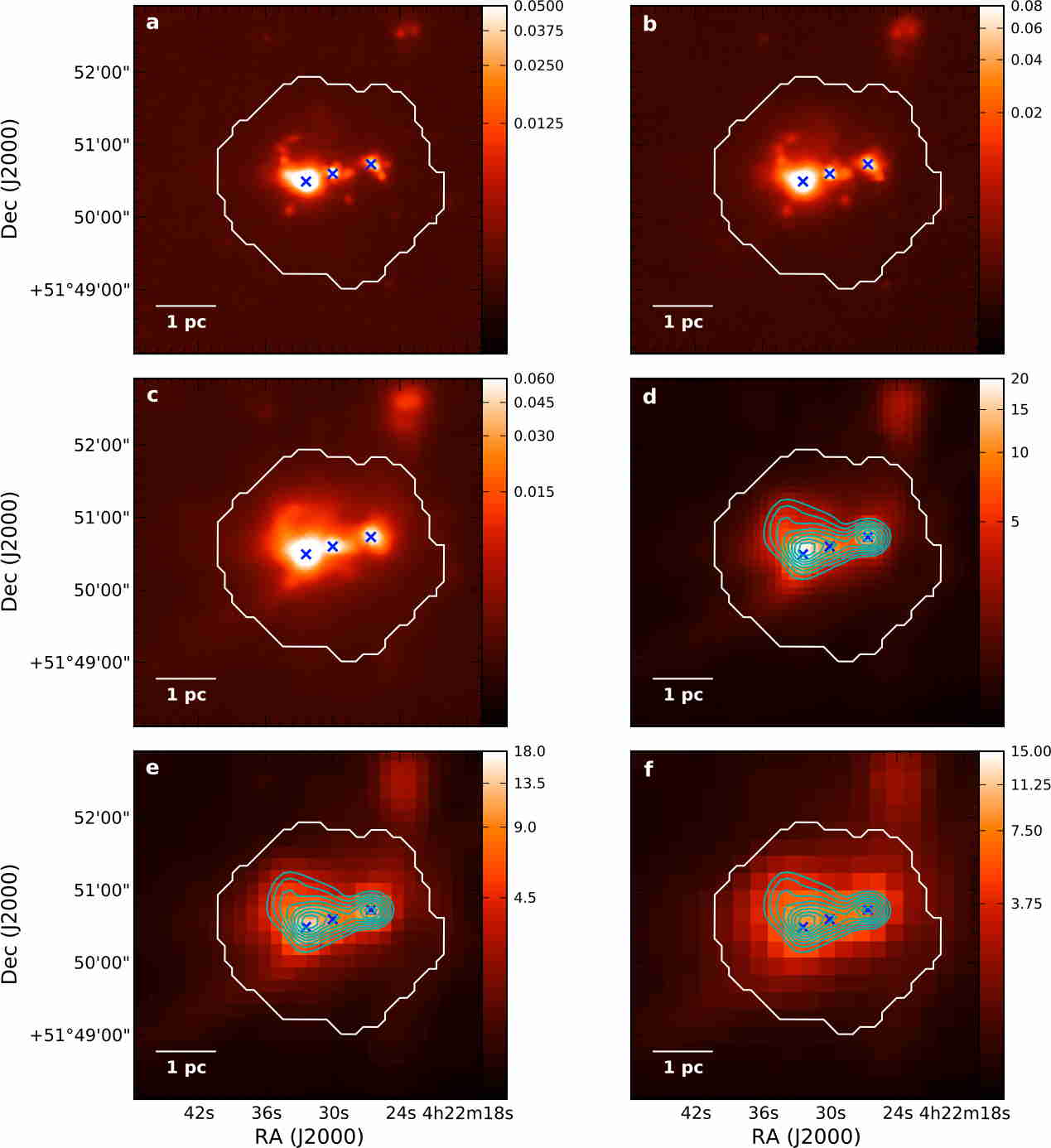}  
\caption{Same image layout as Figure~\ref{fig:hmsc} but for ISOSS J04225+5150. Cyan contours on the SPIRE panels (d, e, and f) show SCUBA 850\,$\mu$m: 0.22, 0.32, 0.43, 0.54, 0.65, 0.76, 0.86, 0.97, 1.08 Jy beam$^{-1}$.}  
\end{figure*}

\clearpage

\begin{figure*}[h]  
\includegraphics[width=0.95\textwidth]{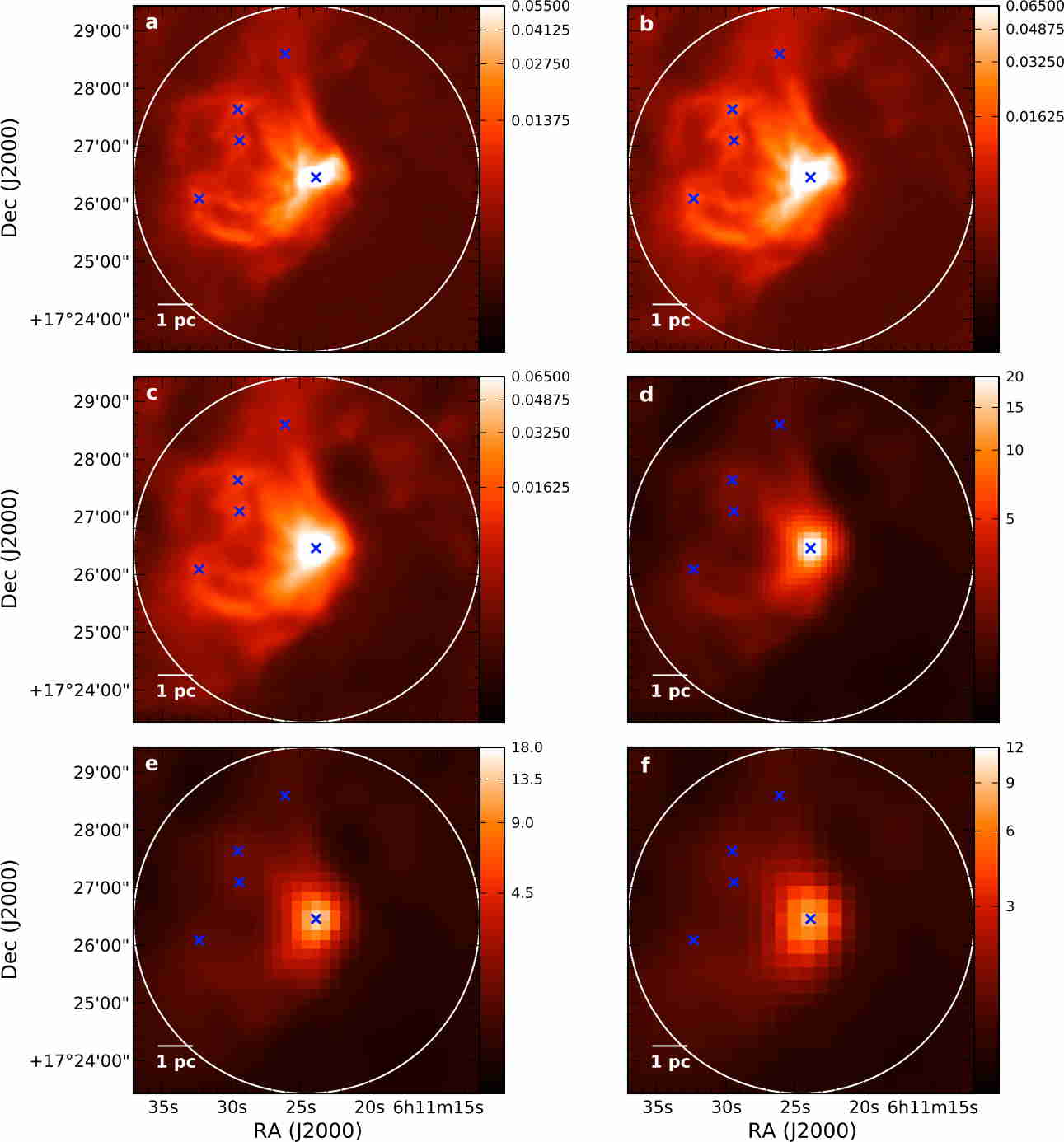}  
\caption{Same image layout as Figure~\ref{fig:hmsc} but for ISOSS J06114+1726. }  
\end{figure*}

\clearpage

\begin{figure*}[h]  
\includegraphics[width=0.95\textwidth]{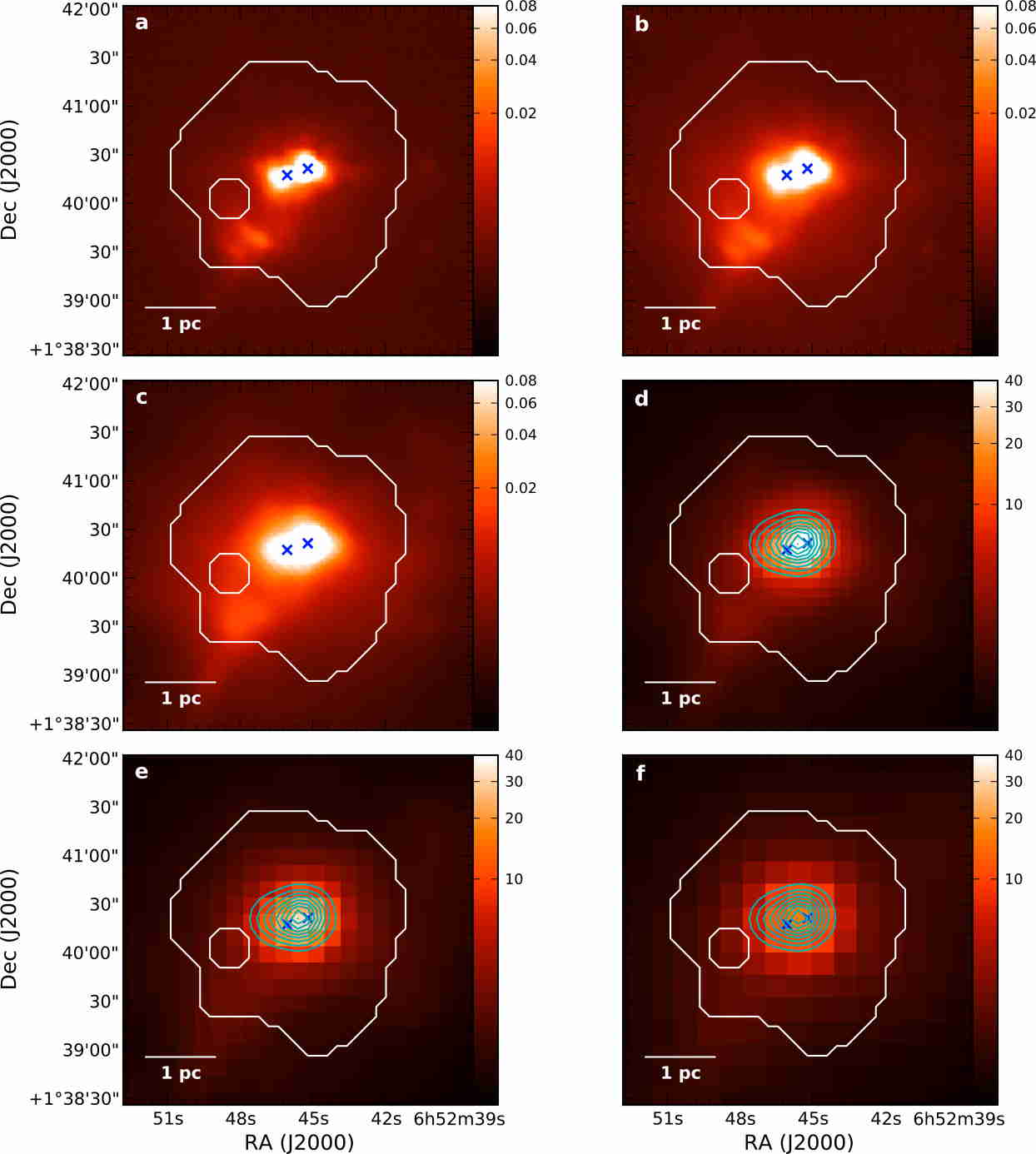}  
\caption{Same image layout as Figure~\ref{fig:hmsc} but for ISOSS J06527+0140. Cyan contours on the SPIRE panels (d, e, and f) show SCUBA 850\,$\mu$m: 0.41, 0.62, 0.82, 1.03, 1.24, 1.44, 1.65, 1.85, 2.06 Jy beam$^{-1}$.}  
\end{figure*}

\clearpage

\begin{figure*}[h]  
\includegraphics[width=0.95\textwidth]{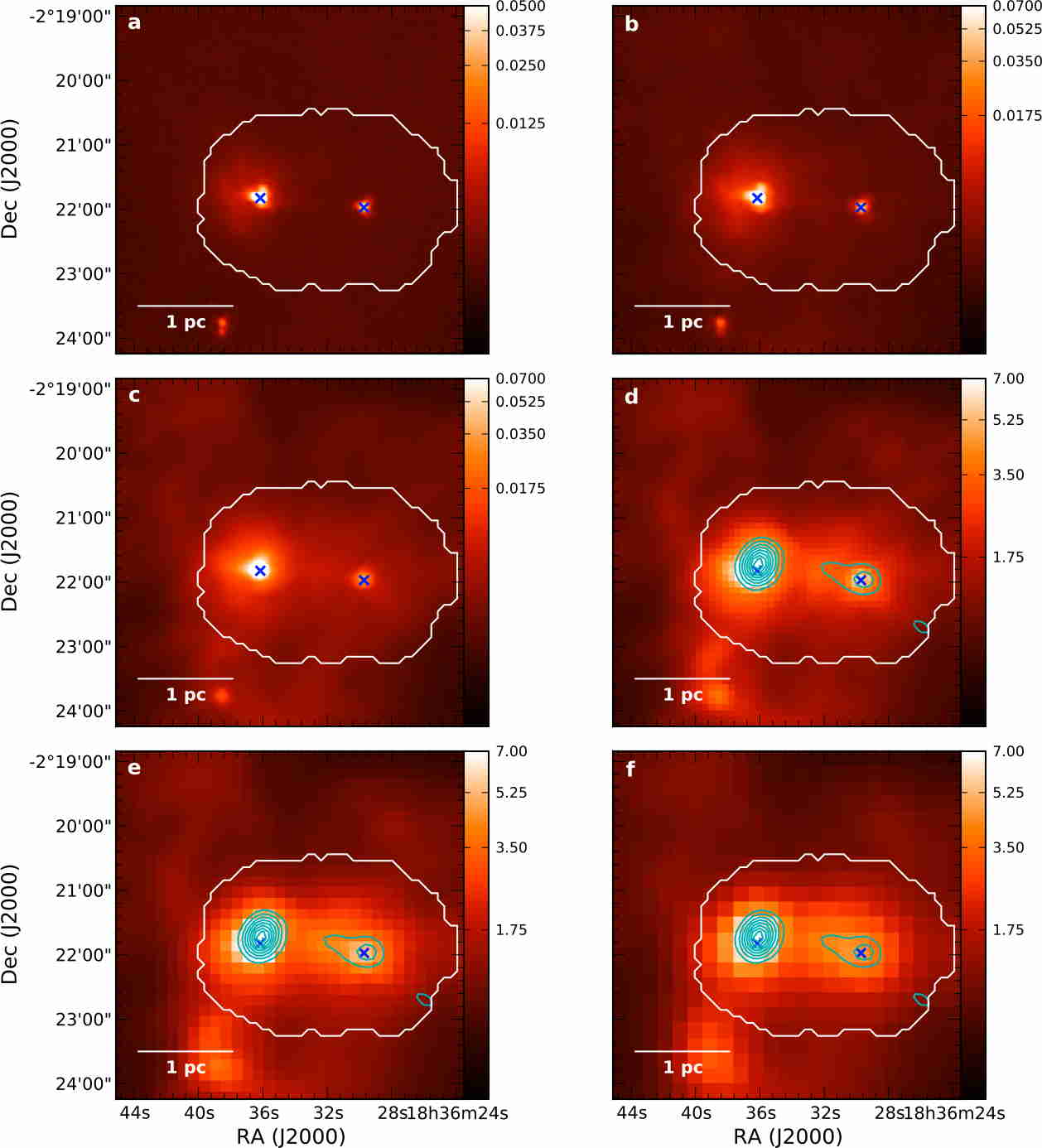}  
\caption{Same image layout as Figure~\ref{fig:hmsc} but for ISOSS J18364-0221. Cyan contours on the SPIRE panels (d, e, and f) show SCUBA 850\,$\mu$m: 0.21, 0.31, 0.42, 0.52, 0.63, 0.73, 0.83, 0.94, 1.04 Jy beam$^{-1}$.}  
\end{figure*}

\clearpage

\begin{figure*}[h]  
\includegraphics[width=0.95\textwidth]{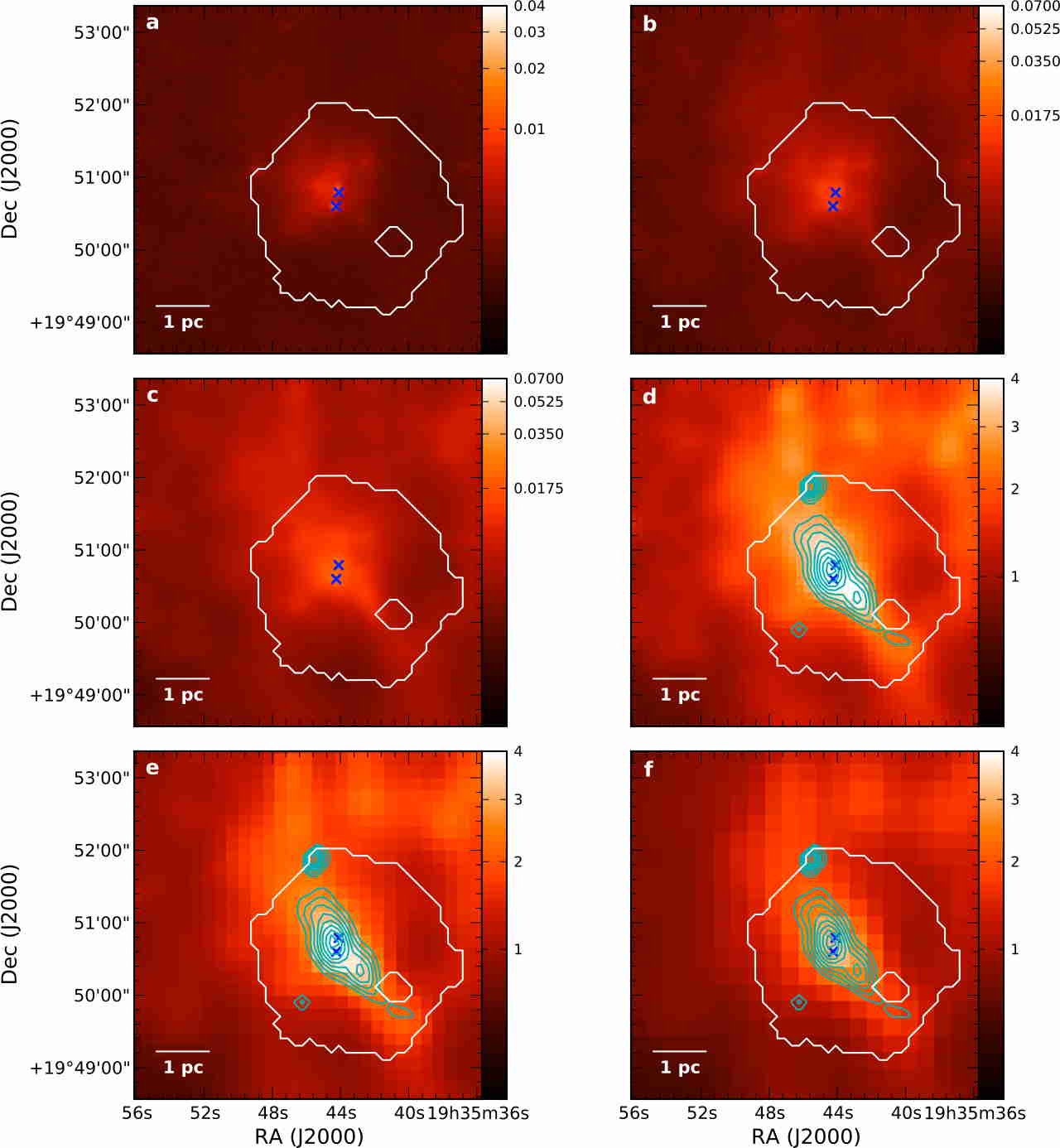}  
\caption{Same image layout as Figure~\ref{fig:hmsc} but for ISOSS J19357+1950. Cyan contours on the SPIRE panels (d, e, and f) show SCUBA 850\,$\mu$m: 0.06, 0.09, 0.12, 0.15, 0.17, 0.20, 0.23, 0.26, 0.29 Jy beam$^{-1}$.}  
\end{figure*}

\clearpage

\begin{figure*}[h]  
\includegraphics[width=0.95\textwidth]{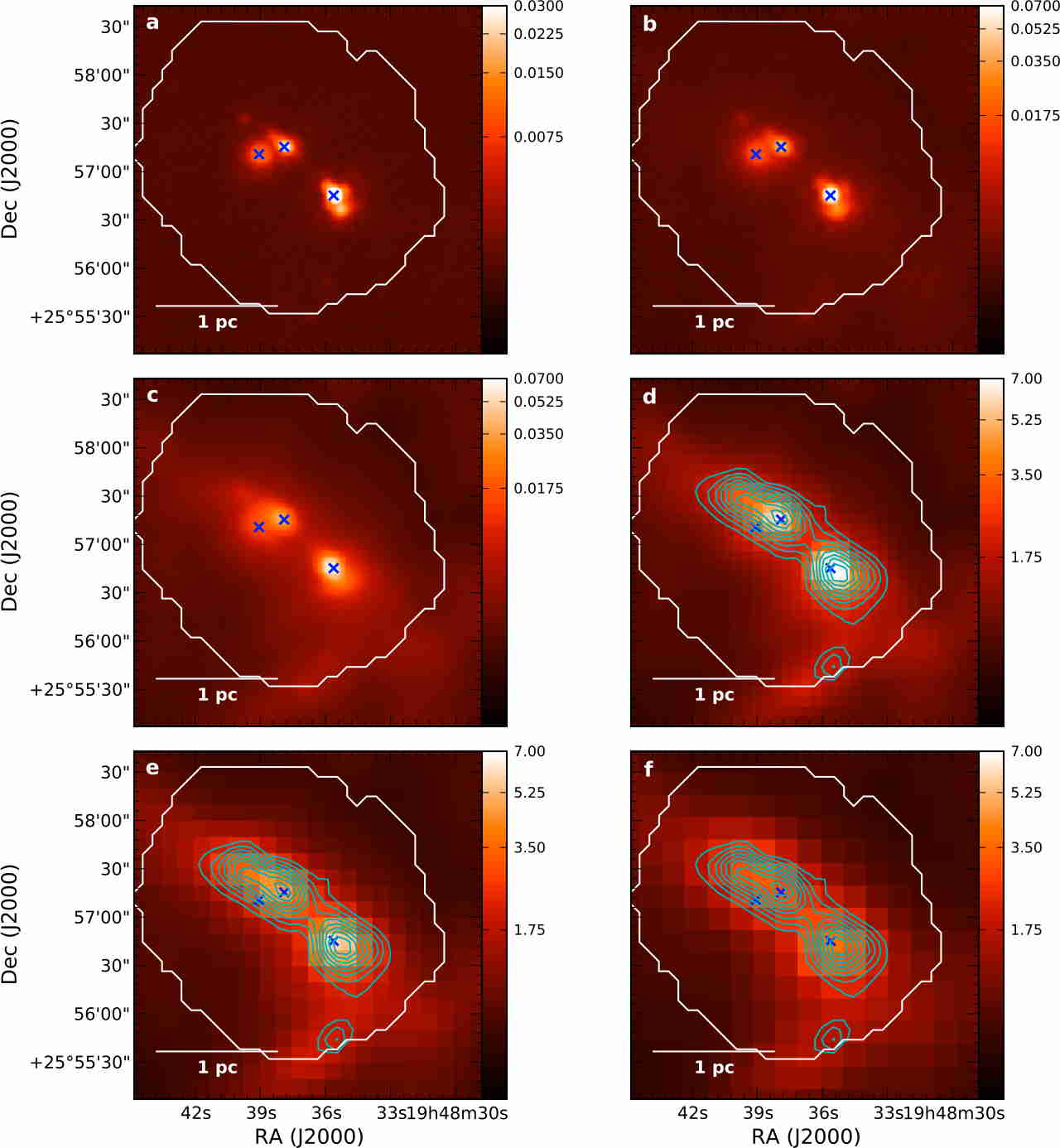}  
\caption{Same image layout as Figure~\ref{fig:hmsc} but for ISOSS J19486+2556. Cyan contours on the SPIRE panels (d, e, and f) show SCUBA 850\,$\mu$m: 0.07, 0.11, 0.14, 0.18, 0.22, 0.25, 0.29, 0.32, 0.36 Jy beam$^{-1}$.}  
\end{figure*}

\clearpage

\begin{figure*}[h]  
\includegraphics[width=0.95\textwidth]{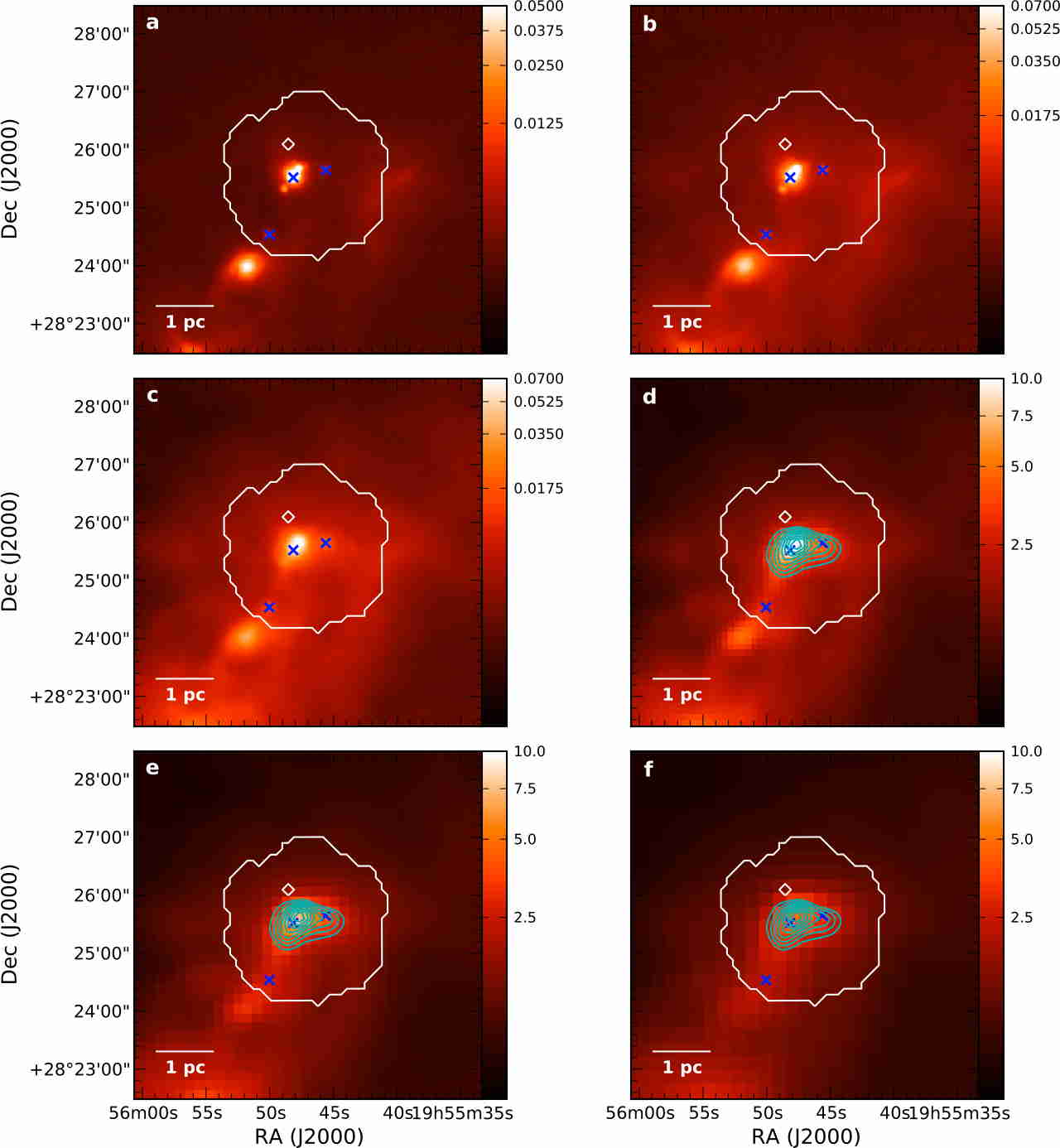}  
\caption{Same image layout as Figure~\ref{fig:hmsc} but for ISOSS J19557+2825. Cyan contours on the SPIRE panels (d, e, and f) show SCUBA 850\,$\mu$m: 0.13, 0.19, 0.26, 0.32, 0.38, 0.45, 0.51, 0.58, 0.64 Jy beam$^{-1}$.}  
\end{figure*}

\clearpage

\begin{figure*}[h]  
\includegraphics[width=0.95\textwidth]{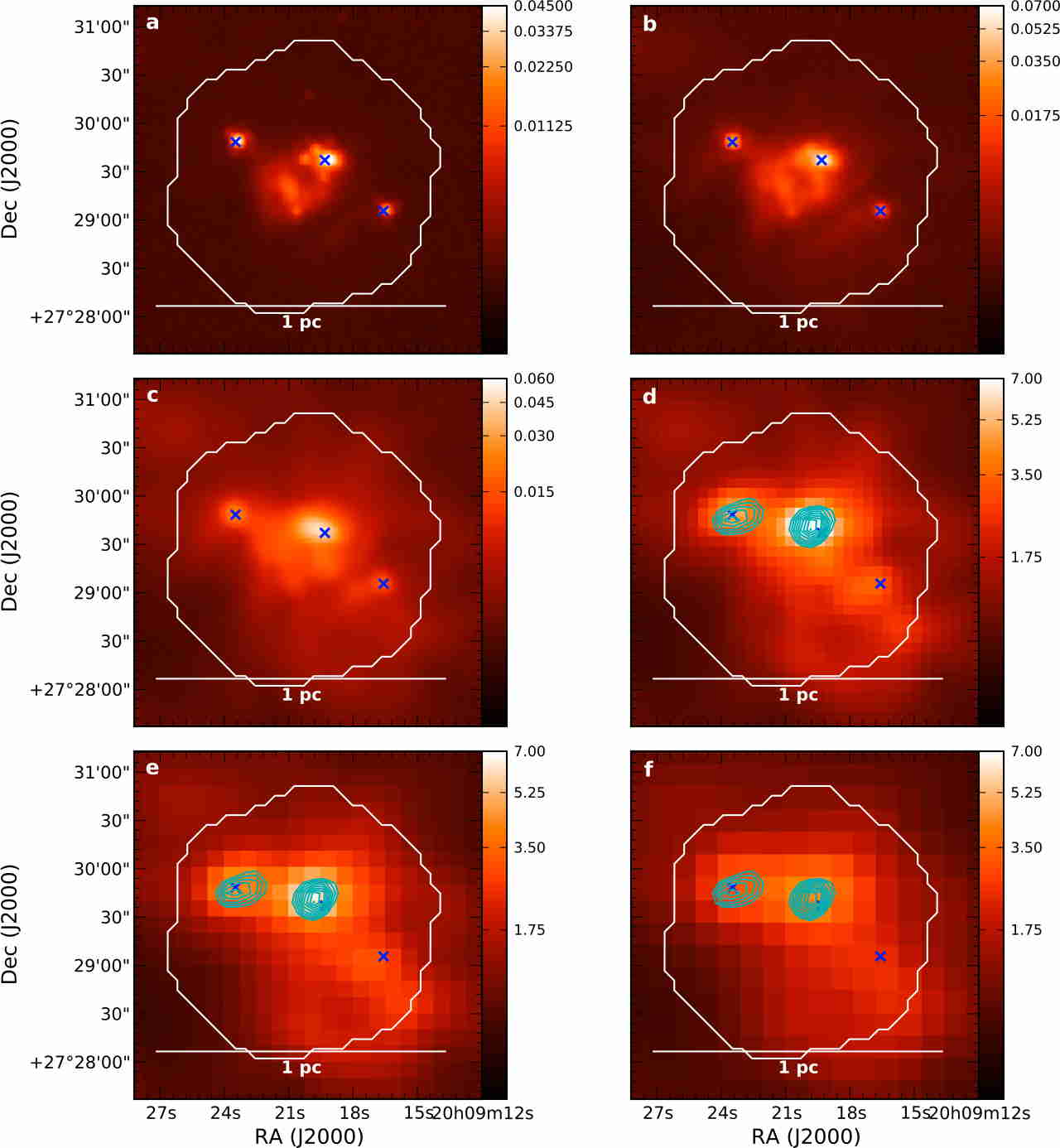}  
\caption{Same image layout as Figure~\ref{fig:hmsc} but for ISOSS J20093+2729. Cyan contours on the SPIRE panels (d, e, and f) show SCUBA 850\,$\mu$m: 0.05, 0.08, 0.10, 0.13, 0.15, 0.18, 0.20, 0.23, 0.25 Jy beam$^{-1}$.}  
\end{figure*}

\clearpage

\begin{figure*}[h]  
\includegraphics[width=0.95\textwidth]{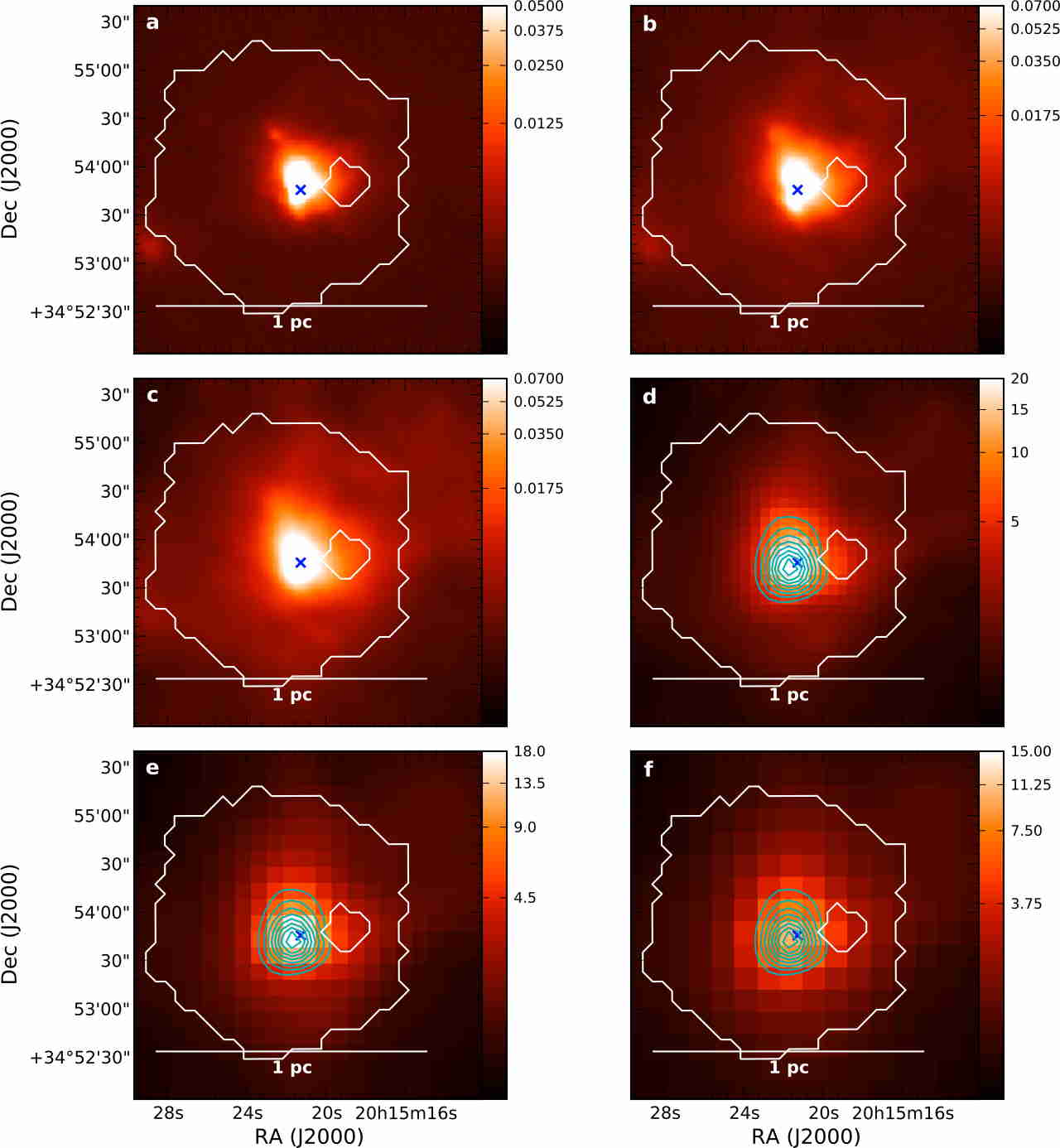}  
\caption{Same image layout as Figure~\ref{fig:hmsc} but for ISOSS J20153+3453. Cyan contours on the SPIRE panels (d, e, and f) show SCUBA 850\,$\mu$m: 0.28, 0.41, 0.55, 0.69, 0.83, 0.97, 1.10, 1.24, 1.38 Jy beam$^{-1}$.}  
\end{figure*}

\clearpage

\begin{figure*}[h]  
\includegraphics[width=0.95\textwidth]{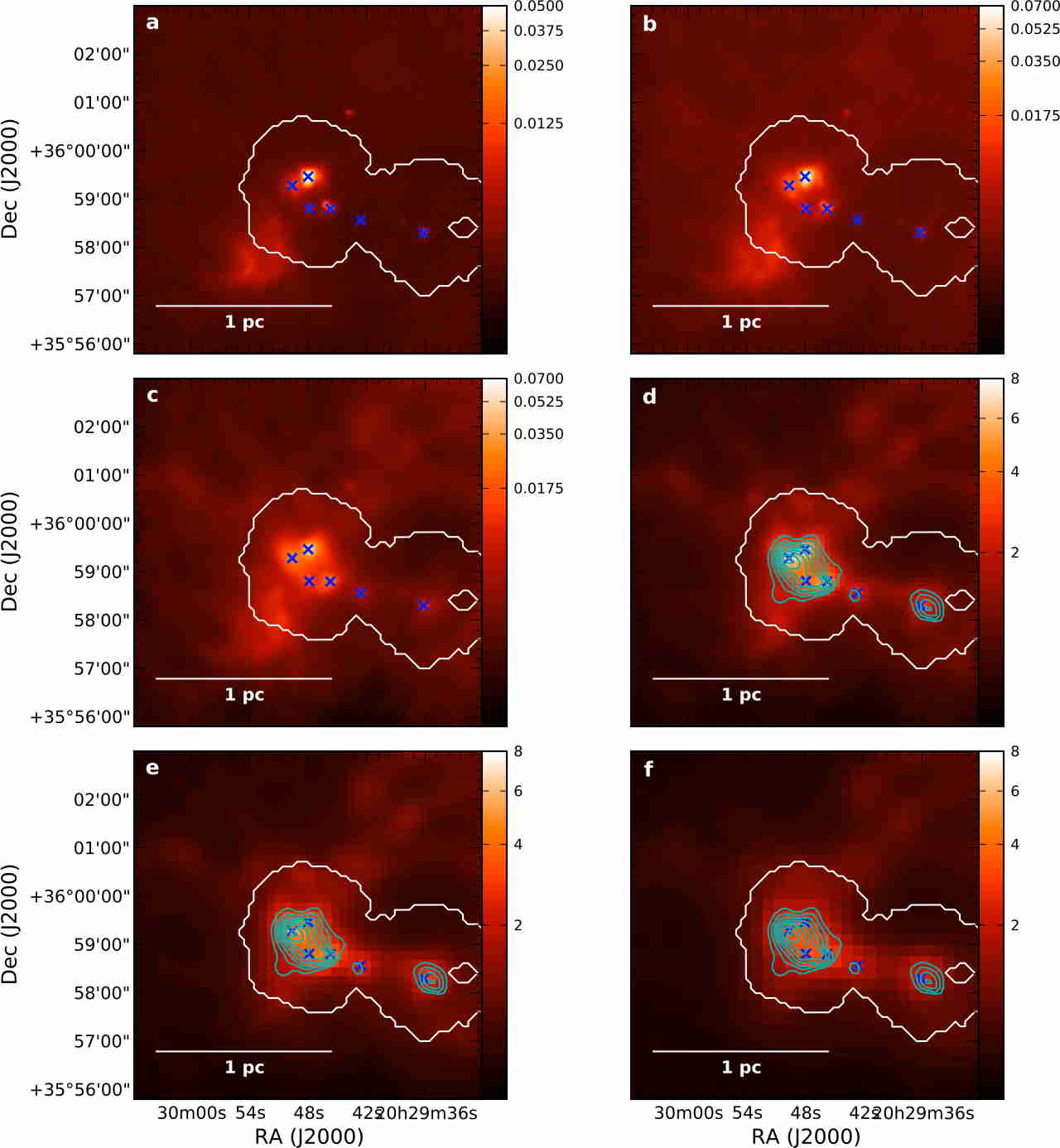}  
\caption{Same image layout as Figure~\ref{fig:hmsc} but for ISOSS J20298+3559. Cyan contours on the SPIRE panels (d, e, and f) show SCUBA 850\,$\mu$m: 0.08, 0.12, 0.20, 0.24, 0.28, 0.32, 0.36, 0.40 Jy beam$^{-1}$.}  
\end{figure*}

\clearpage

\begin{figure*}[h]  
\includegraphics[width=0.95\textwidth]{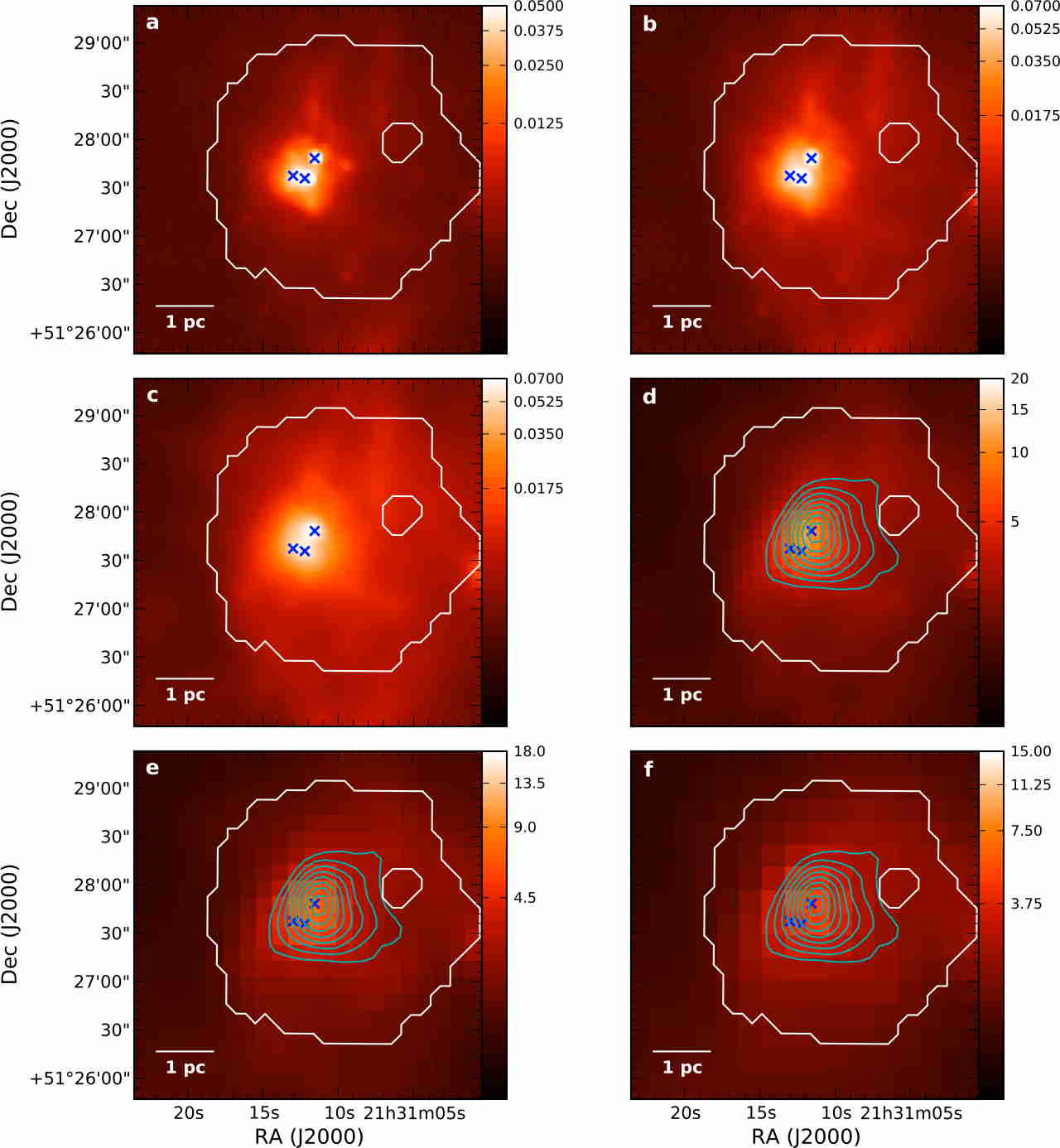}  
\caption{Same image layout as Figure~\ref{fig:hmsc} but for ISOSS J21311+5127. Cyan contours on the SPIRE panels (d, e, and f) show SCUBA 850\,$\mu$m: 0.12, 0.18, 0.24, 0.30, 0.36, 0.42, 0.48, 0.54, 0.60 Jy beam$^{-1}$.}  
\end{figure*}

\clearpage

\begin{figure*}[h]  
\includegraphics[width=0.95\textwidth]{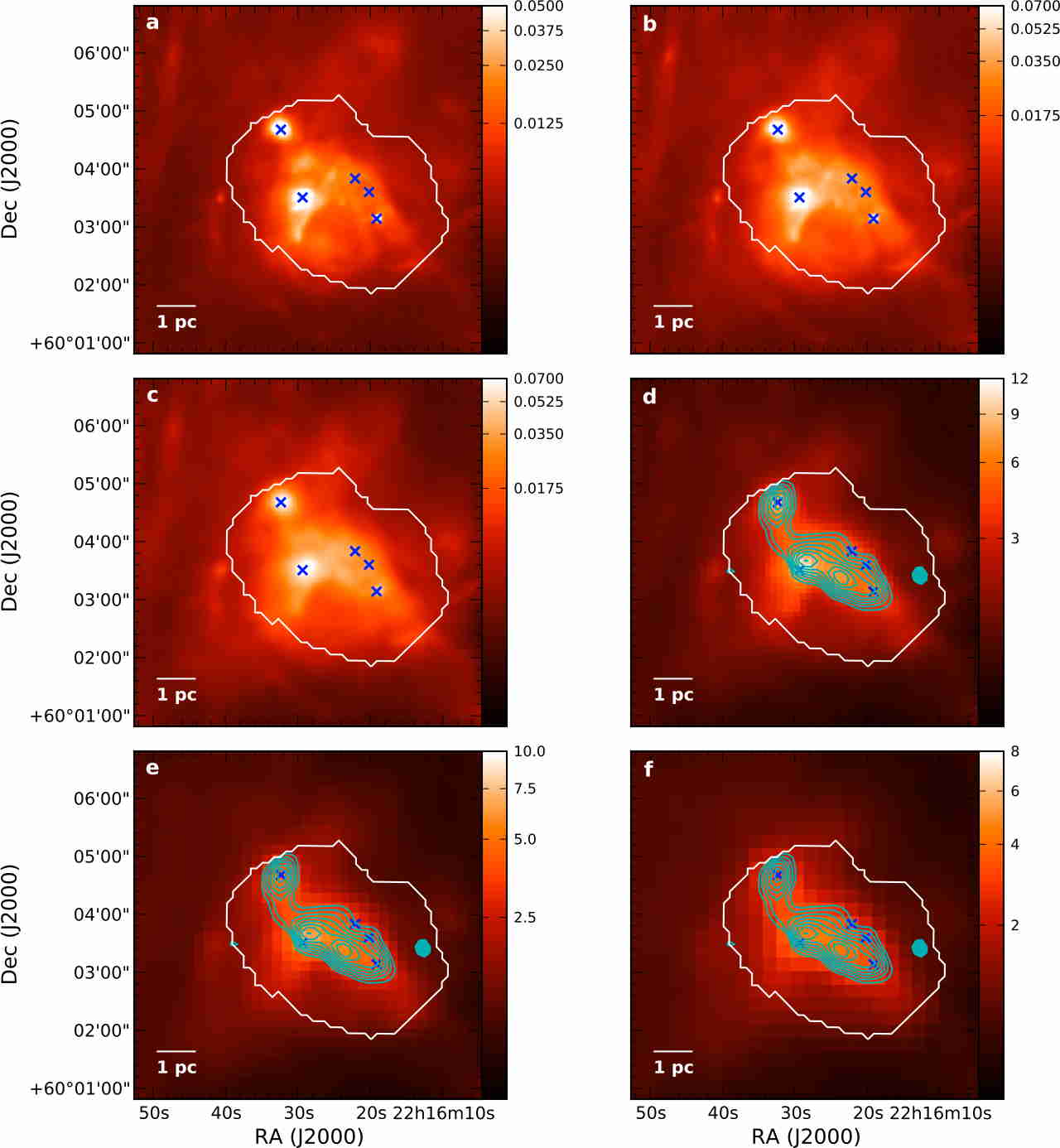}  
\caption{Same image layout as Figure~\ref{fig:hmsc} but for ISOSS J22164+6003. Cyan contours on the SPIRE panels (d, e, and f) show SCUBA 850\,$\mu$m: 0.08, 0.12, 0.20, 0.24, 0.28, 0.32, 0.36, 0.40 Jy beam$^{-1}$.}  
\end{figure*}

\clearpage

\begin{figure*}[h]  
\includegraphics[width=0.95\textwidth]{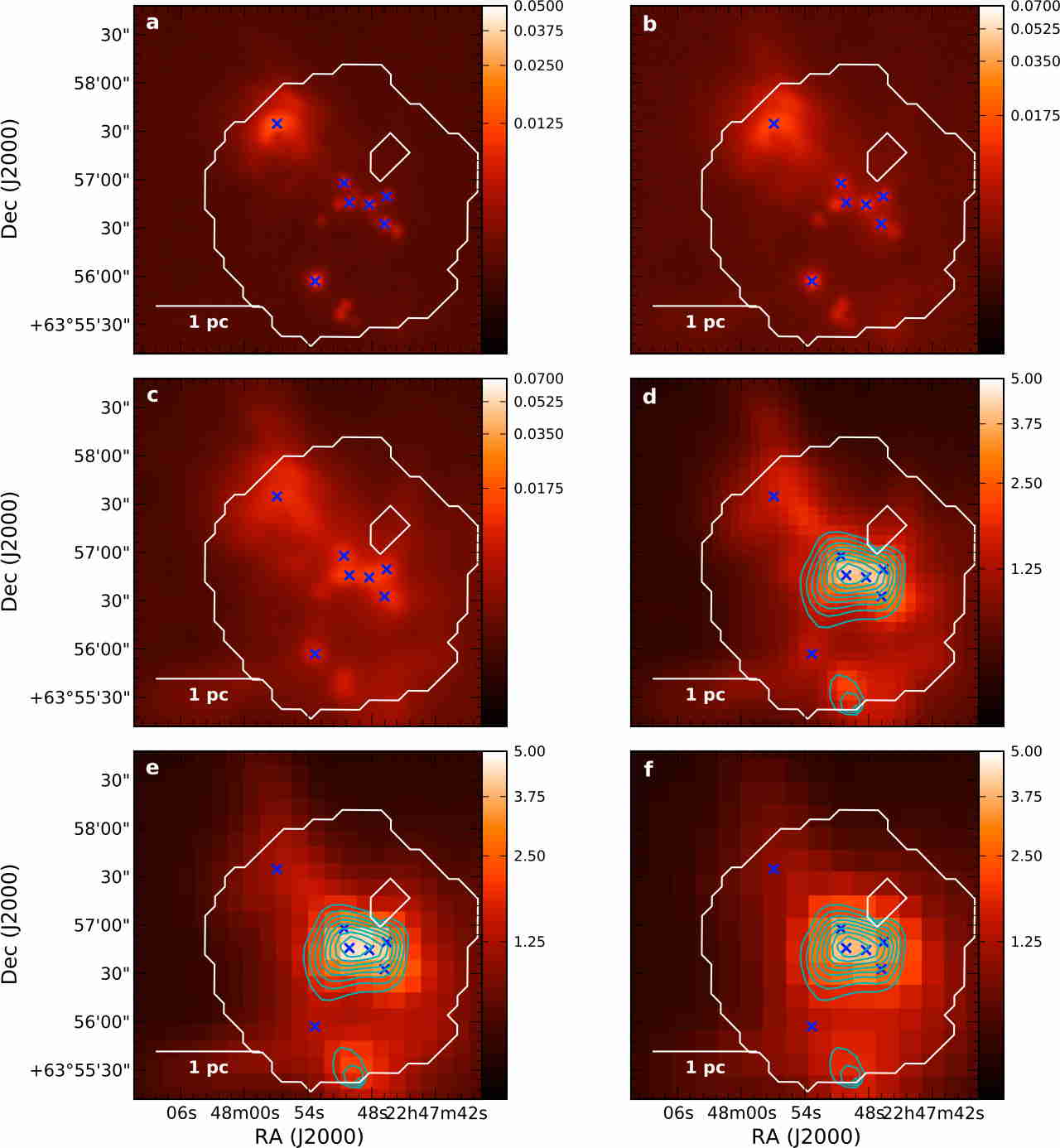}  
\caption{Same image layout as Figure~\ref{fig:hmsc} but for ISOSS J22478+6357. Cyan contours on the SPIRE panels (d, e, and f) show SCUBA 850\,$\mu$m: 0.11, 0.17, 0.22, 0.28, 0.34, 0.39, 0.45, 0.50, 0.56 Jy beam$^{-1}$.}  
\end{figure*}

\clearpage

\begin{figure*}[h]  
\includegraphics[width=0.95\textwidth]{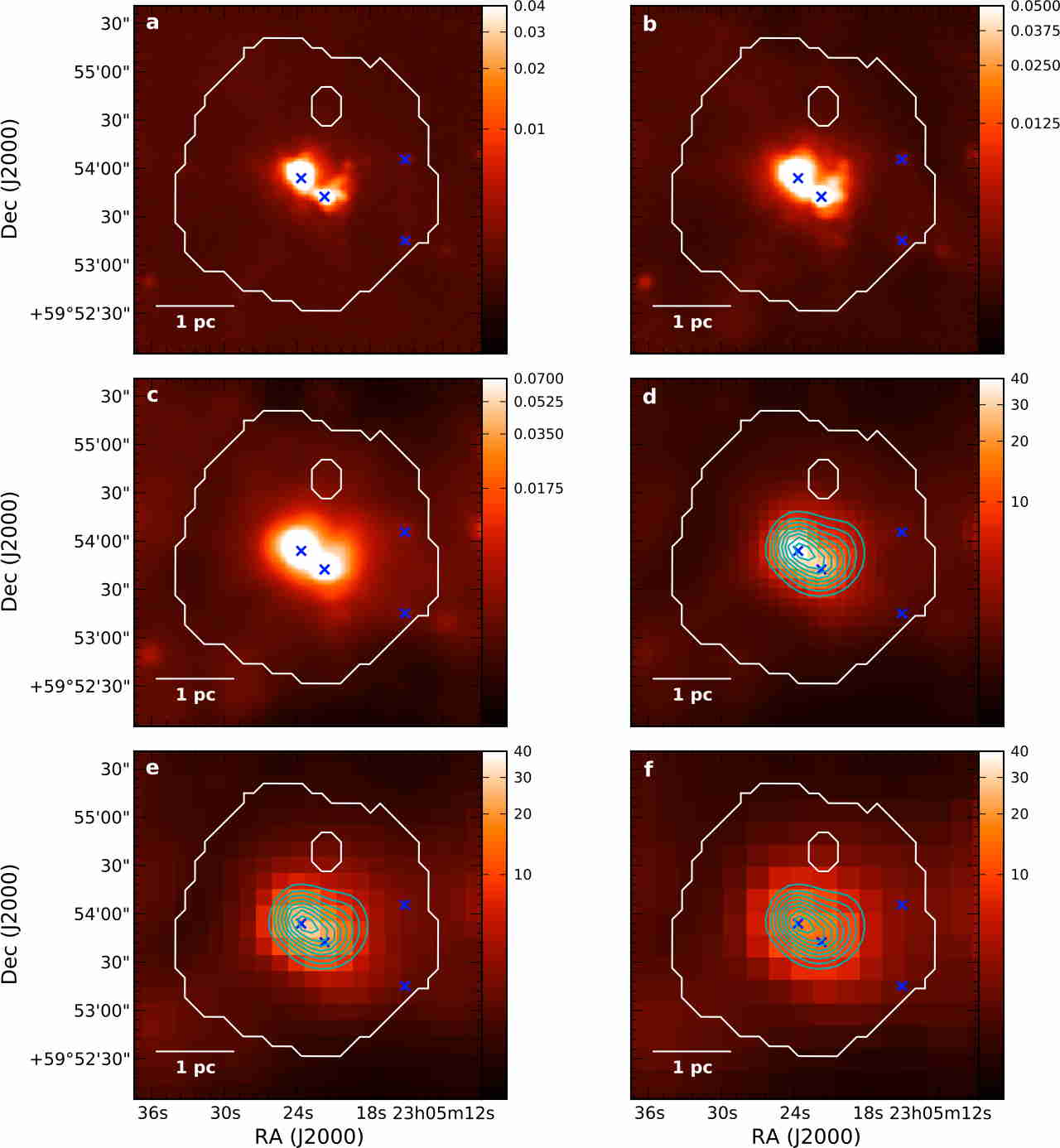}  
\caption{Same image layout as Figure~\ref{fig:hmsc} but for ISOSS J23053+5953. Cyan contours on the SPIRE panels (d, e, and f) show SCUBA 850\,$\mu$m: 0.4, 0.6, 0.8, 1.0, 1.2, 1.4, 1.6, 1.8, 2.0 Jy beam$^{-1}$.}  
\end{figure*}

\clearpage

\begin{figure*}[h]  
\includegraphics[width=0.95\textwidth]{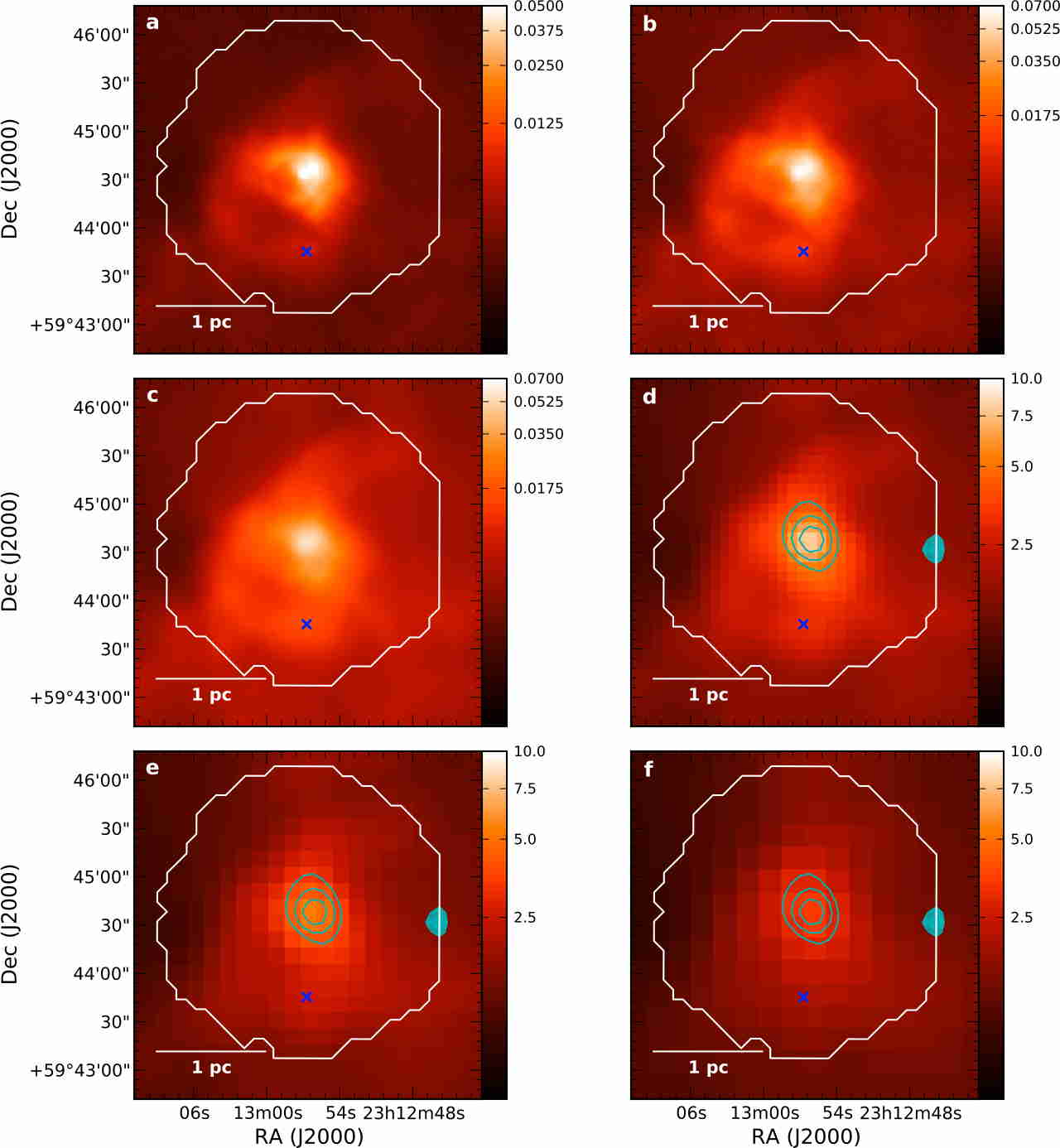}  
\caption{Same image layout as Figure~\ref{fig:hmsc} but for ISOSS J23129+5944. Cyan contours on the SPIRE panels (d, e, and f) show SCUBA 850\,$\mu$m: 0.14, 0.21, 0.28, 0.35, 0.42, 0.49, 0.56, 0.63, 0.70 Jy beam$^{-1}$.}  
\end{figure*}

\clearpage

\begin{figure*}[h]  
\includegraphics[width=0.95\textwidth]{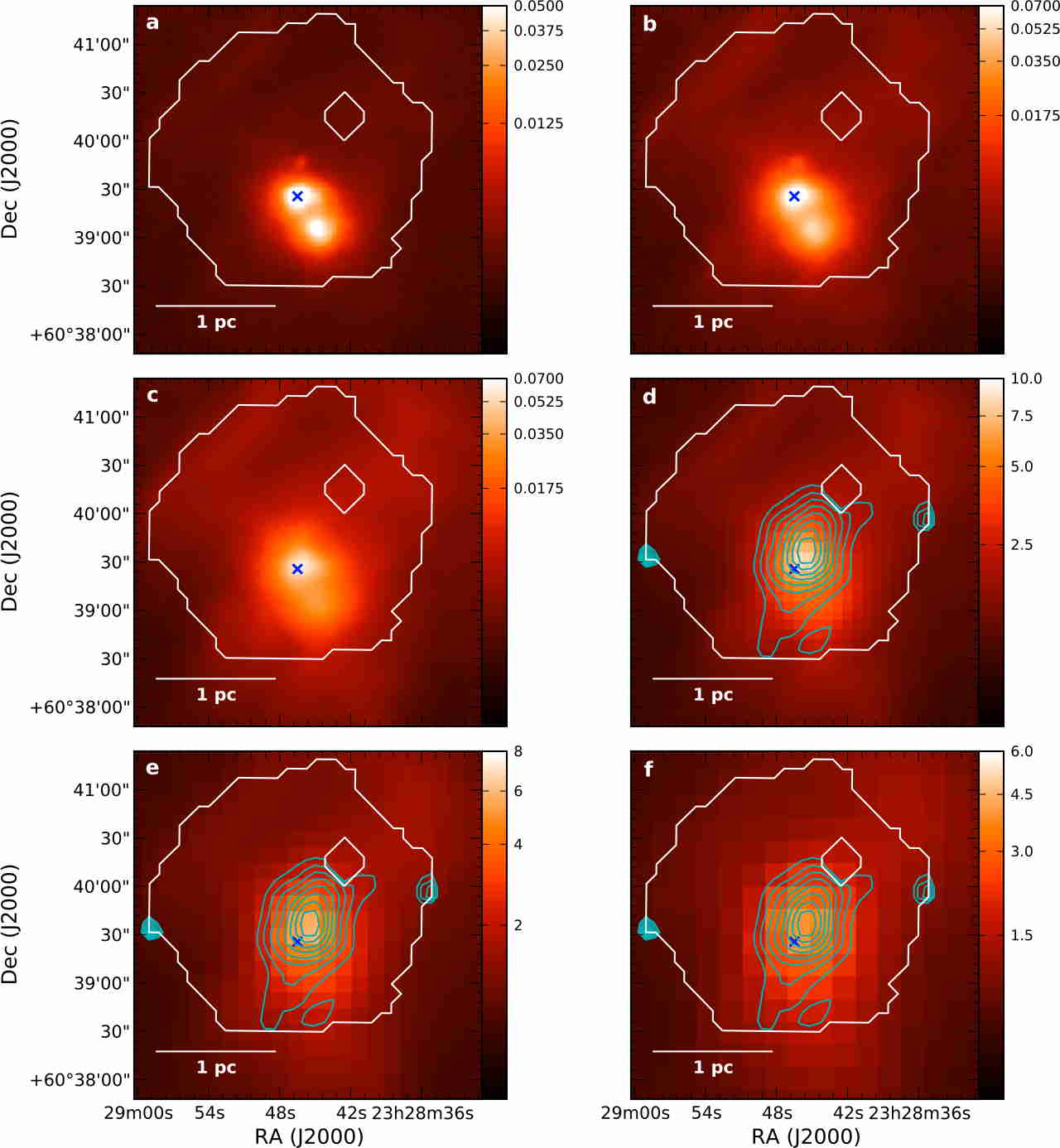}  
\caption{Same image layout as Figure~\ref{fig:hmsc} but for ISOSS J23287+6039. Cyan contours on the SPIRE panels (d, e, and f) show SCUBA 850\,$\mu$m: 0.08, 0.12, 0.20, 0.24, 0.28, 0.32, 0.36, 0.40 Jy beam$^{-1}$.}  
\end{figure*}

\clearpage

\section{Core catalog}
\longtab{1}{

\begin{longtable}{lllrrrrcccl}
\caption{Point sources. \label{pointsrc}}  \\
\hline \hline
ID & RA (J2000) & Dec (J2000) & $S_{24}$ & $S_{70}$ & $S_{100}$ & $S_{160}$ & $T_{dust}$ & Luminosity & Mass & Remark \\
number & ($^{h}$:$^{m}$:$^{s}$) & ($^{\circ}$:$^{'}$:$^{''}$) & (Jy) & (Jy) & (Jy) & (Jy) & (K) &($\lsun$) & ($\msun$) & \\
\hline
\endfirsthead
\caption{continued.} \\
\hline\hline
ID & RA (J2000) & Dec (J2000) & $S_{24}$ & $S_{70}$ & $S_{100}$ & $S_{160}$ & $T_{dust}$ & Luminosity & Mass & Remark\\
number & ($^{h}$:$^{m}$:$^{s}$) & ($^{\circ}$:$^{'}$:$^{''}$) & (Jy) & (Jy) & (Jy) & (Jy) & (K) &($\lsun$) & ($\msun$) & \\
\hline
\endhead
\hline
\endfoot
\hline
HMSC07029-1215 & & & & & & & & & & \\
\hline
  1 & 07:04:57.0 & -12:16:53 &       0.04 &        0.48 &        0.85 &        1.88 &     22 &   0.5 & 0.05 &  \\ 
  2 & 07:04:57.9 & -12:16:47 &       0.04 &        0.14 &        0.20 &        0.82 &     20 &   0.1 & 0.02 &  \\ 
  3 & 07:05:00.5 & -12:16:41 &       0.05 &        0.41 &        0.95 &        1.67 &     22 &   0.4 & 0.05 &  \\ 
  4 & 07:05:03.4 & -12:16:31 &   --   &        0.27 &        1.27 &        3.28 &     18 &   0.7 &   0.2 & \\ 
  5 & 07:05:05.8 & -12:16:49 &       0.02 &        0.39 &        0.54 &        0.93 &     24 &   0.3 & 0.02 &  \\ 
  6 & 07:05:10.6 & -12:21:33 &       0.10 &        0.29 &        0.79 &        1.23 &     21 &   0.3 & 0.04 &  \\ 
  7 & 07:05:10.8 & -12:18:57 &   --   &        9.44 &       31.02 &       52.48 &     20 &           13 &            2 & \\
\hline
\end{longtable}
\tablefoottext{a}{IRAS 25~$\mu$m flux.}

\tablefoottext{b}{MSX 21~$\mu$m flux.}

\tablefoottext{c}{Counterpart exists, but bad fit to the psf.}

}

\end{document}